\documentclass[reprint, showkeys, amsmath, amssymb, aps, prd]{revtex4-2}

\usepackage{graphicx}
\usepackage{dcolumn}
\usepackage{bm}
\usepackage[precent]{overpic}
\usepackage{multirow}
\usepackage{diagbox}
\usepackage{appendix}
\usepackage{lipsum}
\usepackage[unicode=true]{hyperref}
\hypersetup{colorlinks=true, linkcolor=blue, anchorcolor=blue, citecolor=blue}

\newcommand{\tableset}{
\centering \setlength{\abovecaptionskip}{6pt} \setlength{\belowcaptionskip}{6pt} \setlength{\tabcolsep}{6pt} \renewcommand{\arraystretch}{1.0}
}

\begin{document}

\preprint{APS/123-QED}

\title{\boldmath Observation of the $Y(4230)$ and a new structure in $e^+e^- \rightarrow K^+K^-J/\psi$}
\author{
\begin{small}
\begin{center}
M.~Ablikim$^{1}$, M.~N.~Achasov$^{10,b}$, P.~Adlarson$^{68}$, M.~Albrecht$^{4}$, R.~Aliberti$^{28}$, A.~Amoroso$^{67A,67C}$, M.~R.~An$^{32}$, Q.~An$^{64,50}$, X.~H.~Bai$^{58}$, Y.~Bai$^{49}$
% ...
\\
\vspace{0.2cm}
(BESIII Collaboration)\\
\vspace{0.2cm} {\it
$^{1}$ Institute of High Energy Physics, Beijing 100049, People's Republic of China\\
% ...
$^{4}$ Bochum Ruhr-University, D-44780 Bochum, Germany\\
% ...
$^{10}$ G.I. Budker Institute of Nuclear Physics SB RAS (BINP), Novosibirsk 630090, Russia\\
% ...
$^{28}$ Johannes Gutenberg University of Mainz, Johann-Joachim-Becher-Weg 45, D-55099 Mainz, Germany\\
% ...
$^{32}$ Liaoning Normal University, Dalian 116029, People's Republic of China\\
% ...
$^{49}$ Southeast University, Nanjing 211100, People's Republic of China\\
$^{50}$ State Key Laboratory of Particle Detection and Electronics, Beijing 100049, Hefei 230026, People's Republic of China\\
% ...
$^{58}$ University of Jinan, Jinan 250022, People's Republic of China\\
% ...
$^{64}$ University of Science and Technology of China, Hefei 230026, People's Republic of China\\
% ...
$^{67}$ University of Turin and INFN, (A)University of Turin, I-10125, Turin, Italy; (B)University of Eastern Piedmont, I-15121, Alessandria, Italy; (C)INFN, I-10125, Turin, Italy\\
$^{68}$ Uppsala University, Box 516, SE-75120 Uppsala, Sweden\\
% ...
$^{b}$ Also at the Novosibirsk State University, Novosibirsk, 630090, Russia\\
% ...
}
\end{center}
\vspace{0.4cm}
\end{small}
}
\noaffiliation{}
\date{\today}

\begin{abstract}
	The cross sections of $e^+e^- \rightarrow K^+K^-J/\psi$ at center-of-mass energies from 4.127 to 4.600~GeV are measured based on 15.6 fb$^{-1}$ data collected with the BESIII detector operating at the BEPCII storage ring. Two resonant structures are observed in the line shape of the cross sections. The mass and width of the first structure are measured to be ($4225.3\pm2.3\pm21.5$) MeV and ($72.9\pm6.1\pm30.8$)~MeV, respectively. They are consistent with those of the established $Y(4230)$. The second structure is observed for the first time with a statistical significance greater than 8$\sigma$, denoted as  $Y(4500)$. Its mass and width are determined to be ($4484.7\pm13.3\pm24.1$) MeV and ($111.1\pm30.1\pm15.2$) MeV, respectively.  The first presented uncertainties are statistical and the second ones are systematic. The product of the electronic partial width with the decay branching fraction $ \Gamma(Y(4230)\to e^+ e^-) \mathcal{B}(Y(4230) \to K^+ K^- J/\psi)$ is reported.
\end{abstract}

\maketitle

\section{Introduction}
Recent discoveries of charmonium-like states expand our perspective of the hadron spectrum around the $\tau$-charm energy region, and provide excellent laboratories to study perturbative and non-perturbative strong interaction. The dynamic, however, is more complex than the one of conventional mesons due to possible additional degrees of freedom. Therefore, their nature has not yet been established despite many different speculations about them being hybrids, tetra-quarks, molecules, cusp effects, and so on~\cite{BRAMBILLA20201}.

Among the exotic states, the $Y(4230)$ state, previously called $Y(4260)$, is the first observed vector charmonium-like state. It was discovered in $e^+e^- \to \pi^+ \pi^- J/\psi$ channel by the BaBar Collaboration using initial-state-radiation (ISR) technique~\cite{PhysRevLett.95.142001}, and confirmed by CLEO~\cite{PhysRevD.74.091104} and Belle~\cite{PhysRevLett.99.182004}. In addition, the $Y(4230)$ state was observed in various modes by BESIII, including $e^+e^- \to \pi \pi J/\psi$~\cite{PhysRevLett.118.092001, PhysRevD.102.012009, BESIII:2022jsj}, $\pi^+ \pi^- h_c$~\cite{PhysRevLett.118.092002}, $\pi \pi \psi(2S)$~\cite{PhysRevD.96.032004, PhysRevD.97.052001, PhysRevD.104.052012}, $\omega \chi_{c0}$~\cite{PhysRevLett.114.092003, PhysRevD.99.091103}, and $\pi^+D^0 D^{*-} + c.c.$~\cite{PhysRevLett.122.102002}. It was also found to decay into $X(3872)$ via radiative transition~\cite{PhysRevLett.122.232002} and to $Z_c(3900)$ via pion transition~\cite{PhysRevD.102.012009}. A better understanding of its internal structure and quark components is crucial, and it will also be helpful to understand the series of the tetra-quark candidates~\cite{doi:10.1142/S0217751X21501268}.

Two recent measurements on the cross sections of $e^+ e^- \to \eta J/\psi$~\cite{PhysRevD.102.031101} and $e^+ e^- \to \eta' J/\psi$~\cite{PhysRevD.101.012008} indicate a considerable strange quark component in $Y(4230)$. This hypothesis was analyzed in Ref.~\cite{PhysRevD.105.L031506}. However, a concrete conclusion is still missing due to the large uncertainty. Thus a measurement of $Y(4230) \to K \bar{K} J/\psi$ is important to clarify the puzzle by comparison with the results of $Y(4230) \to \pi \pi J/\psi$. The first evidence for $Y(4230) \to K^+K^-J/\psi$ was found by CLEO~\cite{PhysRevLett.96.162003}. Later, the cross sections of $e^+e^- \to K^+ K^- J/\psi$ at center-of-mass (c.m.) energies between threshold and $6.0$ GeV were measured for the first time via ISR process~\cite{PhysRevD.77.011105} and was updated ~\cite{PhysRevD.89.072015} by Belle. No significant signal was observed in both measurements, and an upper limit $\mathcal{B}(Y(4230) \to K^+K^-J/\psi)\Gamma(Y(4230) \to e^+e^-)< 1.7$ eV at 90\% confidence level was obtained. Here and following the natural unit system is adopted, i.e., $\hbar=c=1$. Recently, BESIII has measured the cross sections of $e^+e^- \to K\bar{K}J/\psi$ at c.m. energies from $4.189$ to $4.600$~GeV with an integrated luminosity ($\mathcal{L}_{int}$) of $4.7$ fb$^{-1}$~\cite{PhysRevD.97.071101}, and no significant signal of the $Y(4230)$ is observed too. In Refs.~\cite{PhysRevD.77.011105, PhysRevD.89.072015, PhysRevD.97.071101}, a structure around $\sqrt{s} = 4.5$ GeV was seen, though the statistics is too small to identify its properties. The structure with higher mass is consistent with the calculation by Ref.~\cite{PhysRevD.99.114003}, which suggests a conventional charmonium state $\psi(4500)$ with mass of $4489-4529$ MeV by the 5S-4D mixing scheme. It is also consistent with a virtual state predicted in Ref.~\cite{Dong:2021juy}, where the mass of the heavy-antiheavy hadronic molecule is calculated to be $4483-4503$~MeV that is just below the $D_s \bar{D}_{s1}$ threshold. Meanwhile, Ref.~\cite{PhysRevD.73.094510} predicts an exotic state with $c\bar{c}s\bar{s}$ component in quenched lattice quantum chromodynamics with exact chiral symmetry, and its mass is predicted to be ($4450\pm100$) MeV. 

In this Letter, we present an updated measurement of the Born cross sections of $e^+e^- \to K^+K^-J/\psi$ at the c.m. energies from 4.127 to 4.600 GeV, using data samples composed of twenty-eight c.m. energy points~\cite{Ablikim_2016, BESIII:2020eyu} with $\mathcal{L}_{int}$ = 15.6~fb$^{-1}$~\cite{Ablikim_2015, BESIII:2022xii}, collected at the BESIII detector operating at the BEPCII storage ring~\cite{BESIII:2020nme}. These samples are about three times in luminosity compared with that used in Ref.~\cite{PhysRevD.97.071101},  and they overlap at eleven energy points. The added points are mainly around the $Y(4230)$ mass region, except two points around $4.5$~GeV that are crucial for determining the line shape of $Y(4500)$. Furthermore, although previously in Ref.~\cite{PhysRevD.97.071101} a full reconstruction method is applied, considering the low efficiency of kaon reconstruction with low momentum, a new partial reconstruction method is applied improving significantly the efficiency.

\section{Data Analysis}
The BESIII detector is described in detail in Ref.~\cite{ABLIKIM2010345, BESIII:2020nme}. The {\sc geant4-}based~\cite{AGOSTINELLI2003250} Monte Carlo (MC) simulation software framework {\sc boost}~\cite{2005-0159}, which consists of detector geometry and its response, is used to produce large simulated event samples. These samples are used to optimize the event selection criteria, determine the detection efficiency, evaluate the ISR correction factor, and estimate background contributions. The signal events are generated at each c.m. energy, where signal events include $e^+e^- \to K^+ K^- J/\psi$ (phase space (PHSP) model ), $f_0(980)J/\psi$, and $f_2(1270) J/\psi$. Both $f_0(980)$ and $f_2(1270)$ decay into $K^+K^-$, and $J/\psi$ decays into one lepton pair ($\mu^+\mu^-/e^+e^-$). The simulation includes the beam energy spread and ISR in the $e^+e^-$ annihilation modelled with the generator {\sc kkmc}~\cite{PhysRevD.63.113009, 2014-08-01-083001} and {\sc evtgen}~\cite{2007-0205, LANGE2001152}, where the angular distributions according to spin and parity are considered by the specific models implemented in the generator. The final state radiation effect associated with leptons is handled by the {\sc photos} package~\cite{Golonka2006}. The potential backgrounds are estimated by the inclusive MC sample, that includes the production of open charm processes, the ISR production of vector charmonium(-like) states, and the continuum processes incorporated in {\sc kkmc}. All particle decays are modelled with {\sc evtgen} using branching fractions either taken from the Particle Data Group~\cite{Zyla:2020zbs}, when available, or otherwise estimated with {\sc lundcharm}~\cite{PhysRevD.62.034003, YANG-Rui-Ling:61301}.

A pair of leptons ($e^+e^- / \mu^+\mu^-$) and at least one kaon is required for a signal candidate. For each track, the polar angle $\theta$, with respect to the symmetry axis of the multiple drift chamber, must satisfy $\vert\cos\theta\vert<0.93$, and the point of the closest approach to the $e^+e^-$ interaction point must be less than 10.0 cm in the beam direction and 1.0 cm in the plane perpendicular to the beam direction. Each charged track with momentum larger than 1.0~GeV in the laboratory frame is assumed to be a lepton. The ratio of the energy deposited in the electromagnetic calorimeter over the momentum of each lepton candidate is required to be greater than 0.8 for electrons and less than 0.8 for muons. For particle identification (PID), the energy loss in the main drift chamber and the time measured with the time-of-flight system are combined to calculate the confidence levels ($C.L.$) with kaon and pion hypotheses, and the confidence level is required to satisfy $C.L.(K) > 0.001$ and $C.L.(K) > C.L.(\pi)$ for each kaon candidate. In order to improve resolution and suppress backgrounds, a vertex fit and a kinematic fit with one constraint on the mass of the missing kaon are performed. If there is more than one kaon track candidate, the selected one has the smallest $\chi^2_F$, defined as the sum of the $\chi^2$ of vertex and kinematic fits and required to be less than $20$. To remove radiative Bhabha background, where the radiative photon would convert into an $e^+e^-$ pair, all pairs of oppositely charged tracks must have an opening angle $\cos(\theta_{open}) < 0.98$ for the $e^+e^-$ mode. For the $\mu^+ \mu^-$ mode, the penetration depth for one of the muon candidate in the muon counter is required to be greater than $40$ cm to remove hadron backgrounds.

\begin{figure}[!htbp]
\centering
\includegraphics[scale=.4]{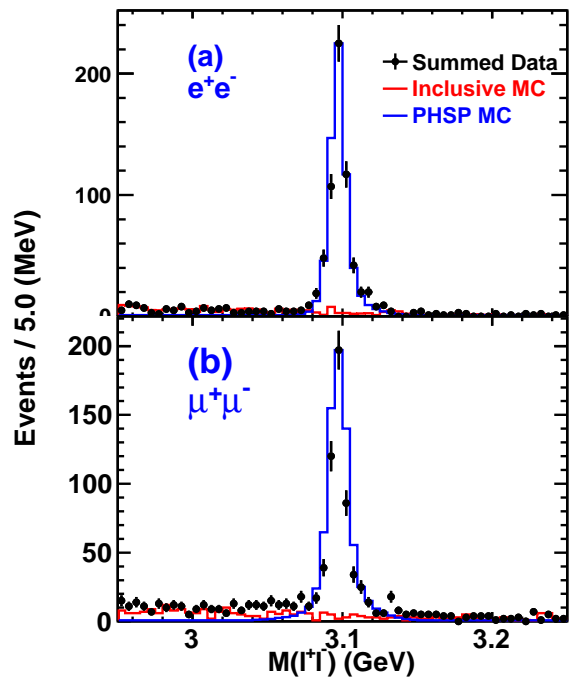}
\caption{The invariant mass distributions of the lepton pairs from different $J/\psi$ decay modes. (a) is $e^+e^-$ mode, (b) is $\mu^+ \mu^-$ mode, where the dots with error bar indicate data, the blue line histograms indicate PHSP signal MC sample, the red line histograms indicate backgrounds from inclusive MC sample.}
\label{Fig-dilepton-mass}
\end{figure}

After applying the aforementioned event selection, the invariant mass distributions of $e^+e^-$ and $\mu^+ \mu^-$ are shown in Fig.~\ref{Fig-dilepton-mass} with all data samples. The signal region of $J/\psi$ is defined as a region with 2.5 times of the mass resolution, and the side band regions are taken as the same size as the signal with 0.01 GeV gap from the signal region. The distributions of $M(\ell^+\ell^-)$ at all the energies are presented in Appendix~\ref{subs_a1}, where $\ell^+\ell^-$ is either $e^+e^-$ or $\mu^+\mu^-$. There are no peaking backgrounds based on the study of inclusive MC samples. The yields of signals ($N^{obs}$) are obtained from background subtraction and the corresponding uncertainties are estimated by the profile likelihood method~\cite{ROLKE2005493}.

\begin{figure}[!htbp]
\centering
\includegraphics[scale=.4]{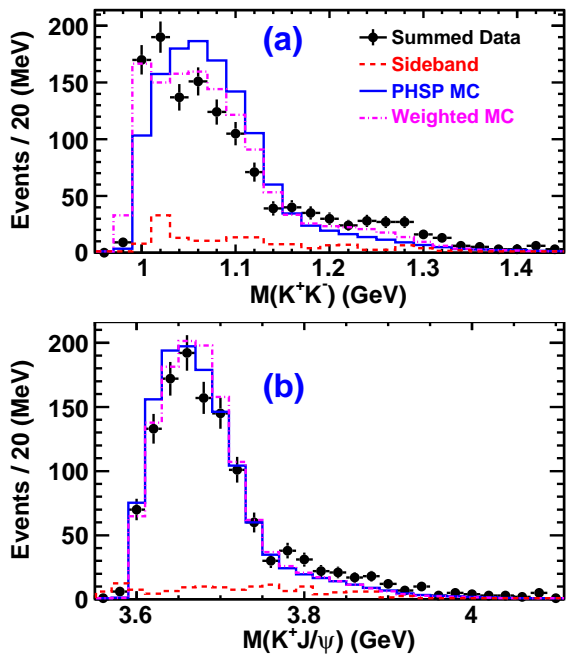}
\caption{(a) and (b) are distributions of $M(K^+K^-)$ and $M(K^+J/\psi$), respectively. The black dots with error bars indicate data from the $J/\psi$ signal region with all data samples, the red dashed curves indicate data from $J/\psi$ side band regions, the blue solid curves indicate PHSP signal MC sample, and the yank dashed-dot curves indicated the weighted signal MC sample.}
\label{Fig-intermediate-states}
\end{figure}

To explore potential intermediate states, the invariant mass distributions of $K^+ K^-$ and $K^+ J/\psi$ are shown in Fig.~\ref{Fig-intermediate-states}, with all data samples, PHSP signal MC samples, and the weighted signal MC samples, where PHSP and weighted signal MC samples are both normalized to data. There is no obvious structure in the $K J/\psi$ invariant mass distributions. The distributions of invariant mass of $K^+K^-$ show signs of mesons $f_0(980)$ and $f_2(1270)$. These distributions roughly match with the theoretical calculation at $\sqrt{s}$ = 4.23 and 4.26 GeV~\cite{PhysRevD.102.016019}, while the corresponding predictions are still missing at higher energy regions. We extract the contributions of different components (PHSP, $f_0(980)$, $f_2(1270)$) by fitting to the data samples with large signal yields ($N^{obs}>55$), then expanding the results to the smaller samples by linear interpolation. The efficiency of event selection ($\varepsilon$) is calculated by the sum of weighted efficiencies of these components.

\section{Born Cross Section and Resonant Parameter}
The Born cross section of $e^+e^- \to K^+K^-J/\psi$ at each energy is obtained by
\begin{equation}
\sigma^B(\sqrt{s}) = \frac{N^{obs}} {\mathcal{L}_{int} \cdot \varepsilon \cdot (1+\delta)_{ISR} \cdot \frac{1}{|1-\Pi|^2} \cdot \mathcal{B}(J/\psi \rightarrow \ell^+\ell^-)},
\label{Eq-cross-section}
\end{equation}
where $\frac{1}{|1-\Pi|^2}$ is the vacuum polarization factor calculated by QED~\cite{Actis2010}, $\mathcal{B}(J/\psi \rightarrow \ell^+\ell^-)$ is the branching fraction that is quoted as $(11.93\pm0.05)\%$~\cite{Zyla:2020zbs}, $(1+\delta)_{ISR}$ is the radiative correction factor obtained by an iteration method, described in details in Ref.~\cite{sun2021iterative}. The results of Born cross sections are listed in Appendix~\ref{subs_a2}. Compared with the previous BESIII measurements~\cite{PhysRevD.97.071101},  the Born cross sections at the same c.m. energies are consistent, and the precision is slightly improved.  The observation of two clear structures in the distribution of the cross sections is due to more data samples used.

\begin{figure}[!htbp]
\centering
\includegraphics[scale=.4]{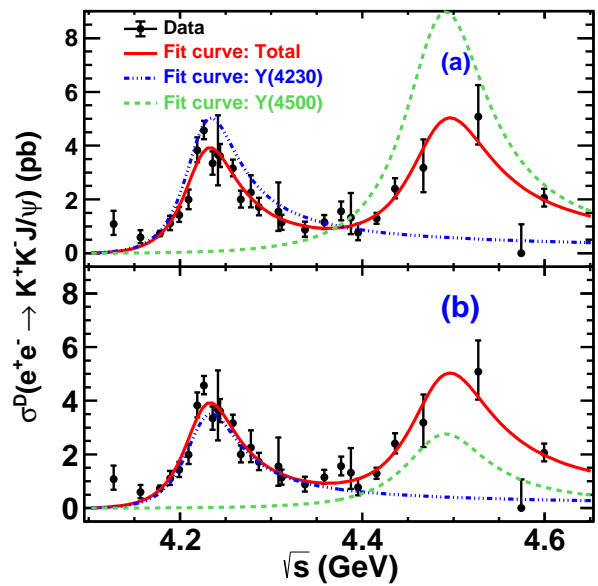}
\caption{Dressed cross sections of $e^+e^- \to K^+K^-J/\psi$, indicated by error bars with only statistical uncertainties. They are fitted by a coherent sum of two Breit-Wigner functions indicated by red solid curves, and the blue and pink dashed curves are the amplitudes describing the resonances $Y(4230)$ and $Y(4500)$, respectively. (a) corresponds to solution I, (b) corresponds to solution II.}
\label{Fig-resonance}
\end{figure}

A maximum likelihood fit is applied to the dressed cross sections ($\sigma^D(\sqrt{s})$, including vacuum polarization effects) of $e^+e^- \rightarrow K^+K^-J/\psi$ to determine the parameters of the resonant structures, and the dressed cross sections are shown in Fig.~\ref{Fig-resonance}. The likelihood is constructed taking the fluctuations of the number of signal events into account. Its definition is described in Appendix~\ref{subs_a3}. The fit function is parameterized as a coherent sum of two relativistic Breit-Wigner ($BW$) functions
\begin{equation}
\sigma^D(\sqrt{s}) = |BW_1(\sqrt{s}) \cdot e^{i\varphi}+ BW_2(\sqrt{s})| ^2,
\end{equation}
where $\varphi$ is the relative phase angle, and
\begin{equation}\label{Eq-func-BW}
BW(\sqrt{s}) = \frac{M}{\sqrt{s}} \cdot \frac{\sqrt{12 \pi \Gamma_{ee} \Gamma_{tot} \mathcal{B}}}{s-M^2 + iM\Gamma_{tot}} \cdot \sqrt{\frac{\Phi(\sqrt{s})}{\Phi(M)}},
\end{equation}
where $M,~\Gamma_{tot},~\Gamma_{ee}$, and $\mathcal{B}$ are the mass, full width, electronic partial width (whose definition includes vacuum polarization effects, that is why the dressed cross sections are fitted to rather than the Born cross sections), and branching fraction of corresponding resonance, respectively. $\Phi(\sqrt{s}) = \int \int \frac{1}{(2\pi)^3 32 (\sqrt{s})^3} dm_{12}^2 dm_{23}^2$ is the three-body phase space~\cite{Zyla:2020zbs}, where $m_{ij}$ is the invariant mass of particles i and j. The fitting curve is shown in Fig.~\ref{Fig-resonance}, and the fit quality is estimated to be $\chi^2/{\rm n.d.f}$ = 37.45 / 21, where n.d.f is the number of degrees of freedom. The resonance with lower mass is consistent with the previously established $Y(4230)$. Its mass and width are determined to be $M(Y(4230)) = (4225.3\pm2.3)$ MeV and $\Gamma(Y(4230)) = (72.9\pm6.1)$ MeV. Since there is no observed state corresponding to the resonance with higher mass, we name it as $Y(4500)$. Its mass and width are determined to be $M(Y(4500)) = (4484.7\pm13.3)$ MeV, $\Gamma(Y(4500)) = (111.1\pm30.1)$ MeV, respectively. The statistical significance of $Y(4230)$ and $Y(4500)$ have been estimated to be 29$\sigma$ and 8$\sigma$, respectively, via the differences of the likelihood values and the degrees of freedom with and without considering the corresponding resonance by the Wilk's theorem~\cite{Wilks:1938dza}. Two solutions with equal goodness-of-fit qualities are found. The masses and total widths are unchanged in the two solutions, while the amplitudes vary significantly due to constructive and destructive interferences between $Y(4230)$ and $Y(4500)$, which are consistent with the mathematical expectation of multiple solutions~\cite{S0217751X11054589}. The products of the electron partial width and branching fraction, of the states $Y(4230)$ and $Y(4500)$, are listed in Table.~\ref{Tab-resonance}.

\begin{table}[!htbp]
	\tableset
	\renewcommand{\arraystretch}{1.2}
	\caption{Fitted parameters of the two resonant structures observed in the cross sections of $e^+e^- \to K^+K^-J/\psi$, where the first uncertainty is statistical and the second one is systematic.}
	\label{Tab-resonance}
	\resizebox{0.48\textwidth}{!}{
	\begin{tabular}{cc cc}
	\hline\hline
	~	&	Parameters      				&	Solution I     			&	Solution II					~\\ \hline
	\multirow{3}{4em}{\centering $Y(4230)$}	&$M$(MeV)	&\multicolumn{2}{c}{$4225.3\pm2.3\pm21.5$}		~\\
	~	&	$\Gamma_{tot}$(MeV)  	&	\multicolumn{2}{c}{$72.9\pm6.1\pm30.8$}			~\\
	~	&	$\Gamma_{ee}\mathcal{B}$(eV)  	&	$0.42\pm0.04\pm0.15$ 	&	$0.29\pm0.02\pm0.10$	~\\ \hline
	\multirow{3}{4em}{\centering $Y(4500)$}	&$M$(MeV)    &\multicolumn{2}{c}{$4484.7\pm13.3\pm24.1$}		~\\
	~	&	$\Gamma_{tot}$(MeV) 		&	\multicolumn{2}{c}{$111.1\pm30.1\pm15.2$}		~\\
	~	&	$\Gamma_{ee}\mathcal{B}$(eV)	&	$1.35\pm0.14\pm0.07$ 	& 	$0.41\pm0.08\pm0.13$ 	~\\ \hline
	Phase angle	&	$\varphi$(rad)			&	$1.72\pm0.09\pm0.52$&	$5.49\pm0.35\pm0.58$ 	~\\ \hline\hline
	\end{tabular}
	}
\end{table}

The main sources of systematic uncertainties of the cross section measurements are: the integrated luminosity $\mathcal{L}_{int}$, the tracking and PID efficiency, the branching fraction of $J/\psi \to \ell^+\ell^-$, kinematic fit,  $(1 + \delta)_{ISR}$,  the intermediate structures of the $K^+K^-$ system, the resolution of $J/\psi$, and the requirement on the penetration depth in the muon counter. The $\mathcal{L}_{int}$ is measured with Bhabha events and the uncertainty is found to be 1.0\%~\cite{Ablikim_2015, BESIII:2022xii}. The differences between data and MC in the tracking and PID efficiencies are studied by using the process $e^+e^- \to K^+ K^- \pi^+ \pi^-$, and 2.5\% per charged kaon is quoted~\cite{PhysRevD.97.071101}. For tracking efficiency of charged leptons, the systematic uncertainty is studied by using the process $\psi(3686) \to \pi^+ \pi^- J/\psi (J/\psi \to \ell^+\ell^-)$, and 1.0\% uncertainty per lepton is quoted~\cite{PhysRevD.91.112005}. The uncertainty of the branching fraction of $J/\psi \to \ell^+\ell^-$ is quoted as 0.4\%~\cite{Zyla:2020zbs}. The uncertainty of kinematic fit is estimated by correcting the corresponding track parameters, and the difference between the efficiencies with (default) and without this correction is quoted as the relevant uncertainty~\cite{PhysRevD.87.012002}. Varied line shape of the input cross sections is constructed by connecting each nearby points with a smooth curve. The difference between the final cross sections with this new line shape and the nominal one is taken as the uncertainty of radiation correction. The uncertainty associated with the intermediate structures is estimated by weighting the PHSP MC samples according to the observed $M(K^+ K^-)$ distribution of data, and the difference between the two efficiencies of the two methods is quoted as the corresponding uncertainty. To estimate the uncertainty according to the difference in the resolutions between data and MC samples, the efficiency is re-obtained by smearing the resolution of the $J/\psi$ signal of the MC sample. The difference is quoted as the associated uncertainty. To consider the uncertainty of the criteria on the penetration depth in the muon counter, the difference between the final results with and without this criterion is adopted. The total uncertainties are calculated by summing all individual items in quadrature. They are energy dependent and vary from $8.6\%$ to $11.1\%$. All the systematic uncertainties, including the individual and total, are listed in Appendix~\ref{subs_a4}.

The systematic uncertainties for the parameters of resonances mainly come from c.m. energy measurements, the form and parameterization of the fit function, and the systematic uncertainties in the cross section measurements that will be discussed later. The c.m. energies were measured with $e^+e^- \rightarrow \mu^+\mu^-$ events and the uncertainties are determined correspondingly for different data samples~\cite{Ablikim_2016, BESIII:2020eyu}. The associated systematic uncertainty is estimated by varying the c.m. energies during the fit. A three-body PHSP shape for a non-resonant component is added to the two coherent $BW$ sum to estimate the uncertainty of the cross section description in the fit, which turns out to be negligible. The uncertainty of the formalism of the full width is estimated by replacing the $\Gamma_{tot}$ with $\Gamma = \Gamma_{tot} \frac{\Phi(\sqrt{s})}{\Phi(M)}$ in the denominator of Eq. (\ref{Eq-func-BW}), where $\Gamma_{tot}$ is the nominal width of the resonance. The systematic uncertainties due to cross section measurements can be divided into two categories. The first one is uncorrelated among the different c.m. energy points including kinematic fit, radiation correction, intermediate structures, and resolution of $J/\psi$. The associated uncertainty is estimated by considering them while doing the fit and comparing to the results obtained only considering statistical uncertainties of the cross sections. The second category of the systematic uncertainties is correlated and common for all data samples (5.1\%), therefore only affects the $\Gamma_{ee}\mathcal{B}$. All of these uncertainties on the parameters of resonances are listed in Appendix~\ref{subs_a4}.

\section{Summary and Discussion}
In summary, the Born cross sections of $e^+e^- \to K^+K^- J/\psi$ at c.m. energy from 4.127 to 4.600 GeV are measured with a new partial reconstruction method and larger data samples compared with Ref~\cite{PhysRevD.97.071101}. Two resonances are observed with high significance. One is consistent to the previous observed $Y(4230)$, and its mass and width are measured to be $M(Y(4230)) = (4225.3 \pm 2.3 \pm 21.5)$~MeV, $\Gamma(Y(4230)) = (72.9 \pm 6.1 \pm 30.8)$~MeV, where the first uncertainties are statistical and the second are systematic. The other one can not be assigned into any experimentally observed resonance, named as $Y(4500)$, and its mass and width are determined to be $M(Y(4500)) = (4484.7 \pm 13.3 \pm 24.1)$ MeV, $\Gamma(Y(4500)) = (111.1 \pm 30.1 \pm 15.2)$ MeV, respectively. There are also evidences of this new structure in the measurements of the cross sections of  $e^+e^- \to \pi \pi J/\psi$~\cite{BESIII:2022jsj}, even in the same channel $K^+ K^- J/\psi$~\cite{PhysRevD.77.011105,PhysRevD.89.072015, PhysRevD.97.071101}. But only with additional data samples at BESIII and improved analysis method, this state has been observed with a significance more than $5\sigma$ for the first time. The mass of $Y(4500)$ is consistent with the prediction of the 5S-4D mixing scheme~\cite{PhysRevD.99.114003}, the heavy-antiheavy hadronic molecules model~\cite{Dong:2021juy} and the lattice quantum chromodynamics result for a $(cs\bar{c}\bar{s})$ state~\cite{PhysRevD.73.094510}, while the width is 2$\sigma$ larger than the prediction of Ref.~\cite{PhysRevD.99.114003}. More experimental measurements and theoretical studies are needed to reveal its nature.

For the first time, the state $Y(4230)$ has been observed in the $K\bar{K}J/\psi$ mode with the significance larger than 5$\sigma$. The product of the electronic partial width and the decay branching fraction is measured to be $\mathcal{B}(Y(4230) \to K^+K^-J/\psi)\Gamma(e^+e^- \to Y(4230))$ = ($0.29 \pm 0.02 \pm 0.10$)~eV or ($0.42 \pm 0.04 \pm 0.15$) eV, according to different interferences, respectively. The ratio between the branching fractions of the $Y(4230)$ decaying into $K\bar{K}J/\psi$ and $\pi \pi J/\psi$~\cite{BESIII:2022jsj}  are calculated and shown in Table.~\ref{Tab-Ratio}. Author of Ref.~\cite{QIAO2006263} predicts the $K\bar{K}$ mode should be suppressed if the $Y(4230)$ is $\Lambda_c$ baryonium. Even at present, no conclusion can be drawn due to the multiple solutions. However, once the physics solution is determined as done in Ref.~\cite{PhysRevD.105.L031506}, it will provide very useful information for understanding the nature of $Y(4230)$.

\begin{table}[!htbp]
\centering\setlength{\abovecaptionskip}{6pt}\setlength{\belowcaptionskip}{6pt}\setlength{\tabcolsep}{6pt}\renewcommand{\arraystretch}{1.3}
	\caption{The ratios between the branching fractions of the $Y(4230)$ decaying into $K\bar{K}J/\psi$ and $\pi \pi J/\psi$, depending on the various combinations of the multiple solutions.}
	\label{Tab-Ratio}
	\resizebox{0.48\textwidth}{!}{
	\begin{tabular}{c|c|c}
	\hline
     ~              & $K\bar{K}J/\psi$ Sol. I  & $K\bar{K}J/\psi$ Sol. II  ~\\ \hline
 	$\pi \pi J/\psi$ Sol. I    & $0.17 \pm 0.02$  & $0.25 \pm 0.04$  ~\\ \hline
 	$\pi \pi J/\psi$ Sol. II   & $0.097\pm 0.017$ & $0.14 \pm 0.03$  ~\\ \hline
 	$\pi \pi J/\psi$ Sol. III  & $0.035\pm 0.004$ & $0.051\pm 0.007$ ~\\ \hline
 	$\pi \pi J/\psi$ Sol. IV   & $0.020\pm 0.002$ & $0.028\pm 0.004$ ~\\ \hline
	\end{tabular}
	}
\end{table}

\section{Acknowledgement}
The BESIII collaboration thanks the staff of BEPCII and the IHEP computing center for their strong support. This work is supported in part by National Key R\&D Program of China under Contracts Nos. 2020YFA0406300, 2020YFA0406400; National Natural Science Foundation of China (NSFC) under Contracts Nos. 11625523, 11635010, 11735014, 11822506, 11835012, 11935015, 11935016, 11935018, 11961141012, 12022510, 12025502, 12035009, 12035013, 12061131003; the Chinese Academy of Sciences (CAS) Large-Scale Scientific Facility Program; Joint Large-Scale Scientific Facility Funds of the NSFC and CAS under Contracts Nos. U1732263, U1832207; CAS Key Research Program of Frontier Sciences under Contract No. QYZDJ-SSW-SLH040; 100 Talents Program of CAS; INPAC and Shanghai Key Laboratory for Particle Physics and Cosmology; ERC under Contract No. 758462; European Union Horizon 2020 research and innovation programme under Contract No. Marie Sklodowska-Curie grant agreement No 894790; German Research Foundation DFG under Contracts Nos. 443159800, Collaborative Research Center CRC 1044, FOR 2359, GRK 214; Istituto Nazionale di Fisica Nucleare, Italy; Ministry of Development of Turkey under Contract No. DPT2006K-120470; National Science and Technology fund; Olle Engkvist Foundation under Contract No. 200-0605; STFC (United Kingdom); The Knut and Alice Wallenberg Foundation (Sweden) under Contract No. 2016.0157; The Royal Society, UK under Contracts Nos. DH140054, DH160214; The Swedish Research Council; U. S. Department of Energy under Contracts Nos. DE-FG02-05ER41374, DE-SC-001206.

\begin{appendix}
\section*{Appendix}
\appendix
\setcounter{table}{0}
\setcounter{figure}{0}
\setcounter{equation}{0}
\renewcommand{\thetable}{A\arabic{table}}
\renewcommand{\thefigure}{A\arabic{figure}}
\renewcommand{\theequation}{A\arabic{equation}}
\subsection{Distributions of $M(\ell^+\ell^-)$ and $M(K^+K^-)$}
\label{subs_a1}
\begin{figure*}[htbp]
	\centering
	\includegraphics[angle=0,width=0.24\textwidth]{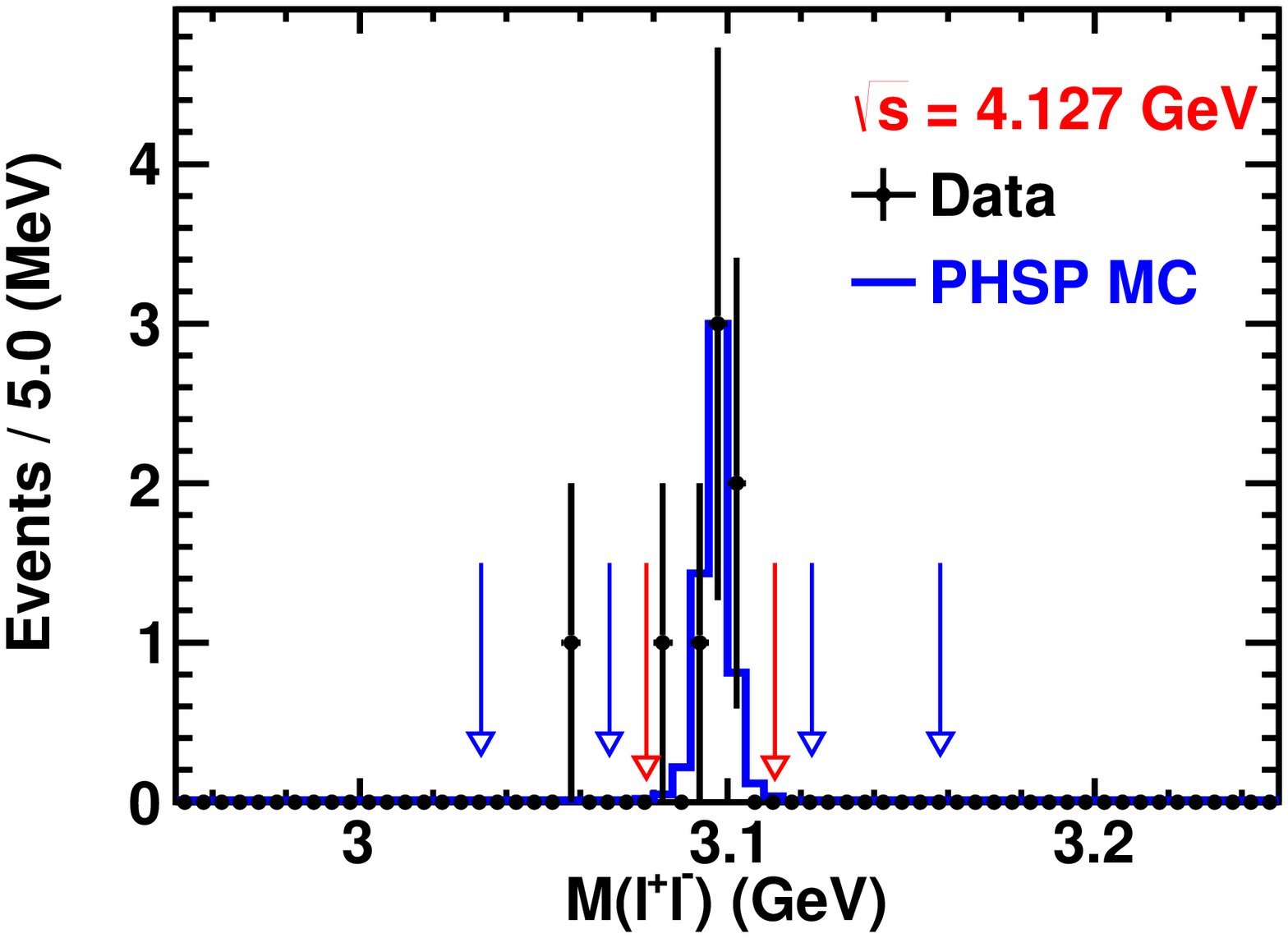}
	\includegraphics[angle=0,width=0.24\textwidth]{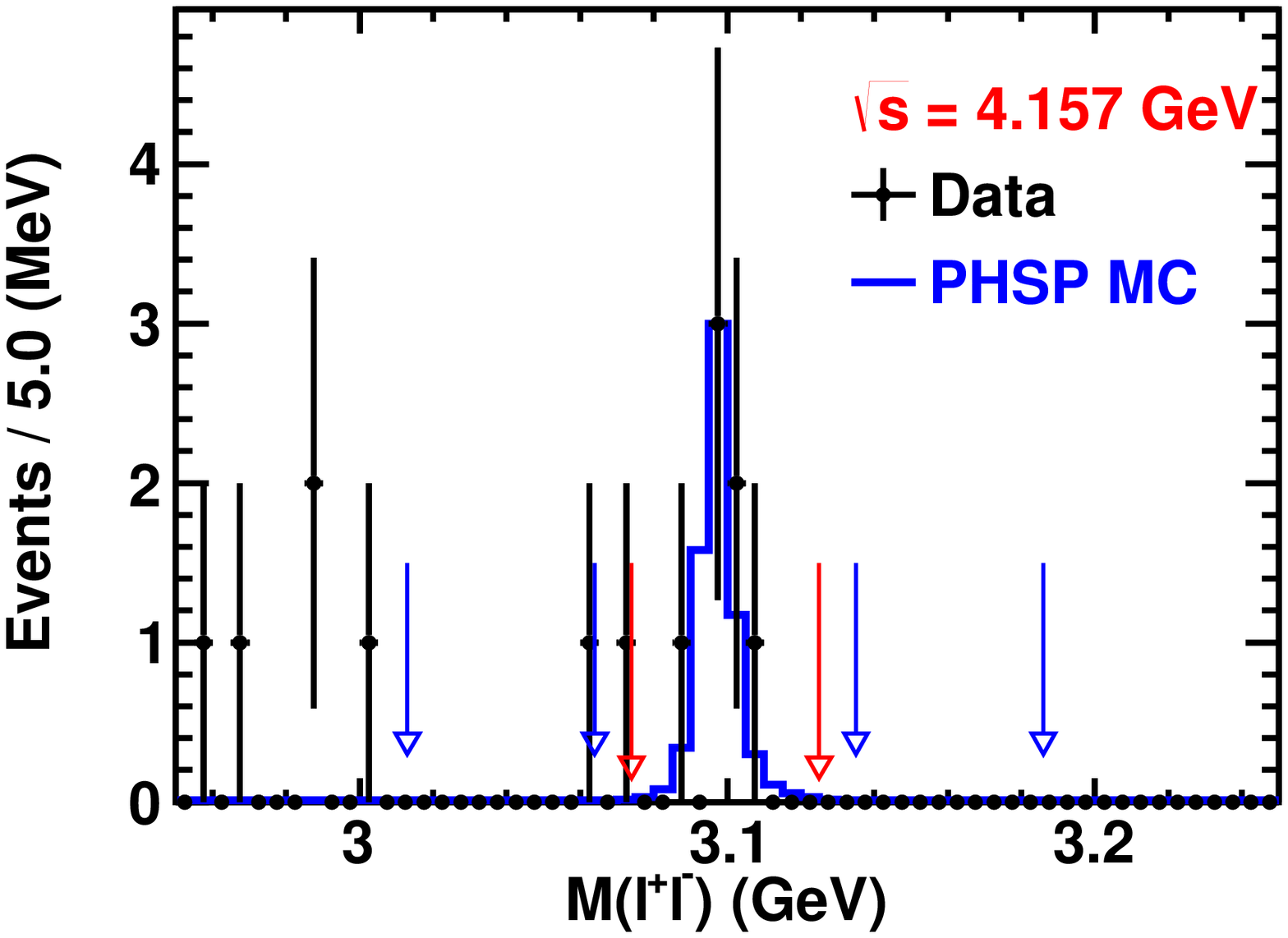}
	\includegraphics[angle=0,width=0.24\textwidth]{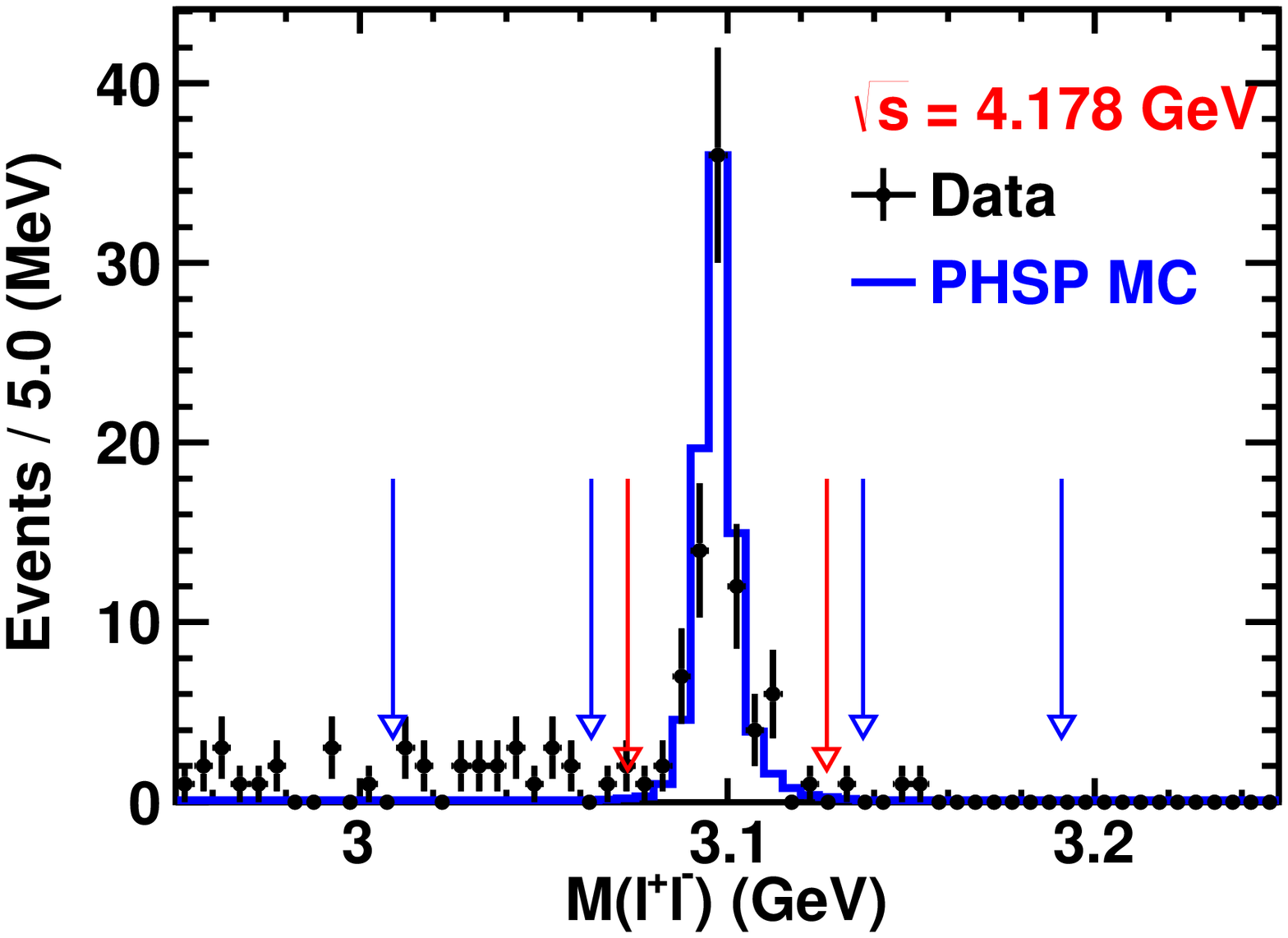}
	\includegraphics[angle=0,width=0.24\textwidth]{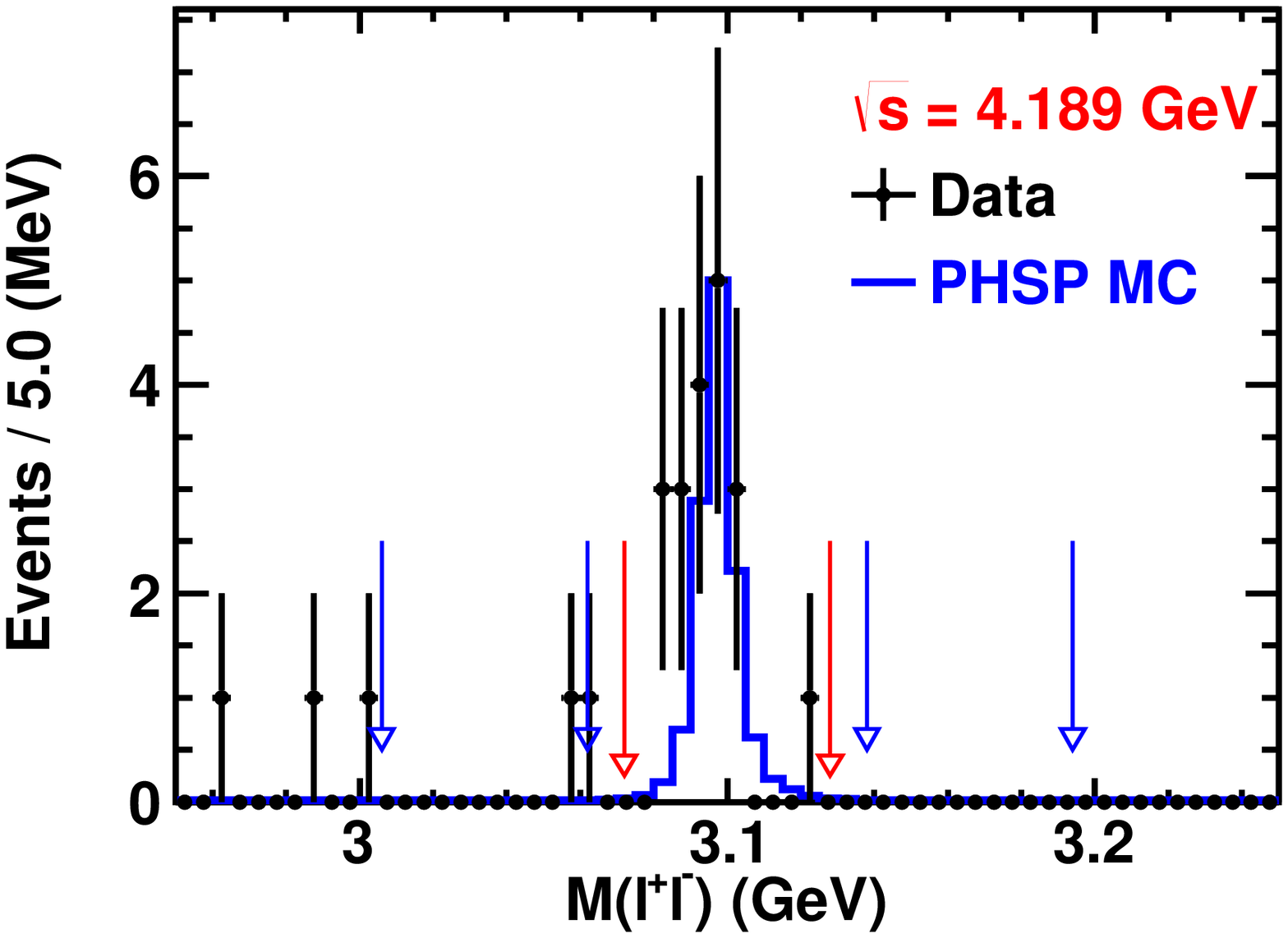}
	\includegraphics[angle=0,width=0.24\textwidth]{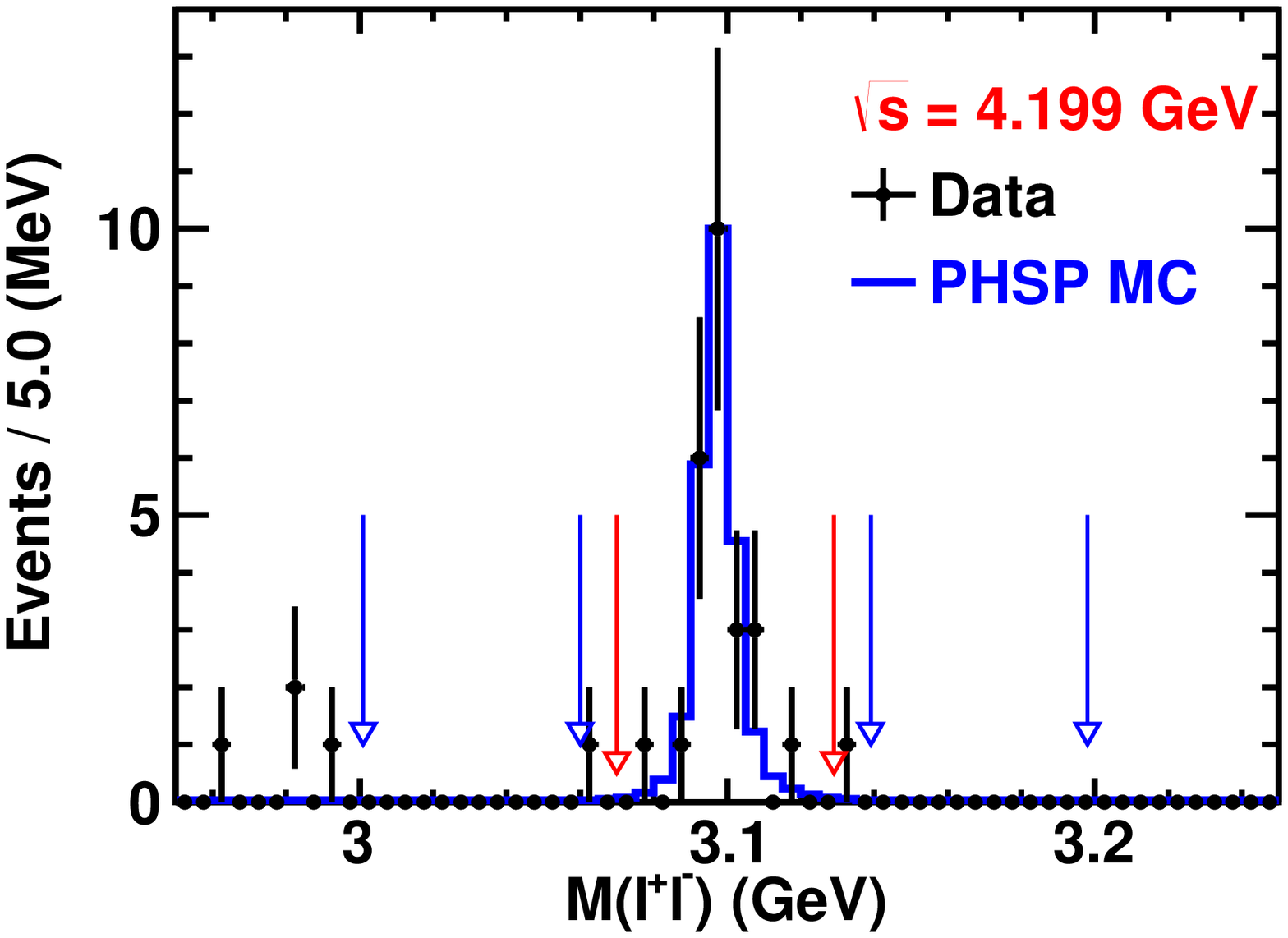}
	\includegraphics[angle=0,width=0.24\textwidth]{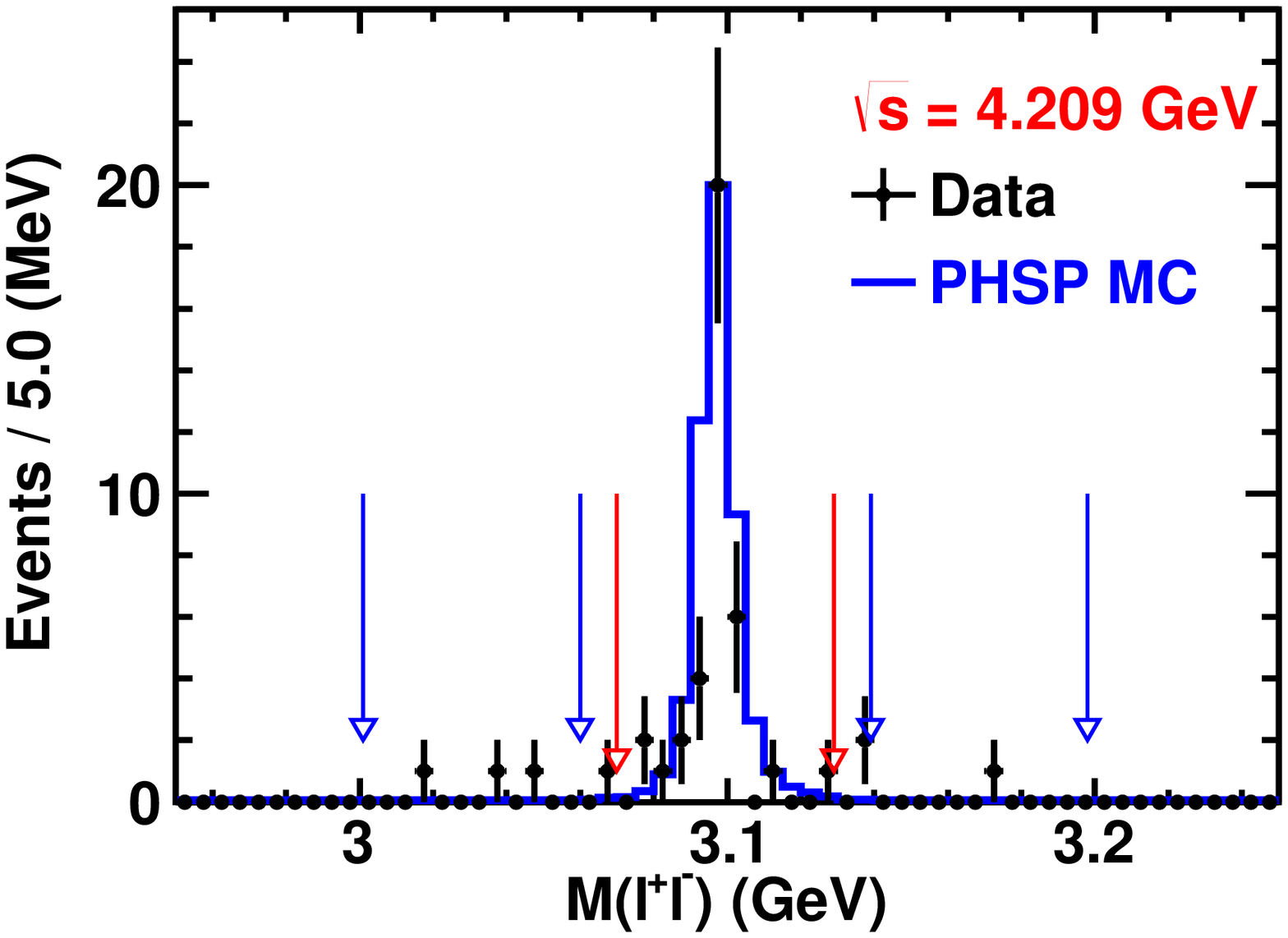}
	\includegraphics[angle=0,width=0.24\textwidth]{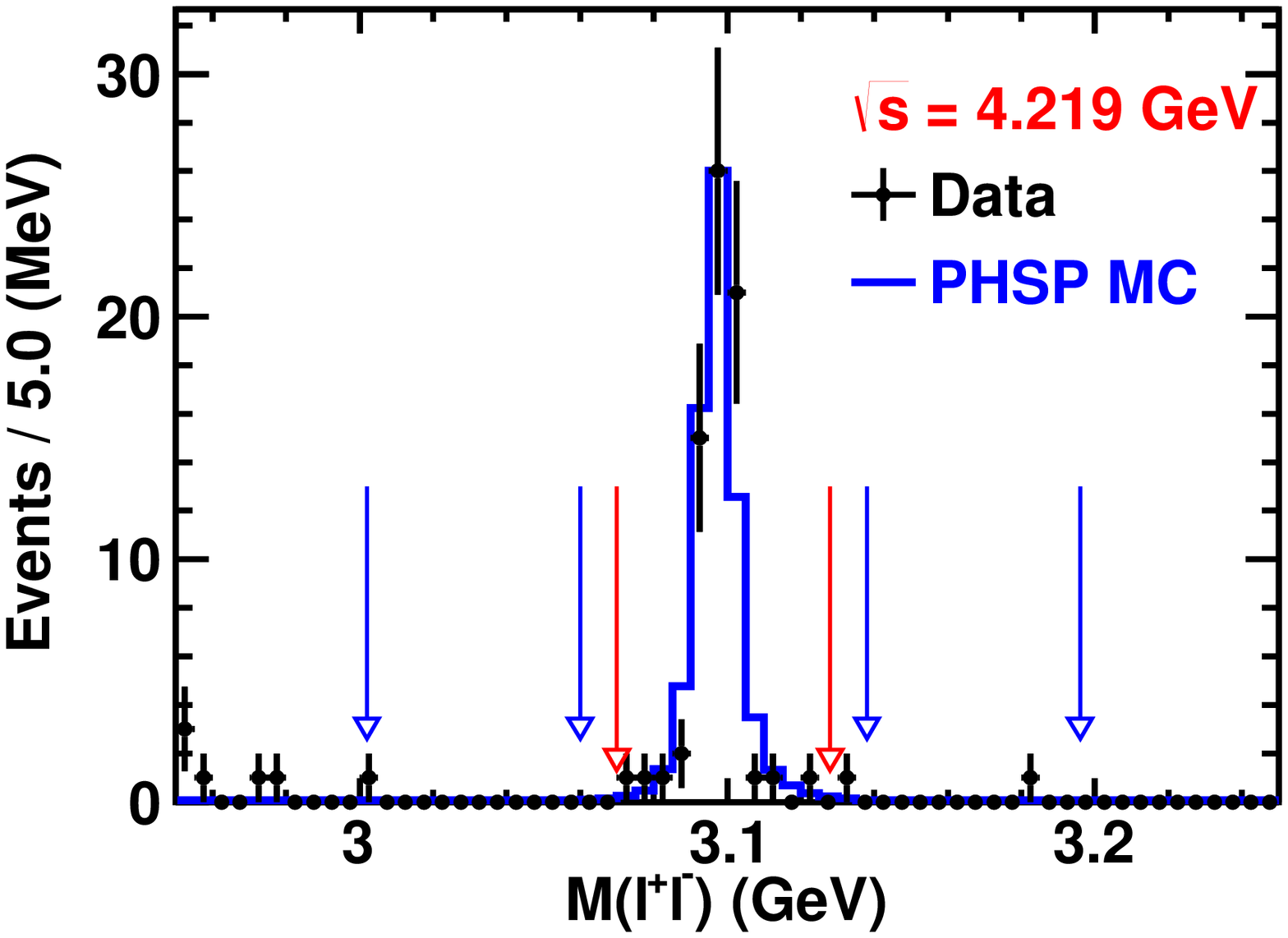}
	\includegraphics[angle=0,width=0.24\textwidth]{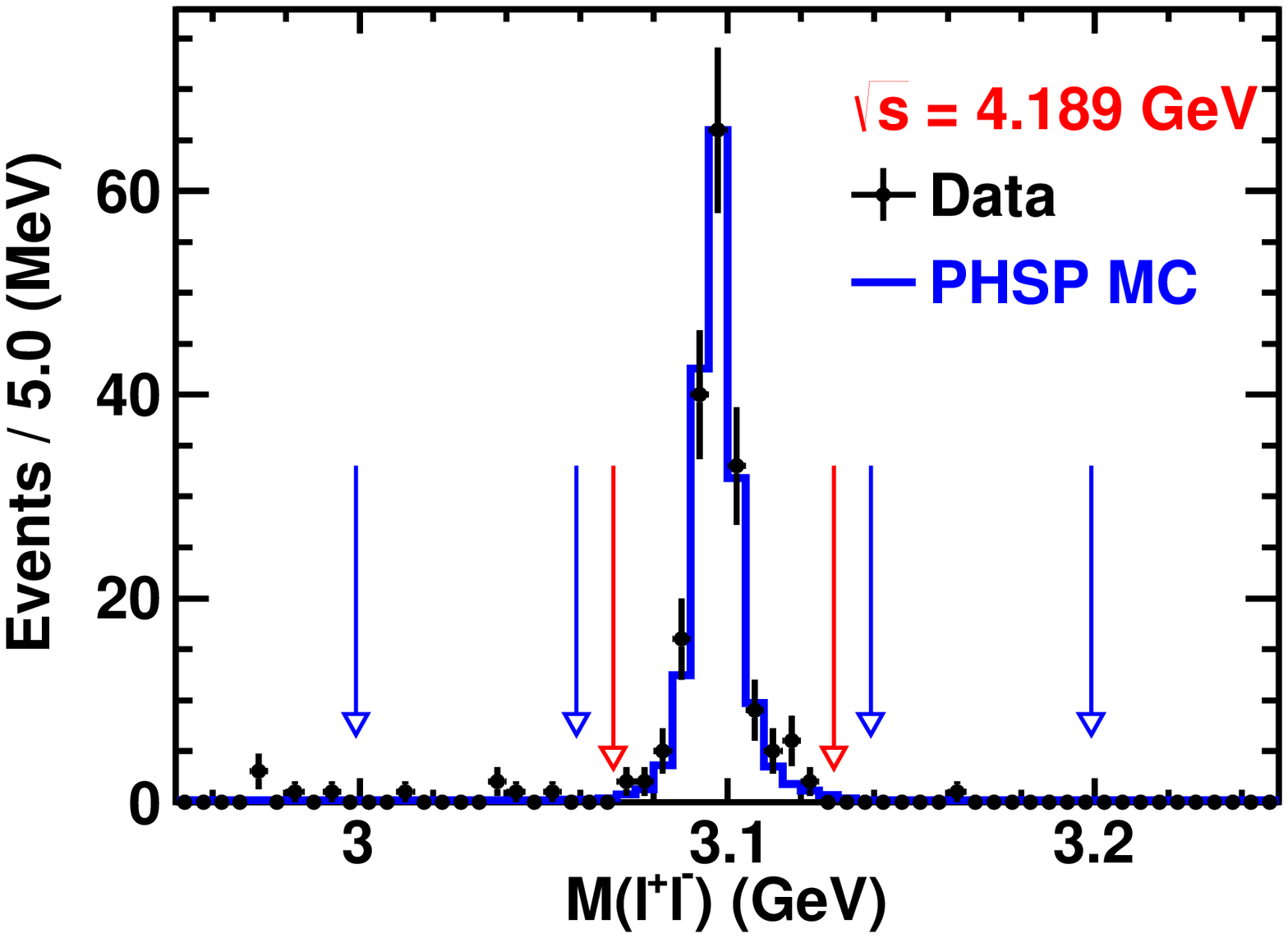}
	\includegraphics[angle=0,width=0.24\textwidth]{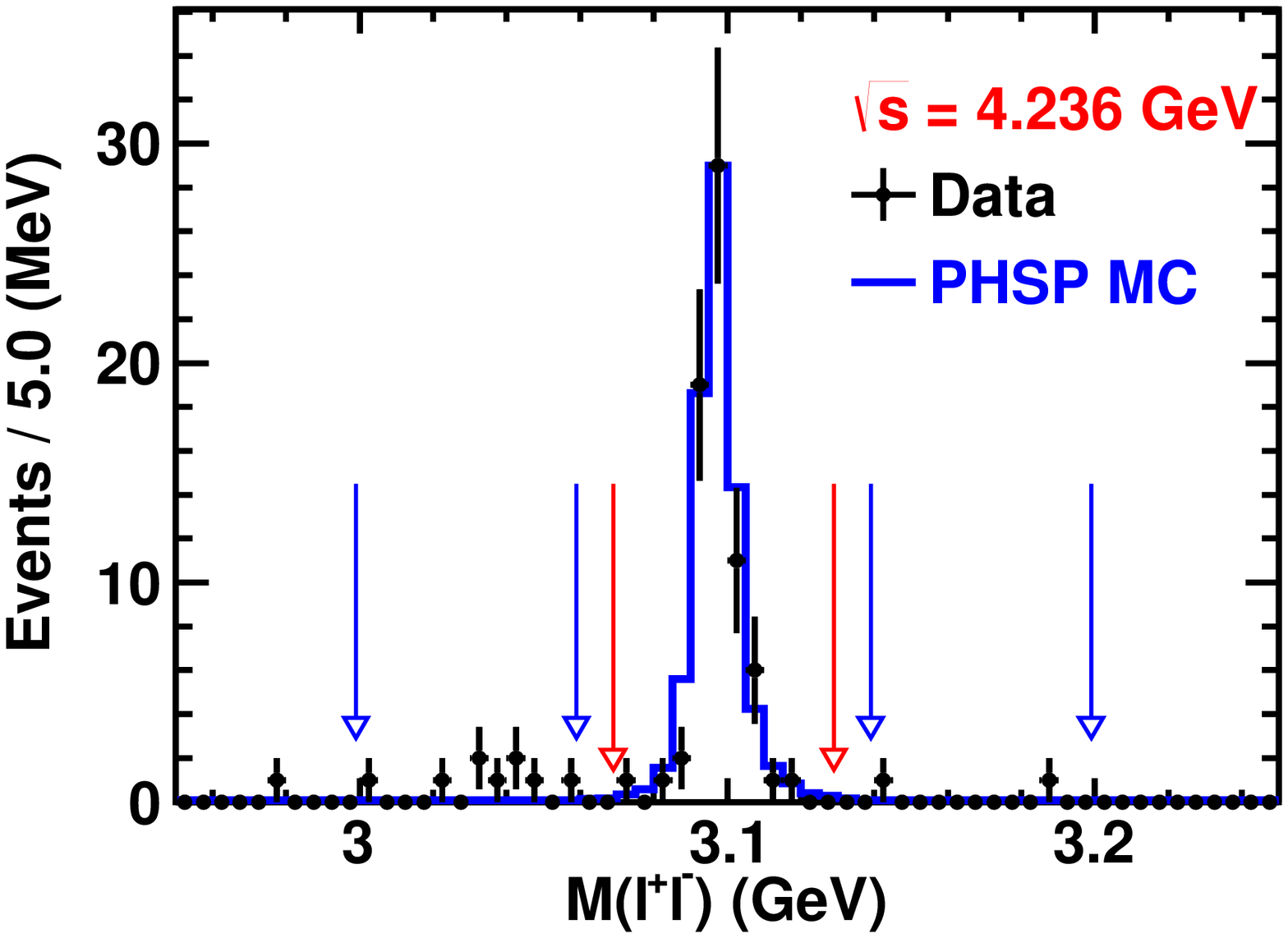}
	\includegraphics[angle=0,width=0.24\textwidth]{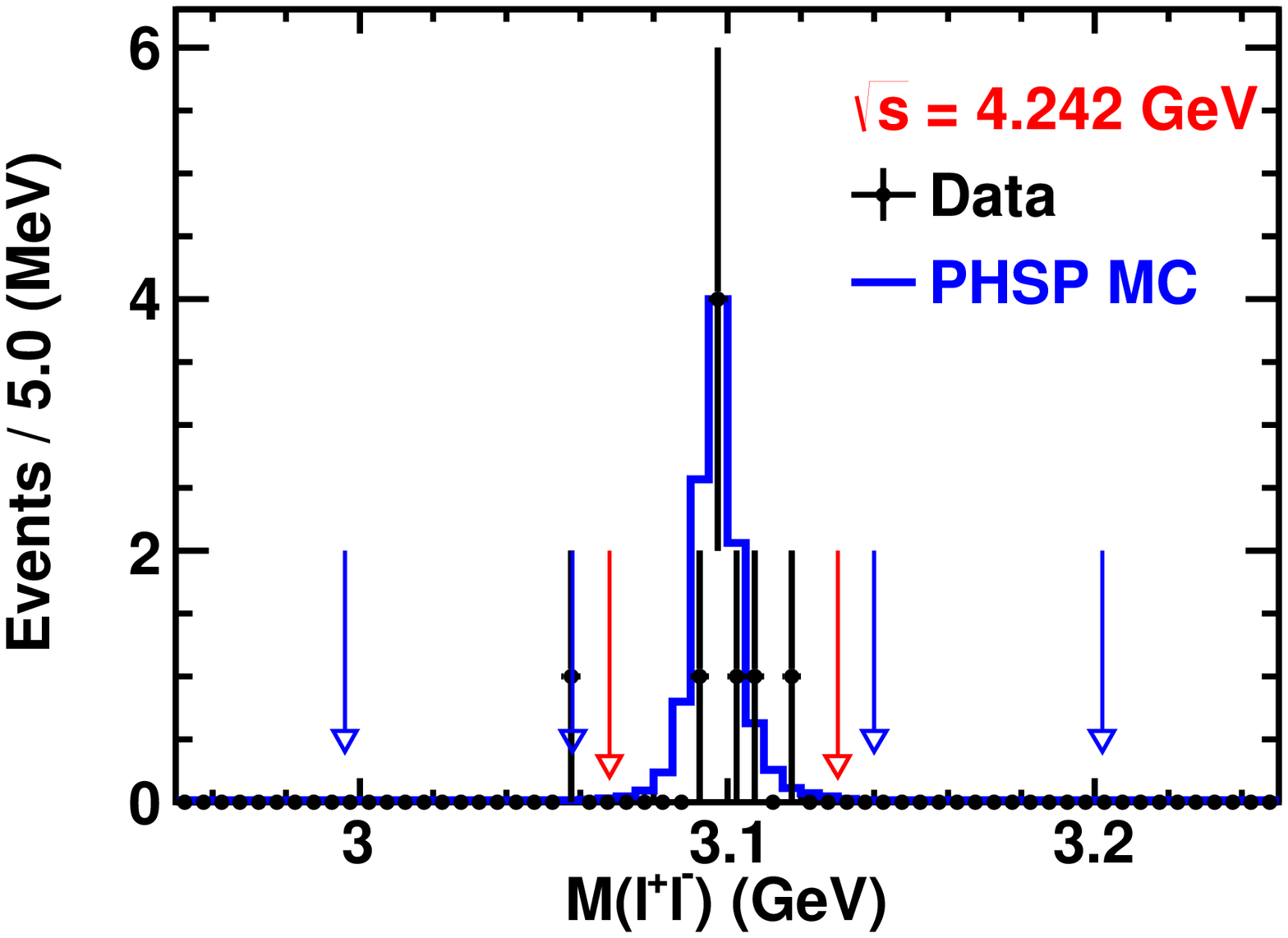}
	\includegraphics[angle=0,width=0.24\textwidth]{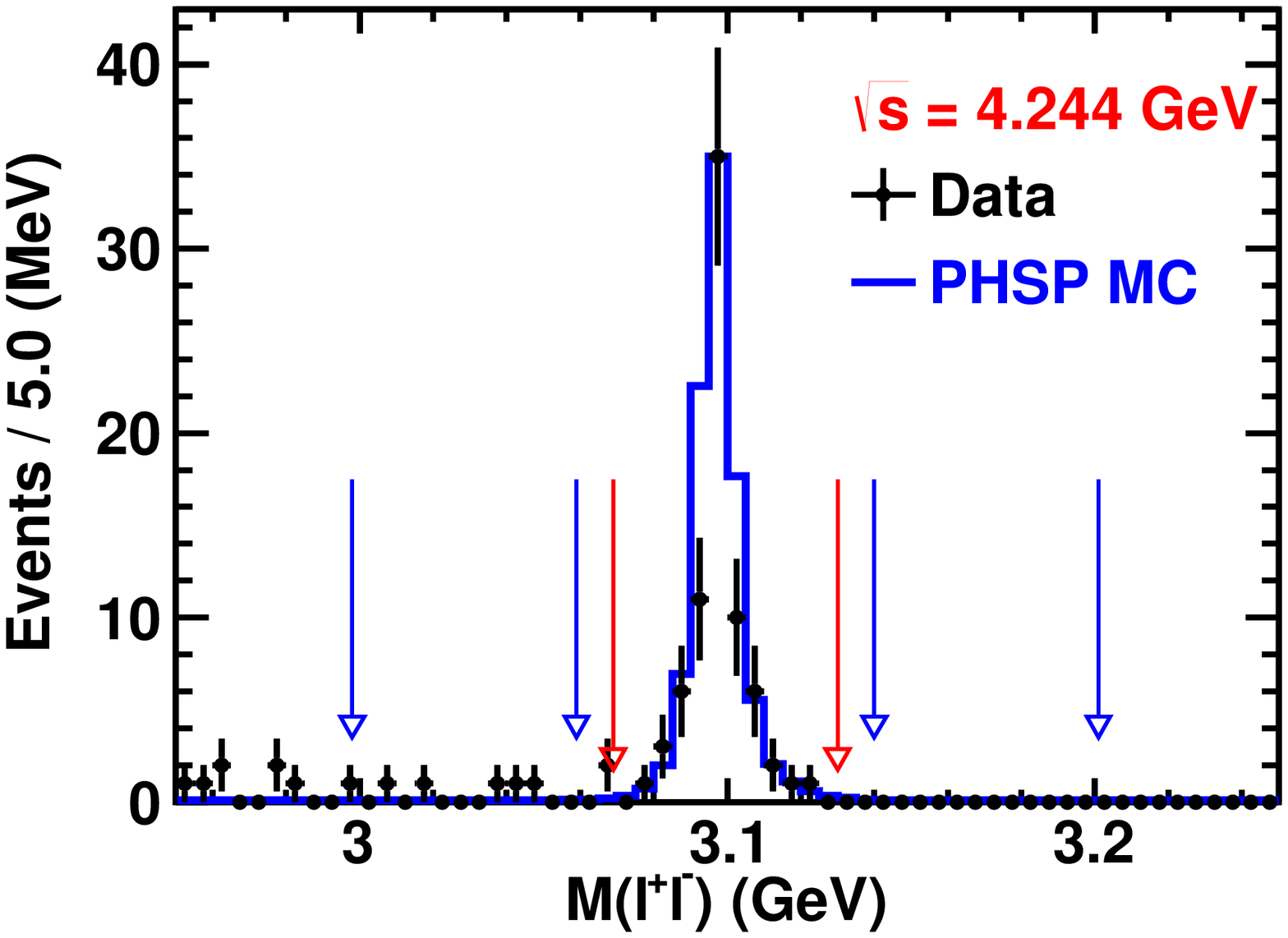}
	\includegraphics[angle=0,width=0.24\textwidth]{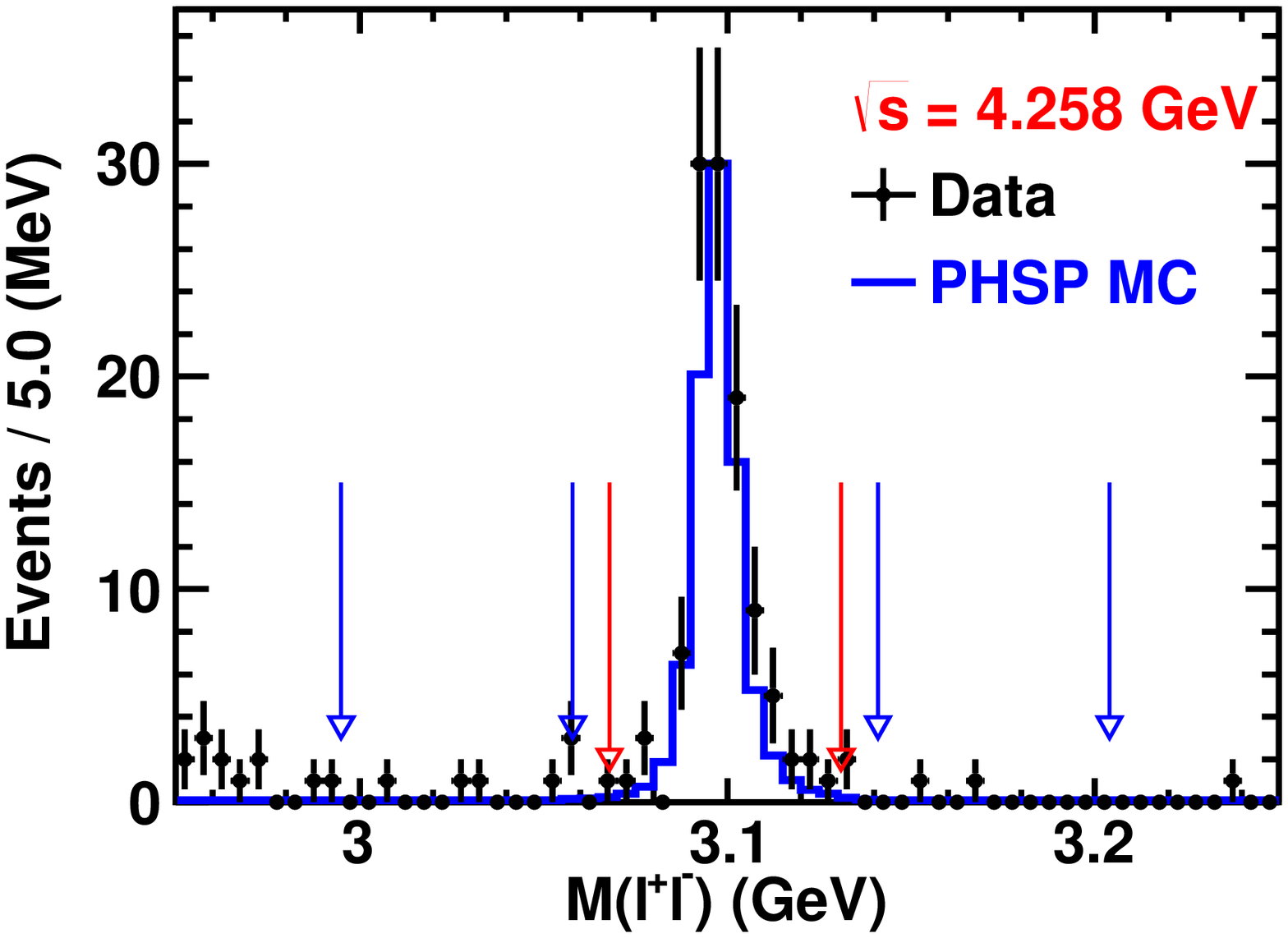}
	\includegraphics[angle=0,width=0.24\textwidth]{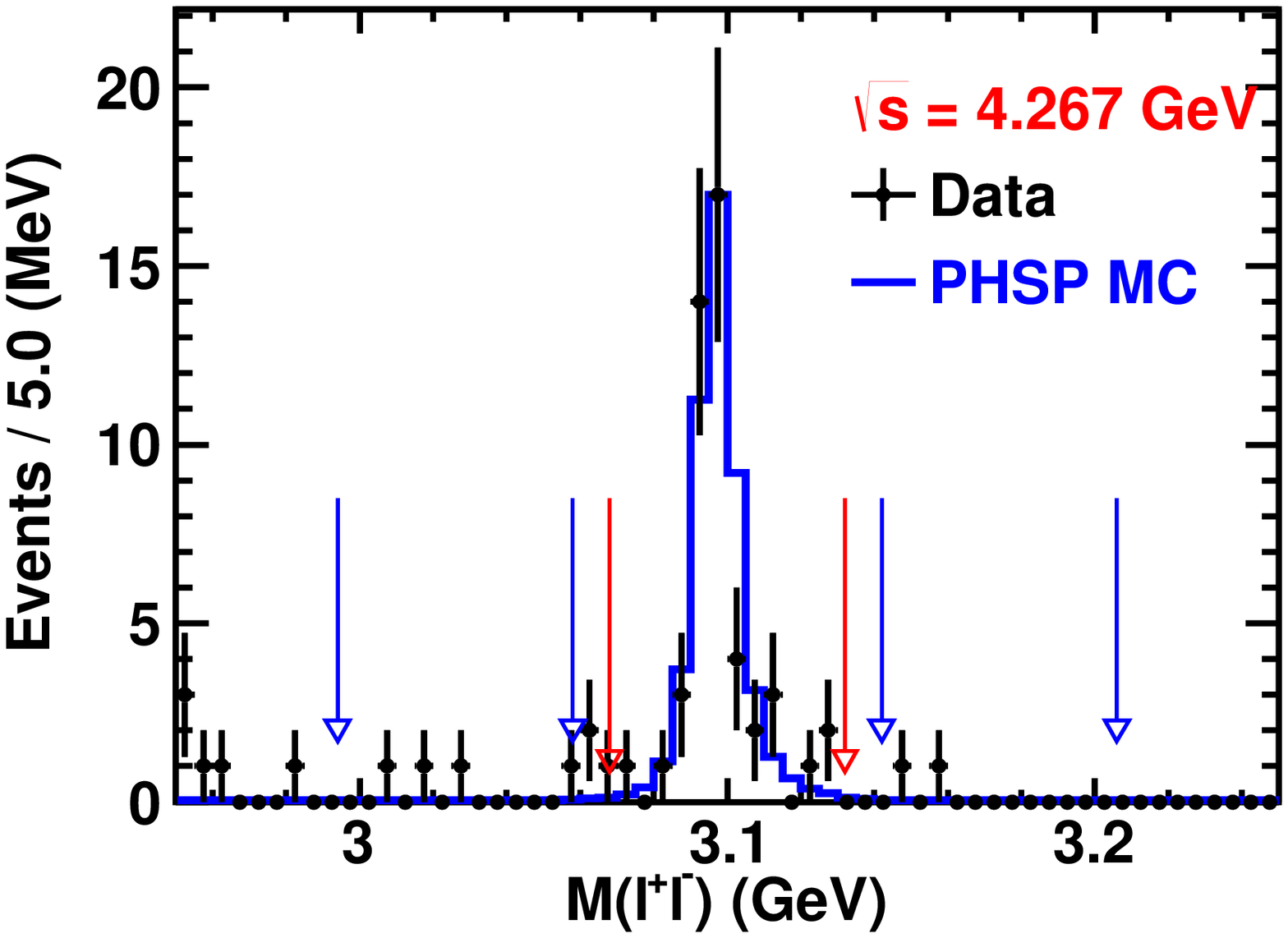}
	\includegraphics[angle=0,width=0.24\textwidth]{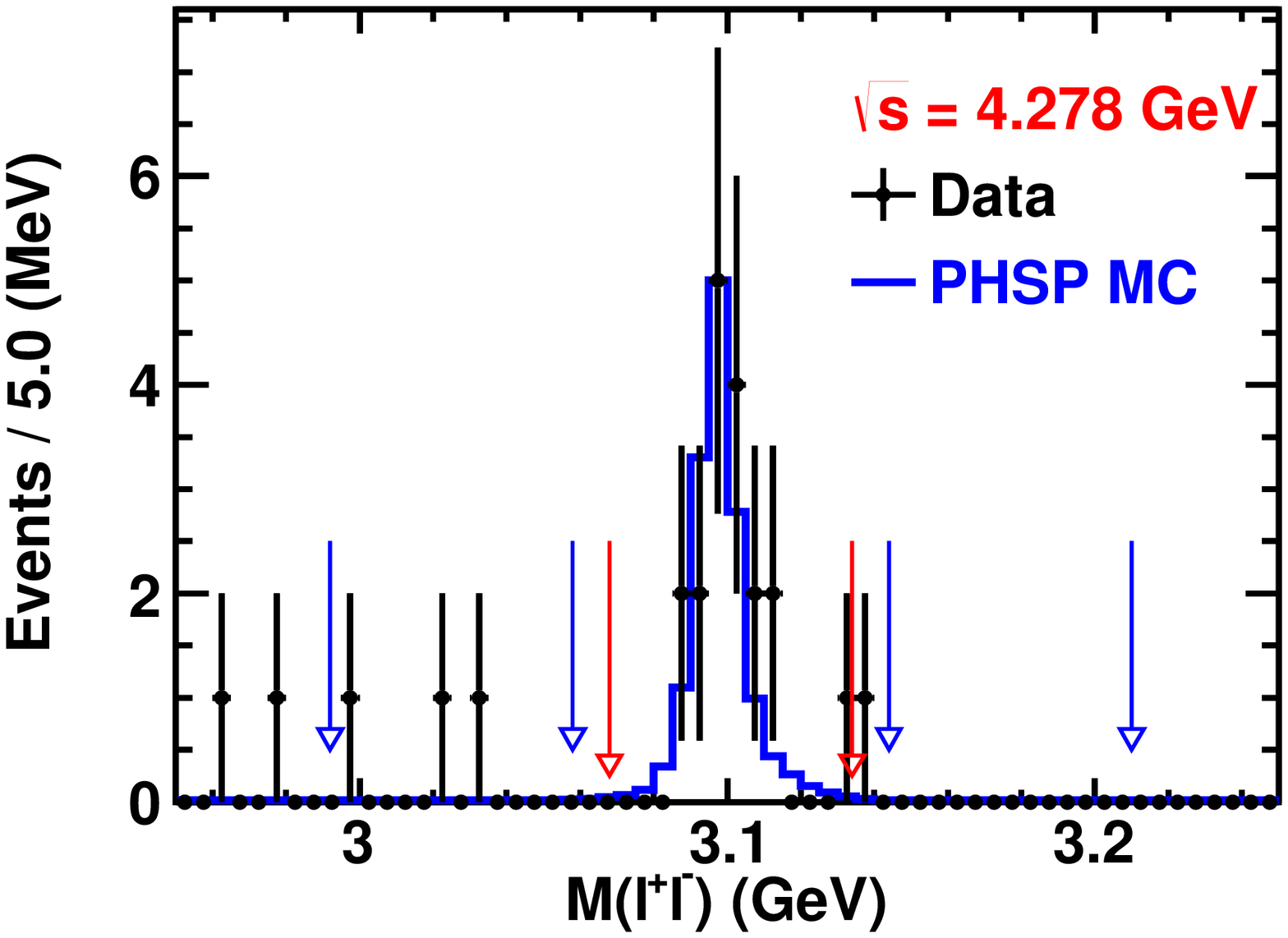}
	\includegraphics[angle=0,width=0.24\textwidth]{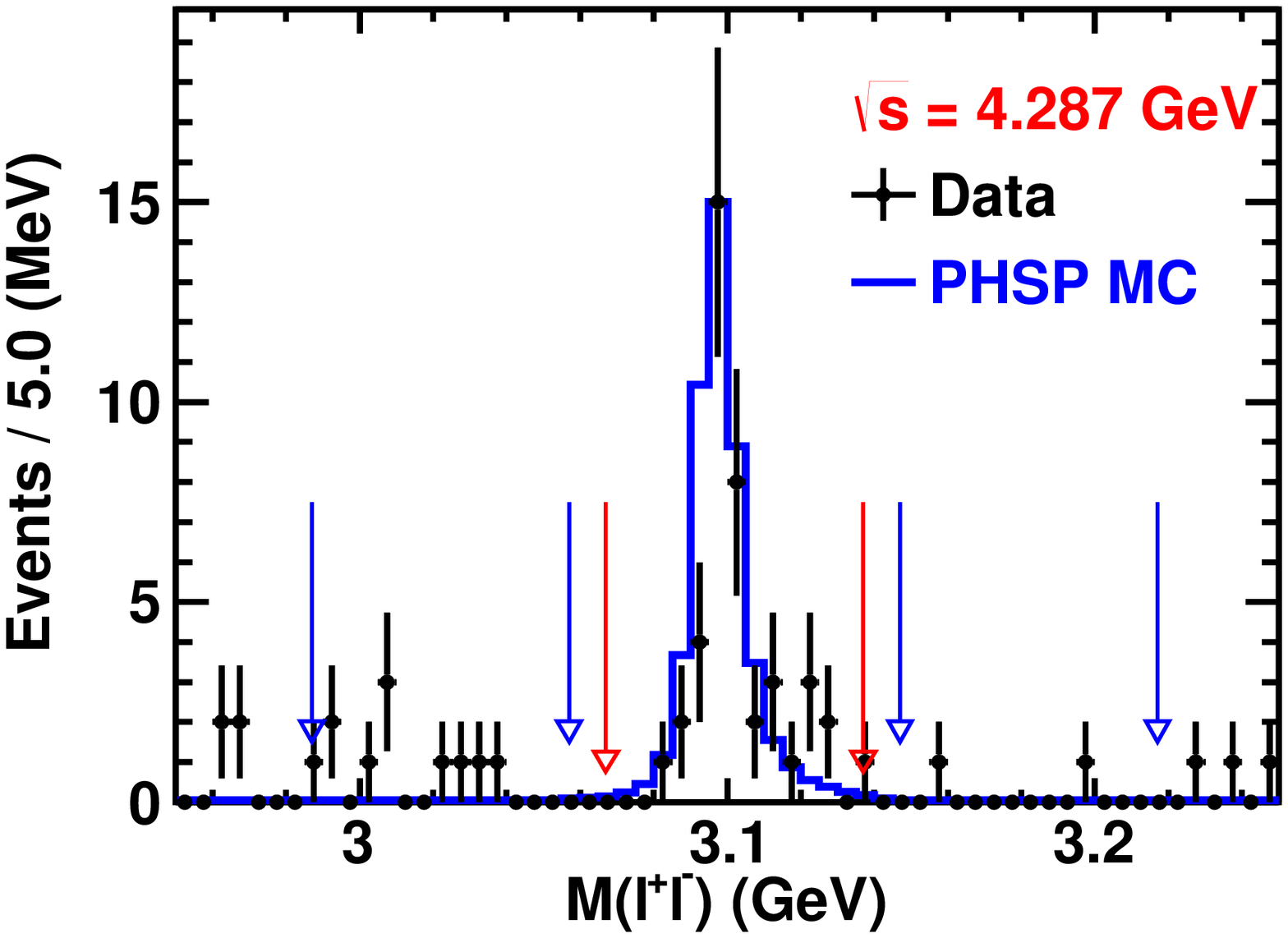}
	\includegraphics[angle=0,width=0.24\textwidth]{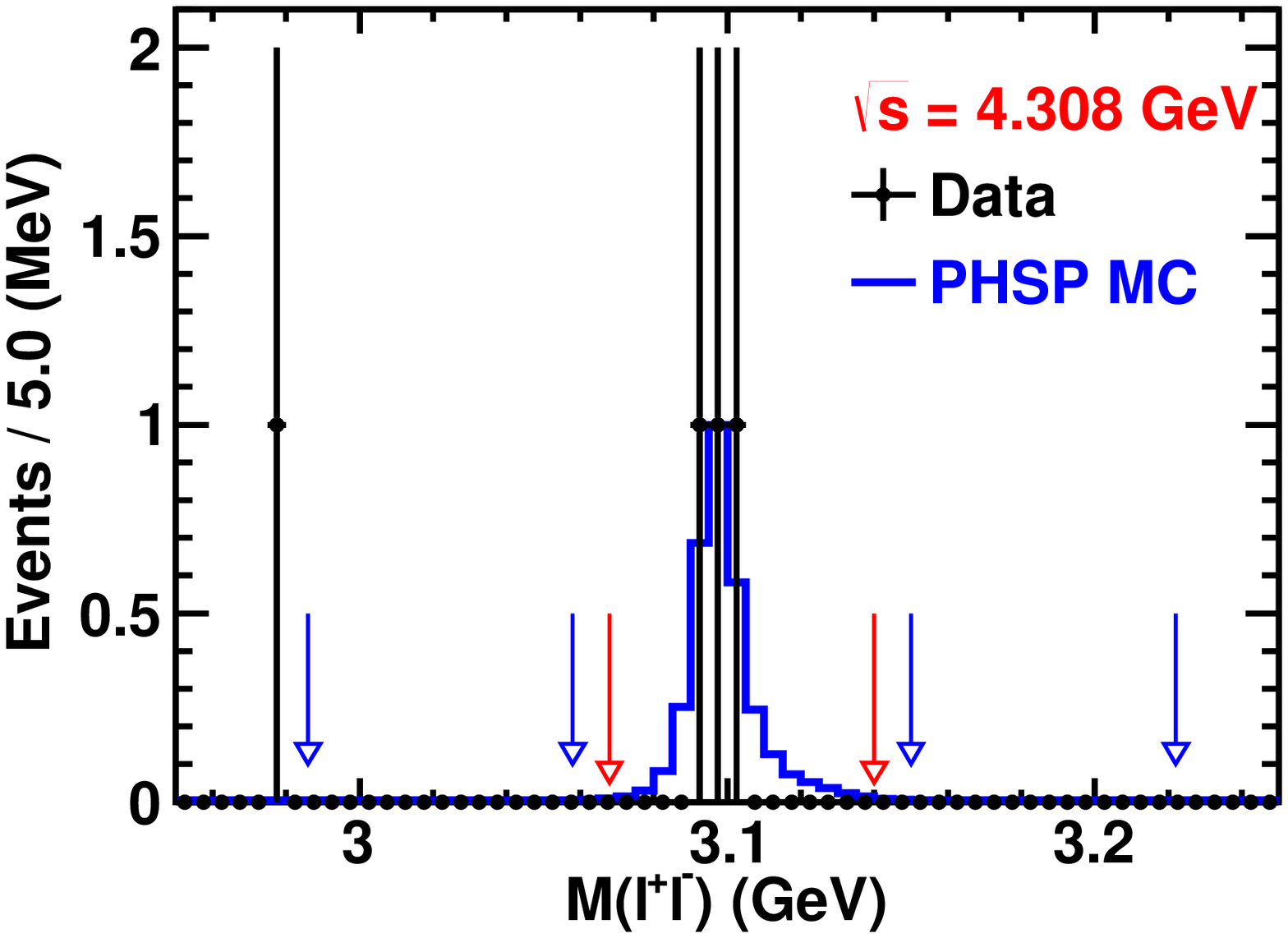}
	\includegraphics[angle=0,width=0.24\textwidth]{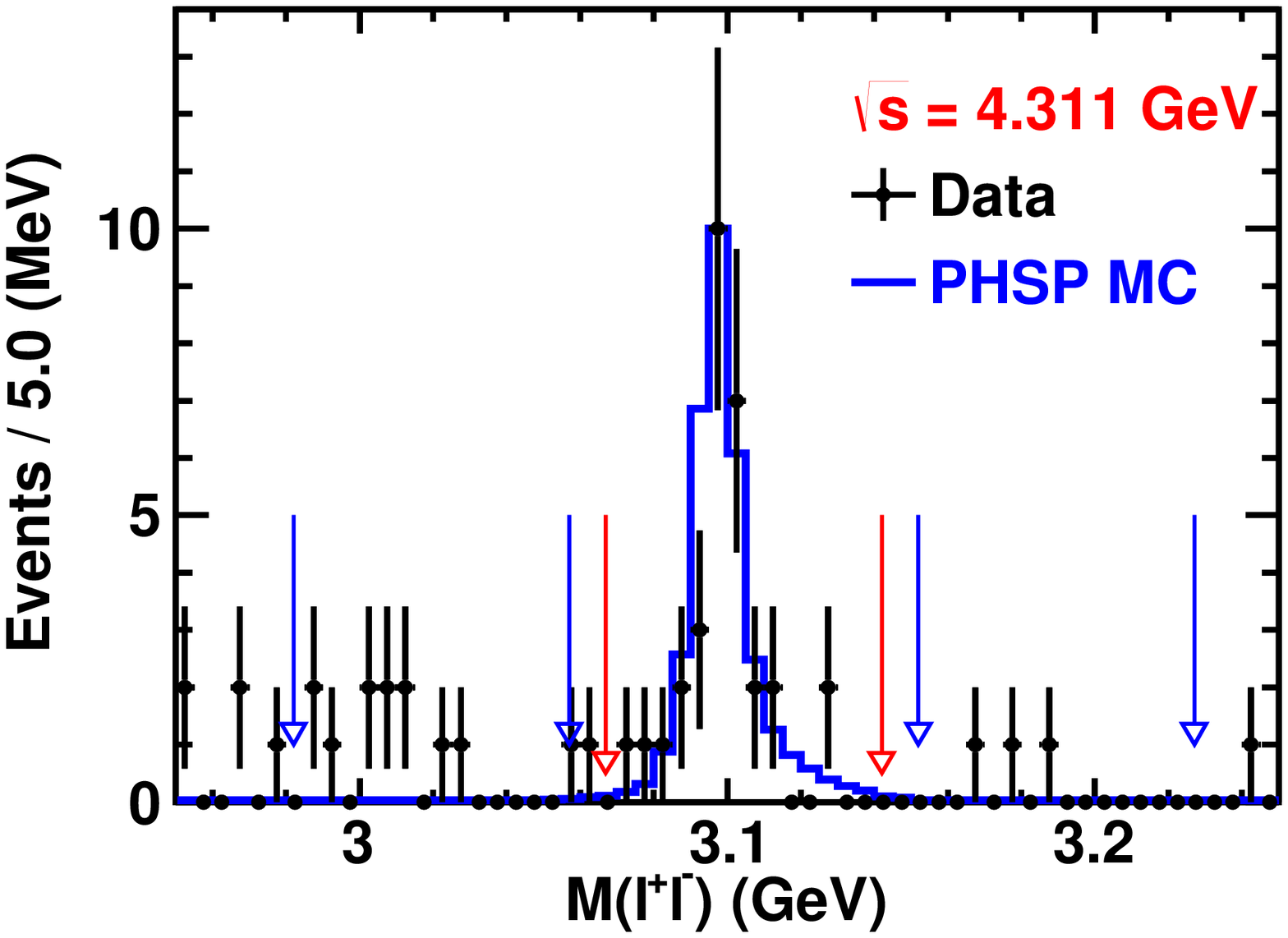}
	\includegraphics[angle=0,width=0.24\textwidth]{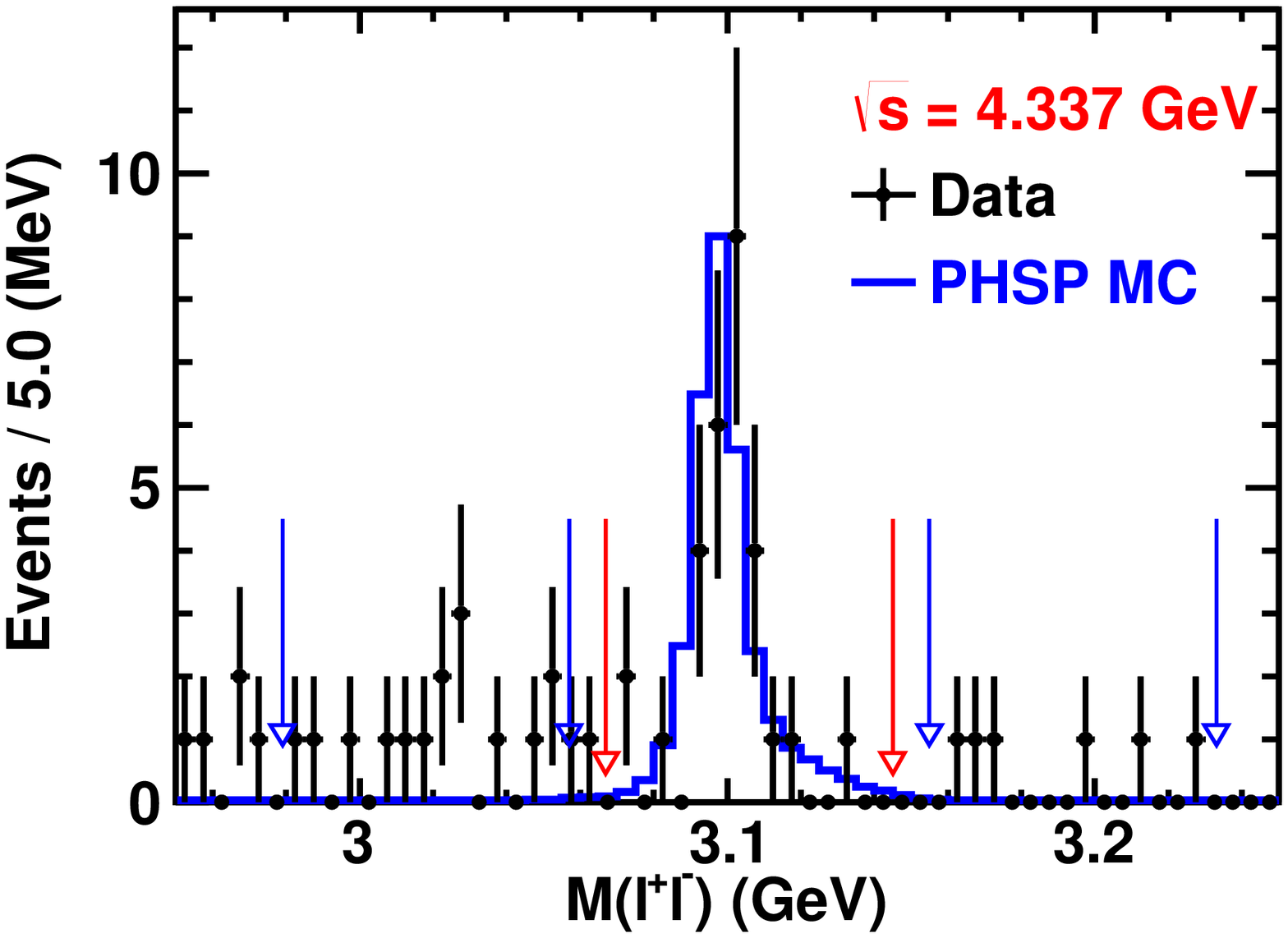}
	\includegraphics[angle=0,width=0.24\textwidth]{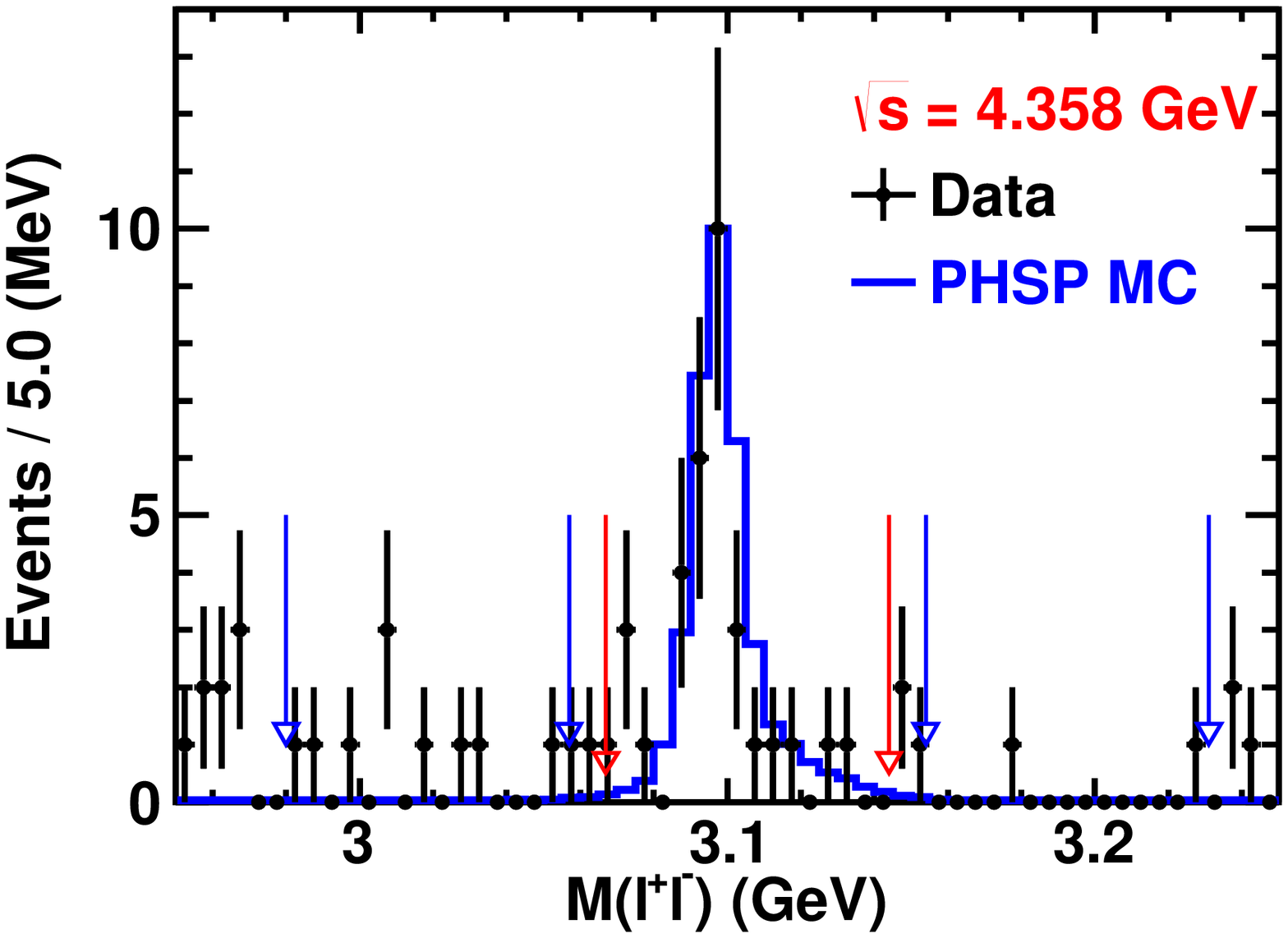}
	\includegraphics[angle=0,width=0.24\textwidth]{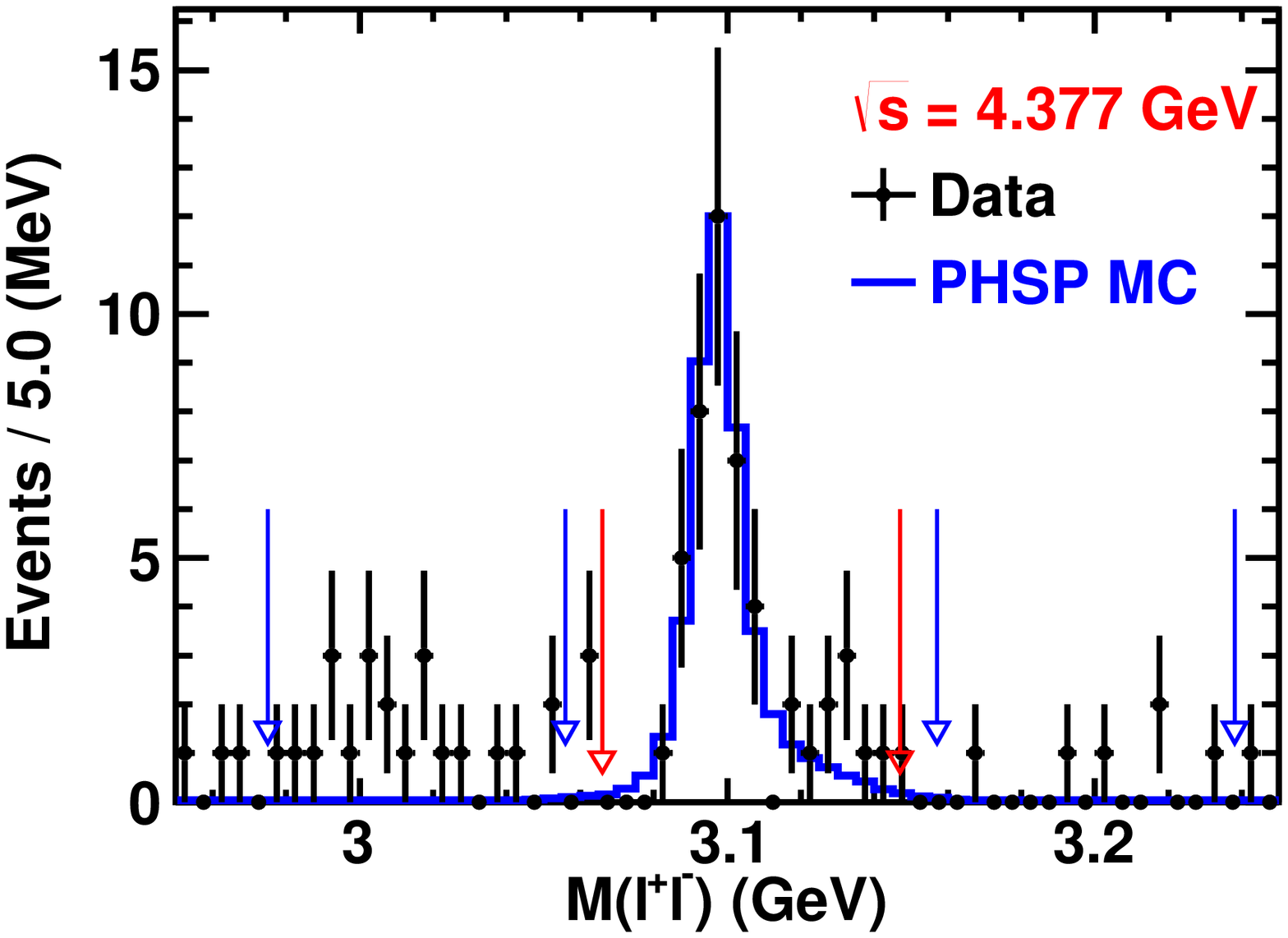}
	\includegraphics[angle=0,width=0.24\textwidth]{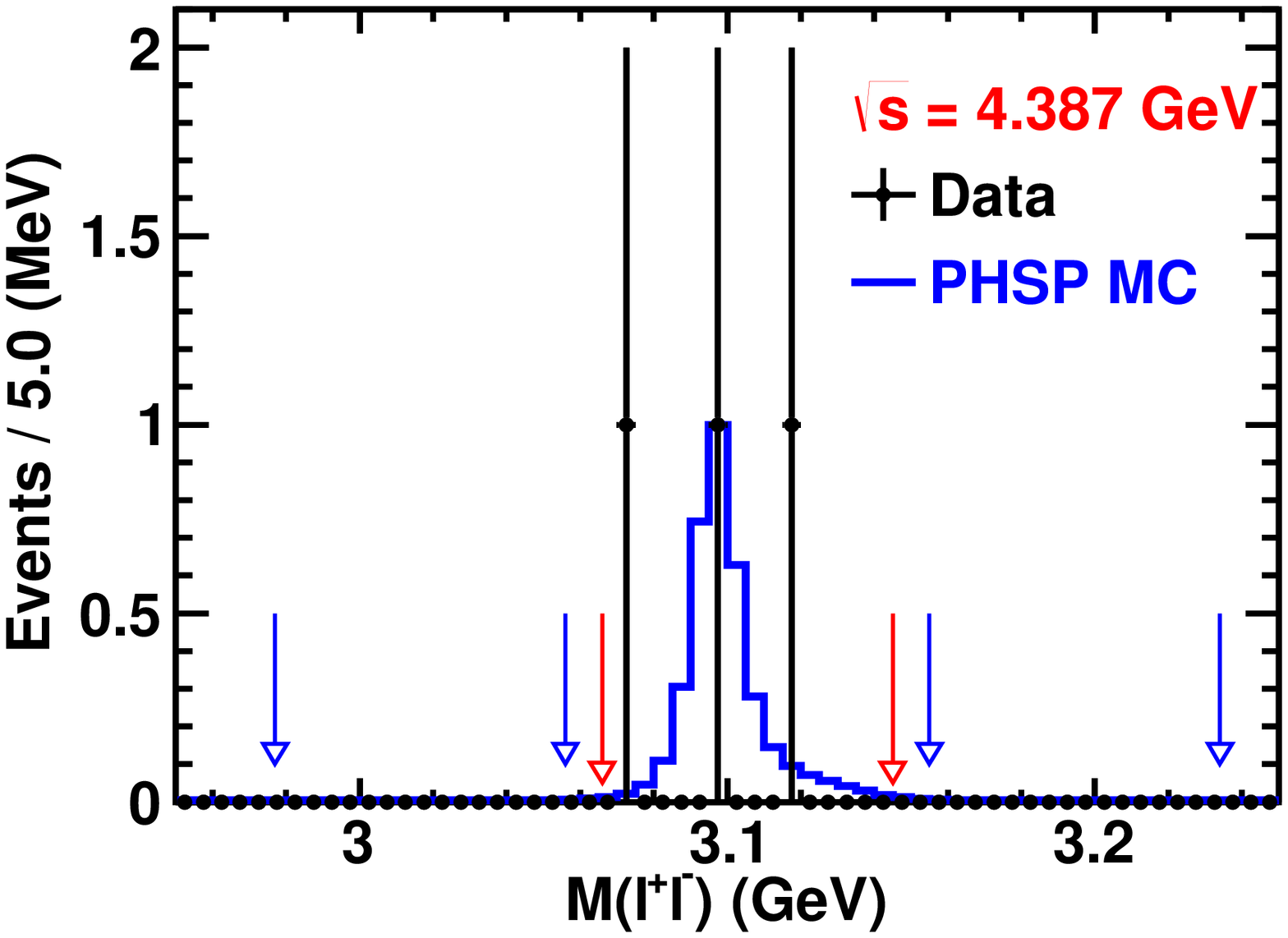}
	\includegraphics[angle=0,width=0.24\textwidth]{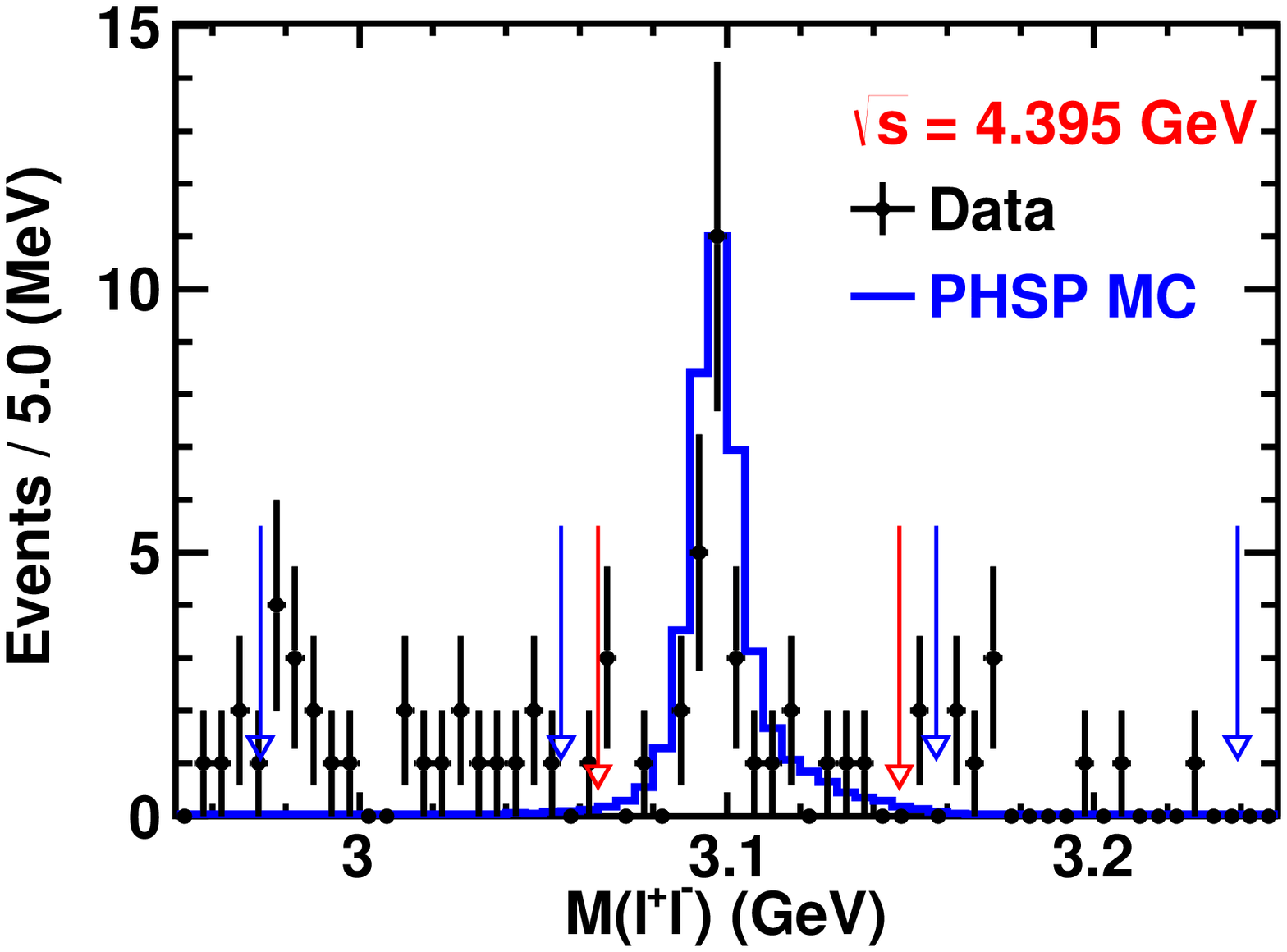}
	\includegraphics[angle=0,width=0.24\textwidth]{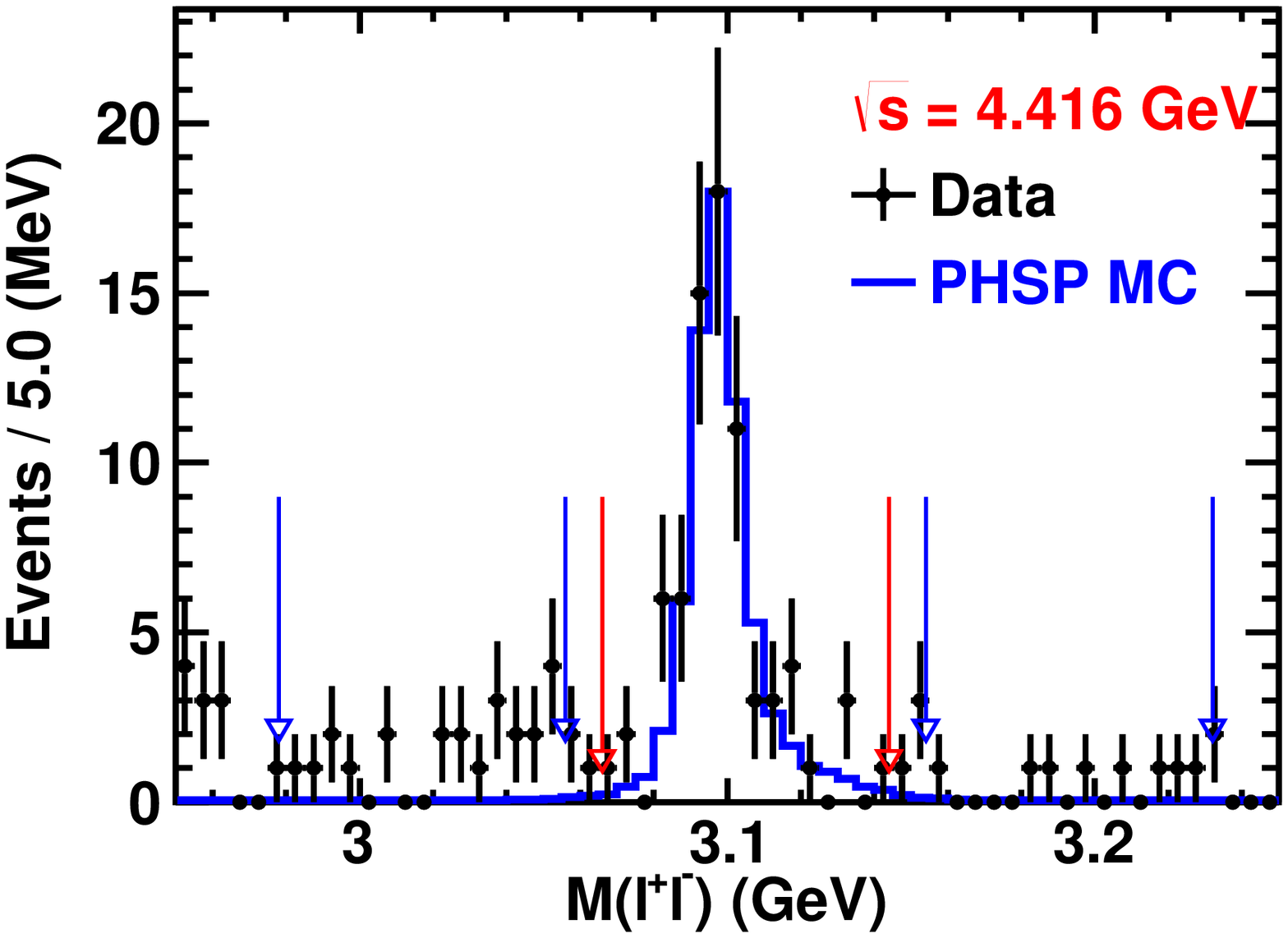}
	\includegraphics[angle=0,width=0.24\textwidth]{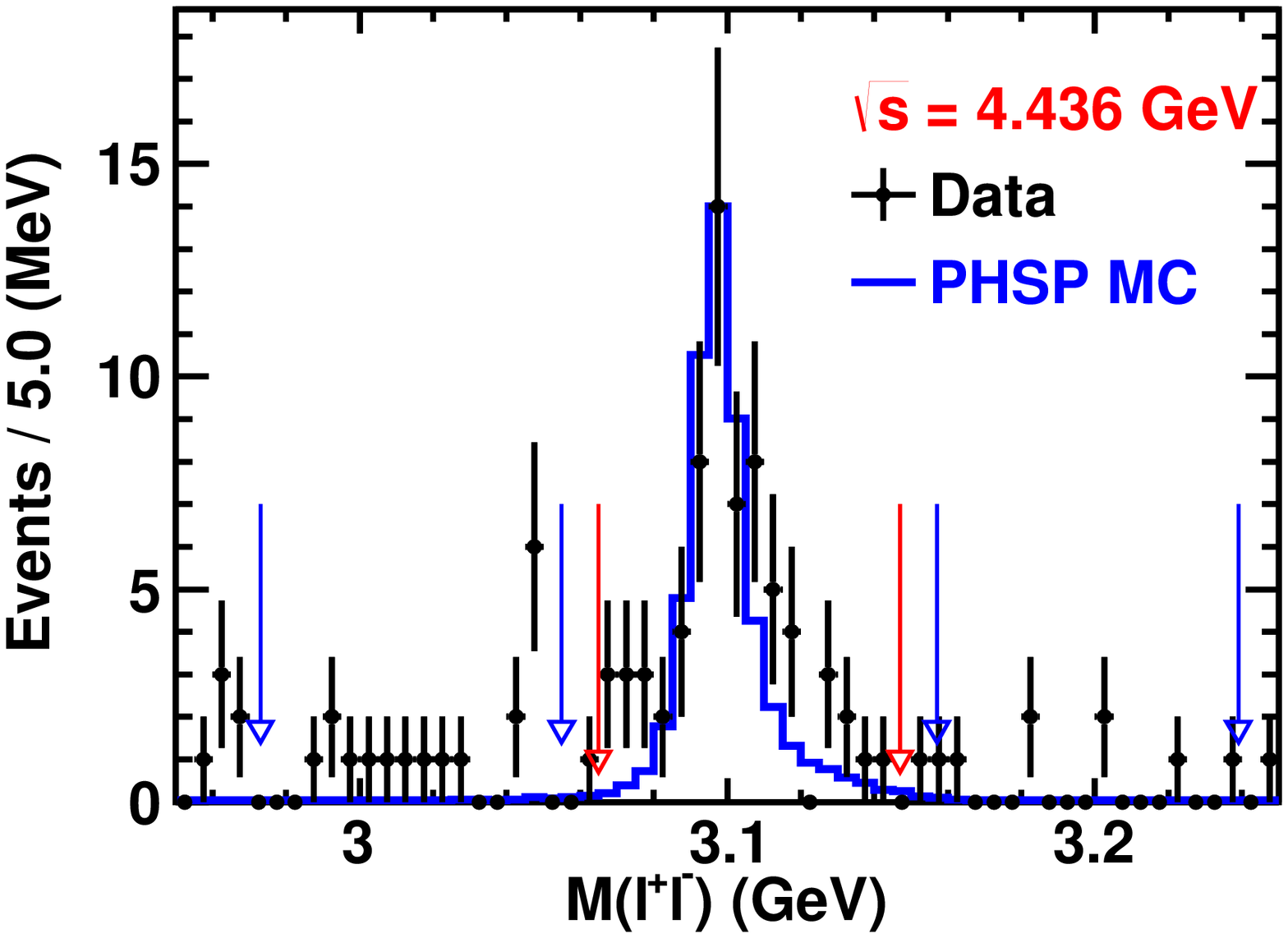}
	\includegraphics[angle=0,width=0.24\textwidth]{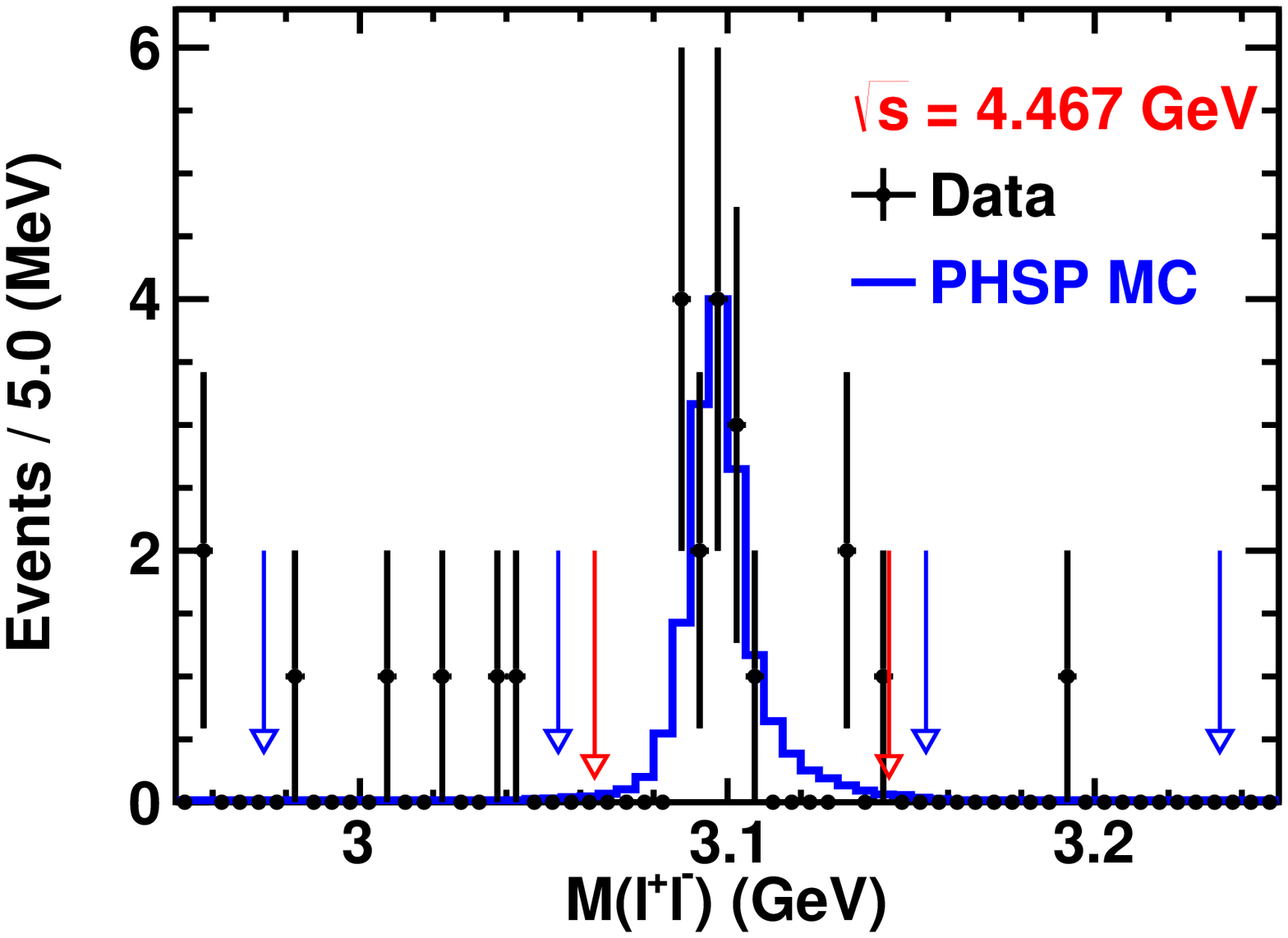}
	\includegraphics[angle=0,width=0.24\textwidth]{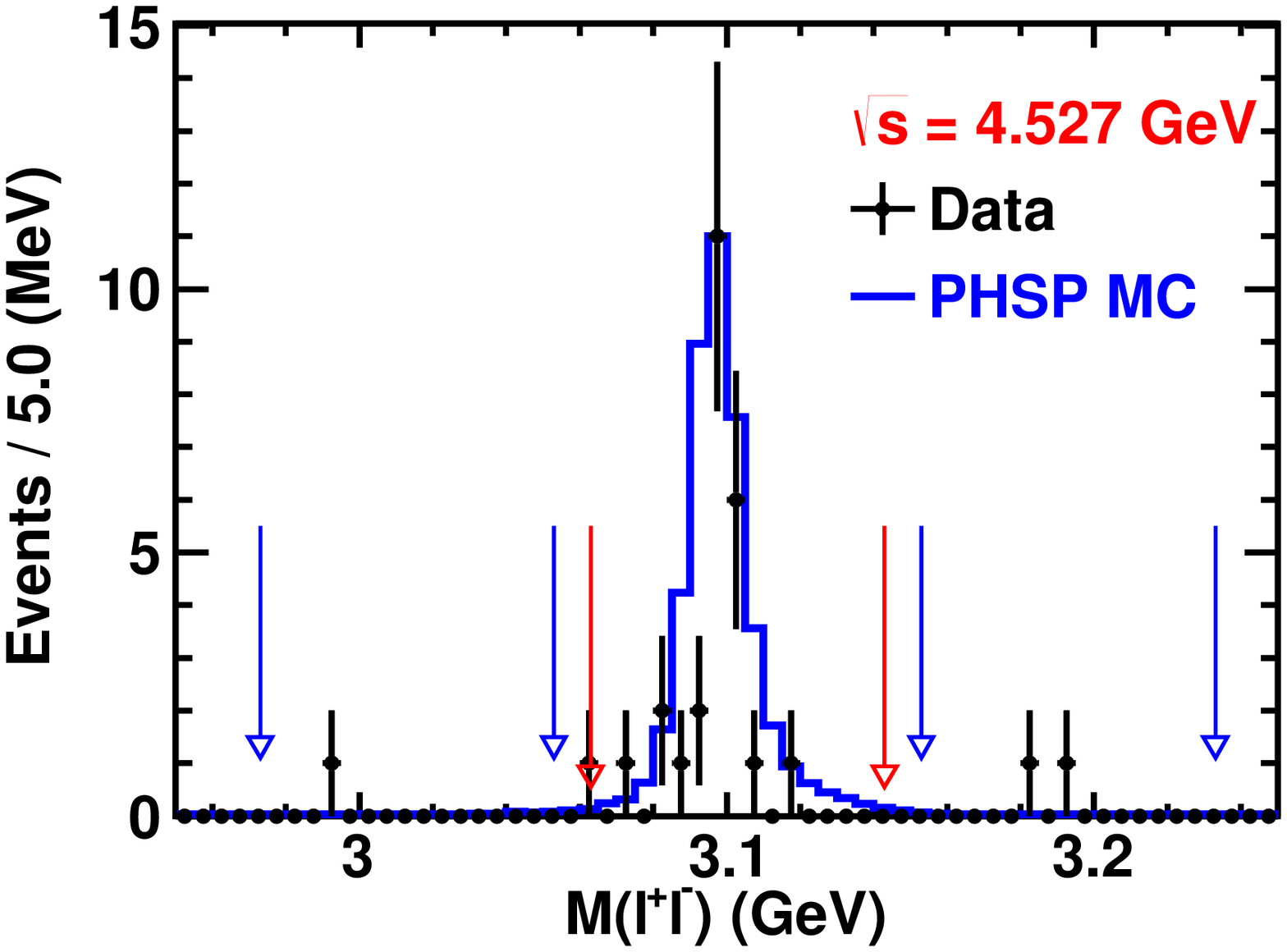}
	\includegraphics[angle=0,width=0.24\textwidth]{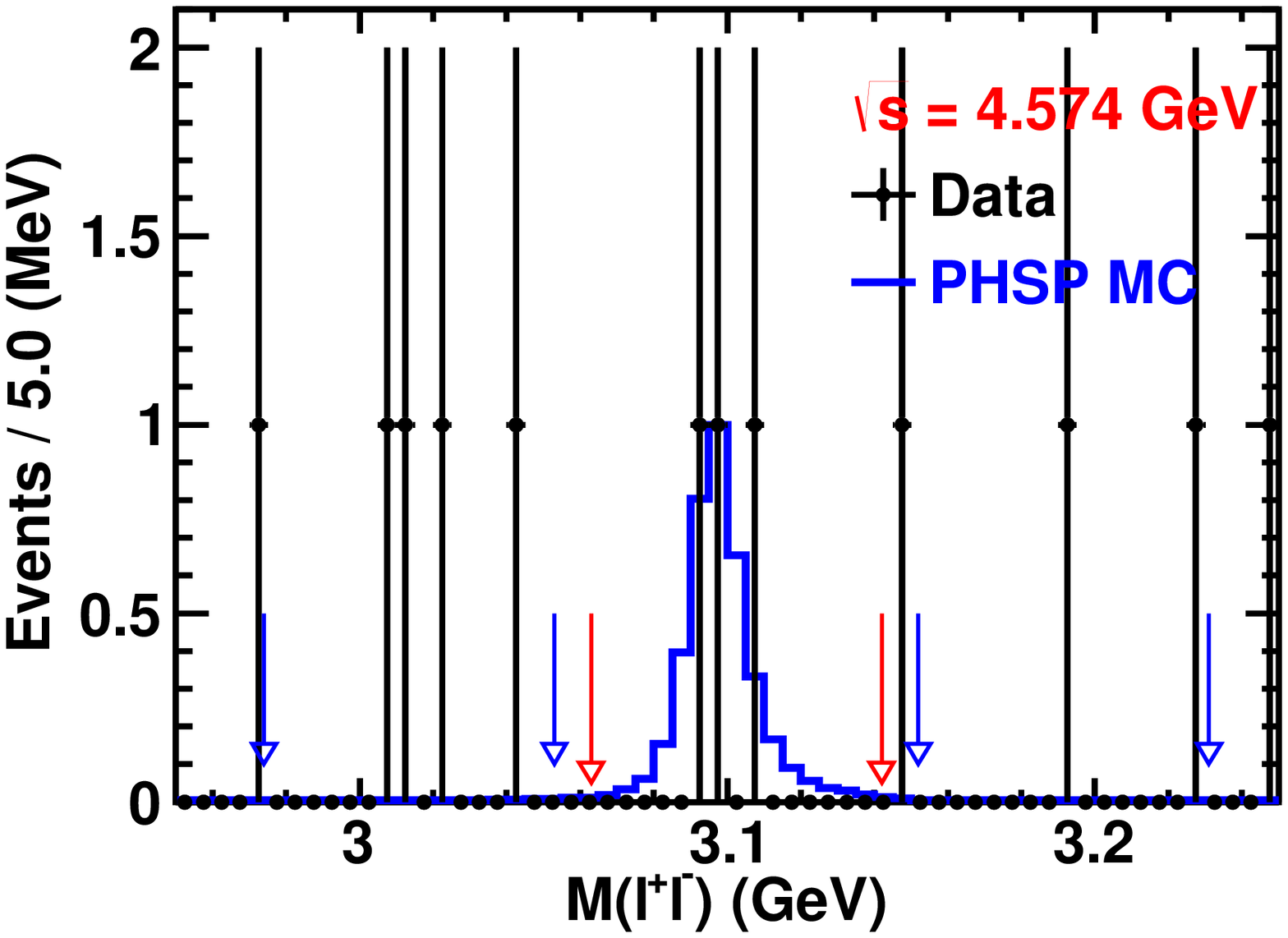}
	\includegraphics[angle=0,width=0.24\textwidth]{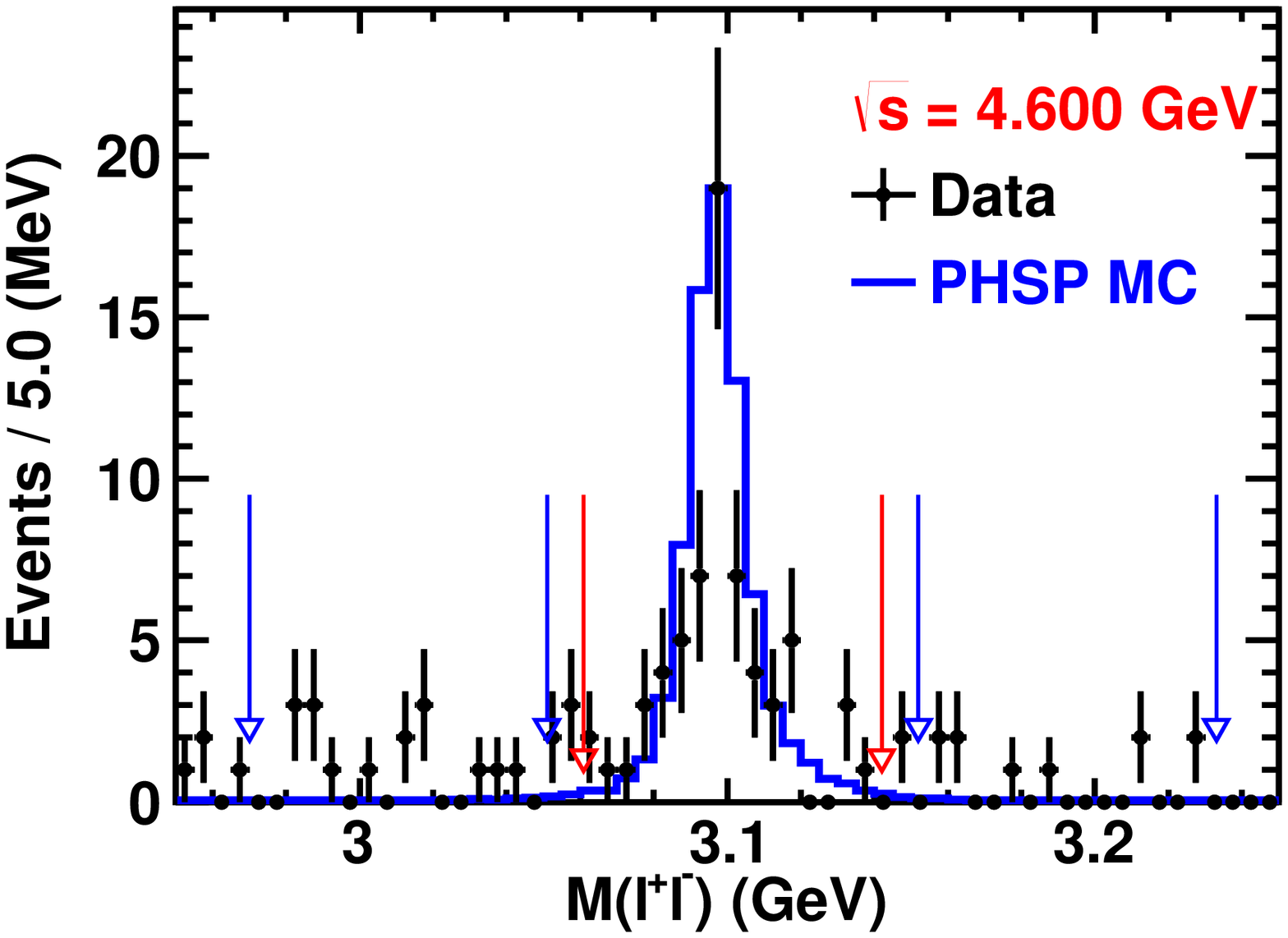}
	\caption{The distributions of invariant mass of lepton pairs $M(\ell^+\ell^-)$, where the dots with error bars are data, the blue histogram is PHSP signal MC sample (normalized to data.), the red and blue arrows identify the $J/\psi$ signal and sideband regions, respectively.}
\label{Fig-sig-yield}
\end{figure*}

\begin{figure*}[htbp]
 	\centering
 	\includegraphics[angle=0,width=0.24\textwidth]{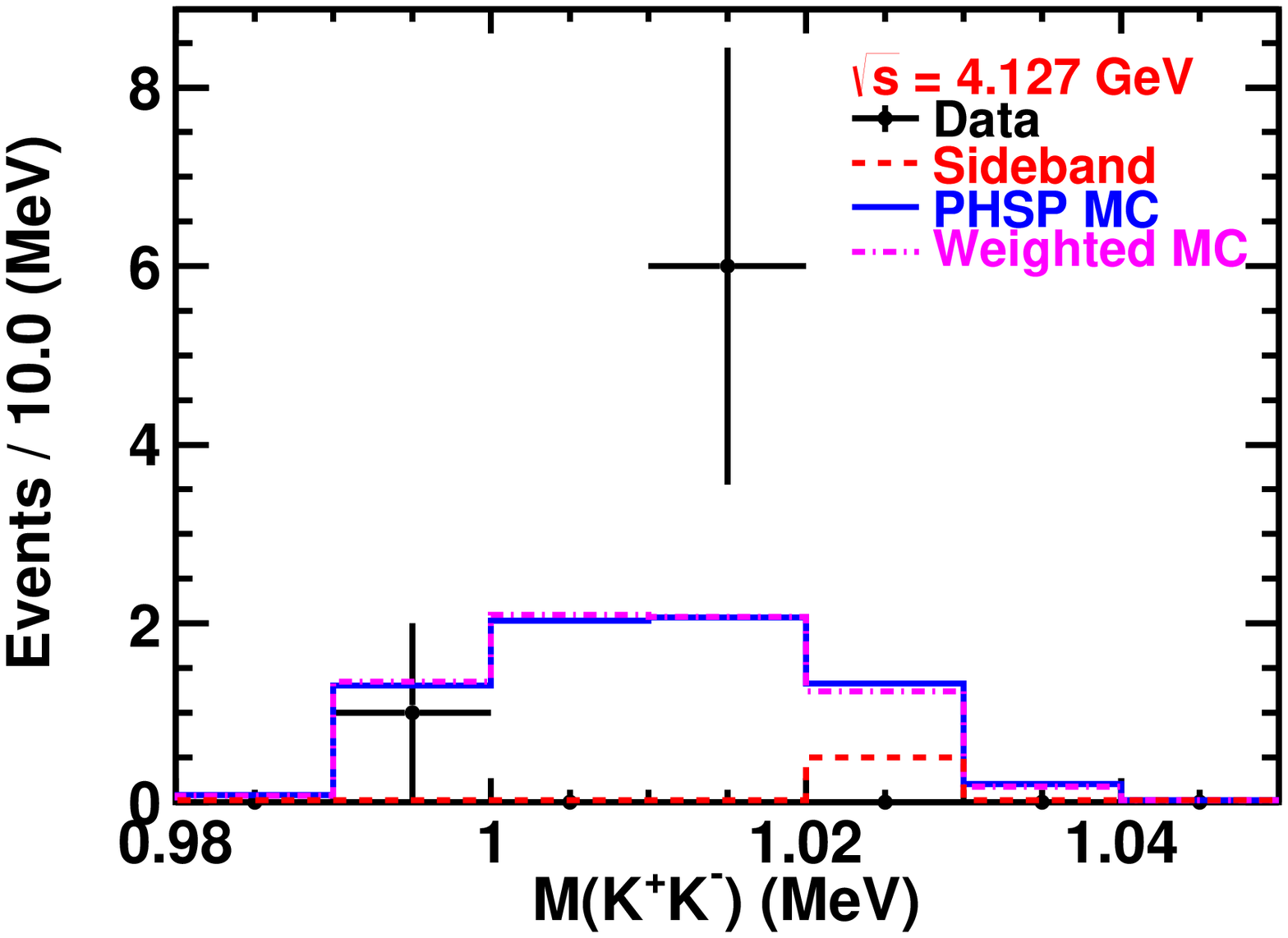}
	\includegraphics[angle=0,width=0.24\textwidth]{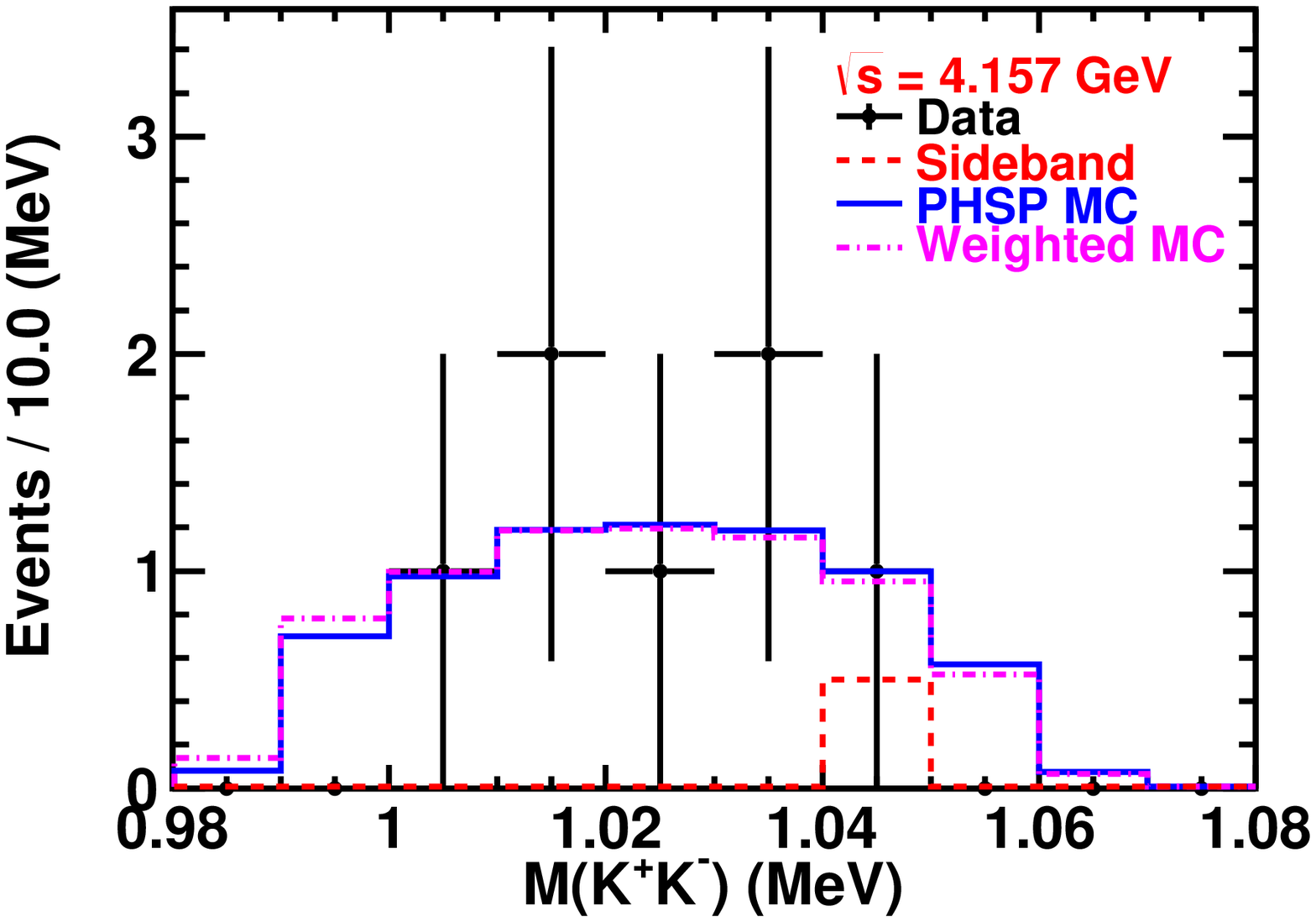}
	\includegraphics[angle=0,width=0.24\textwidth]{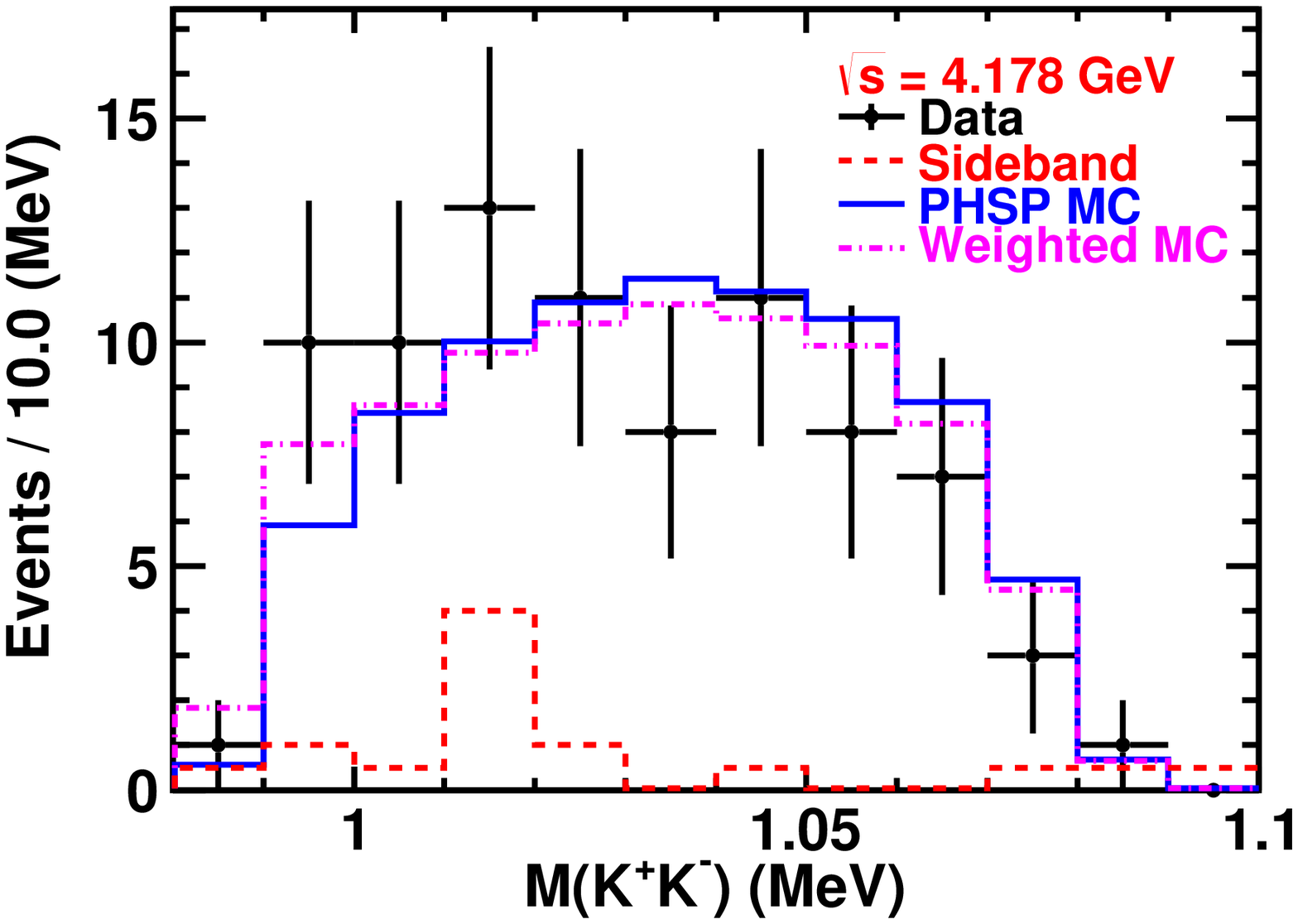}
	\includegraphics[angle=0,width=0.24\textwidth]{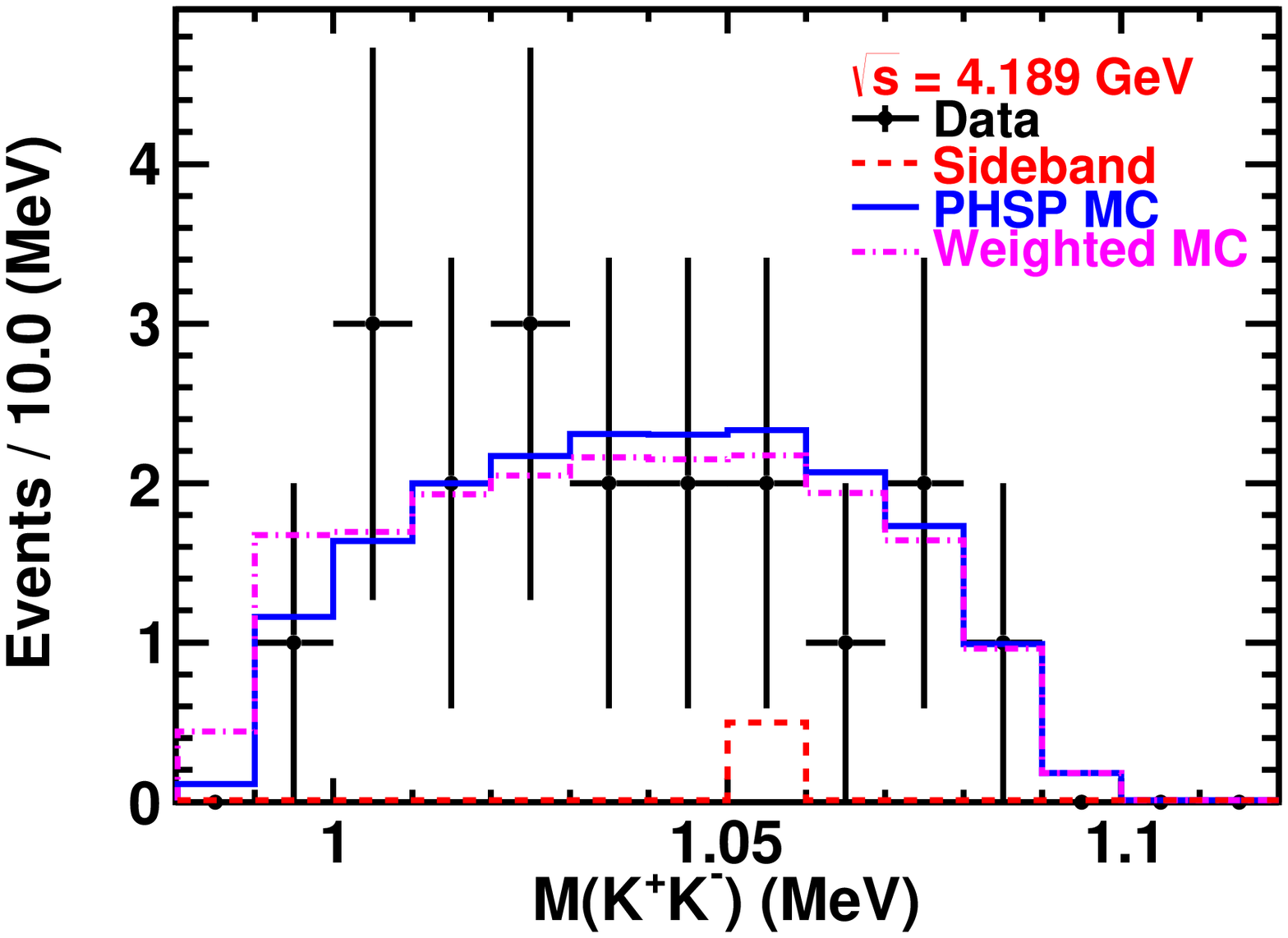}
	\includegraphics[angle=0,width=0.24\textwidth]{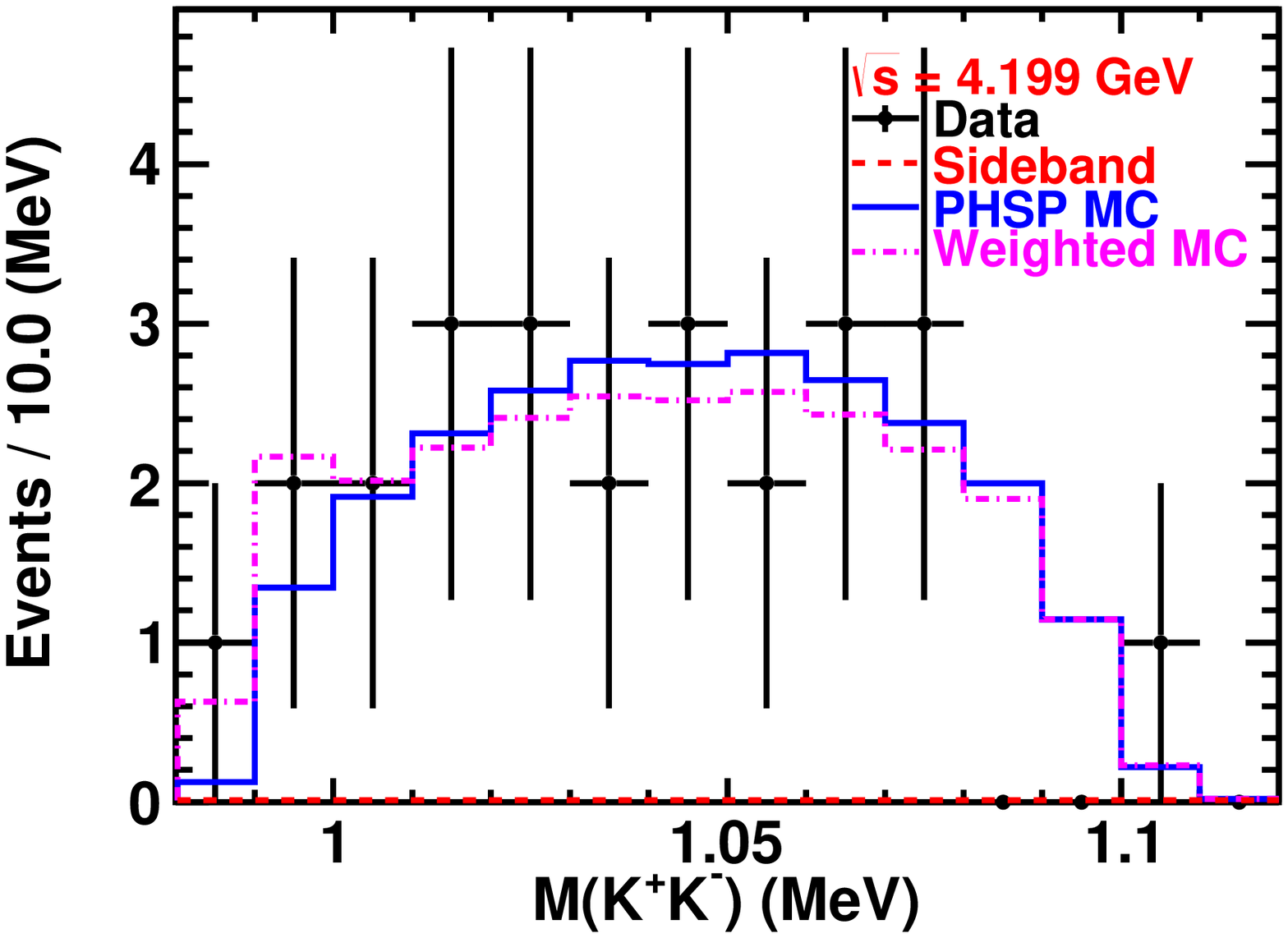}
	\includegraphics[angle=0,width=0.24\textwidth]{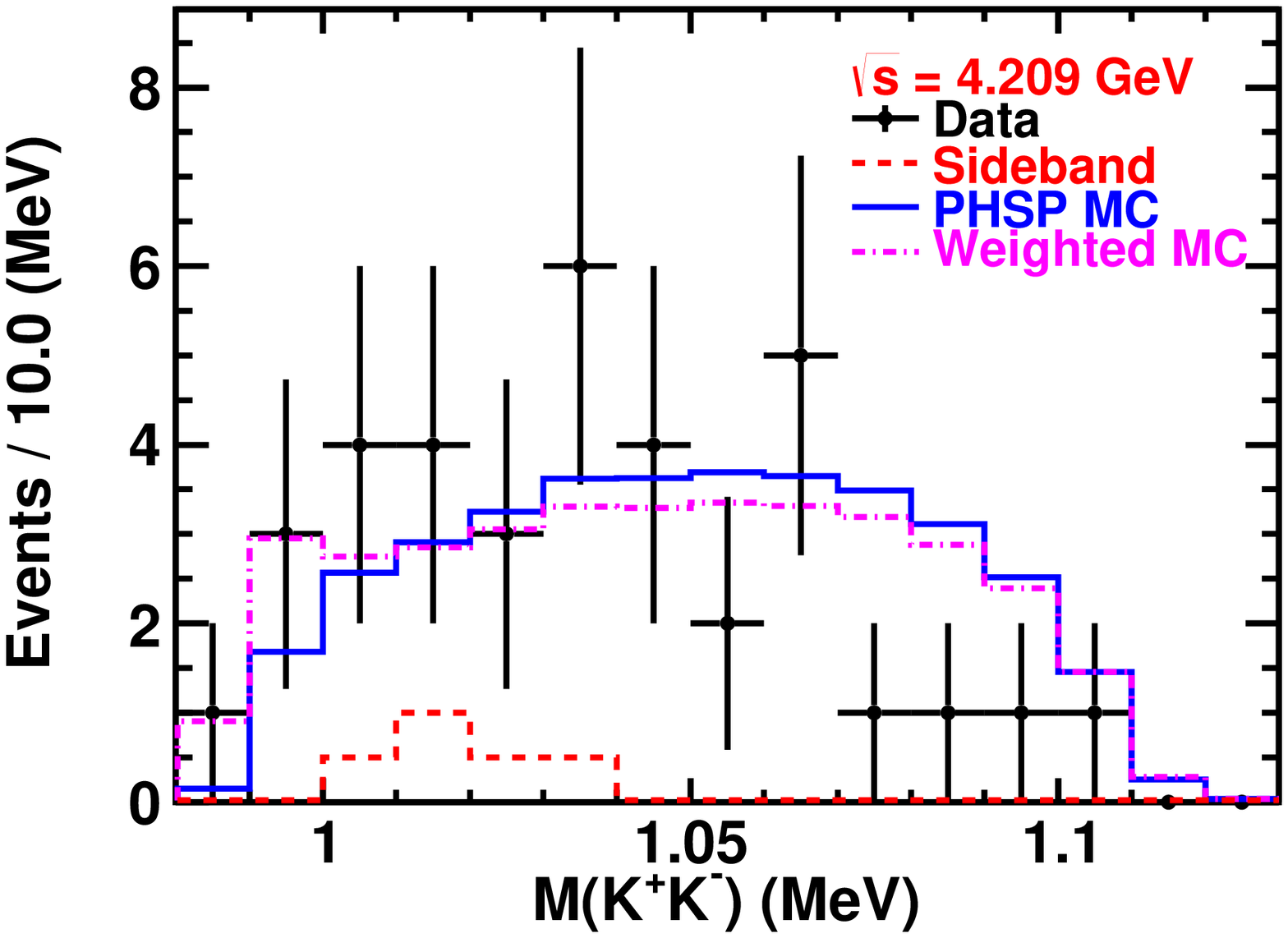}
 	\includegraphics[angle=0,width=0.24\textwidth]{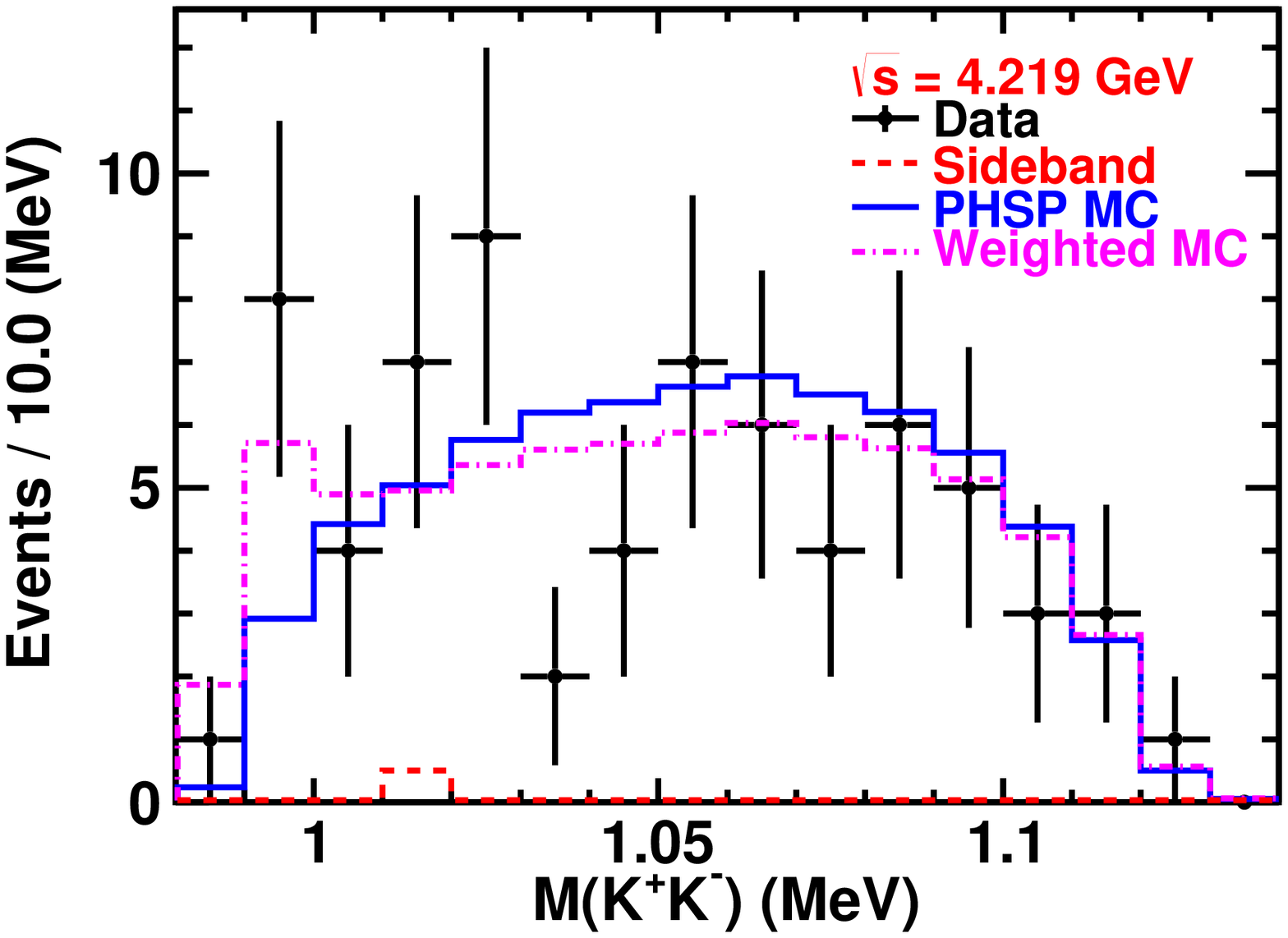}
	\includegraphics[angle=0,width=0.24\textwidth]{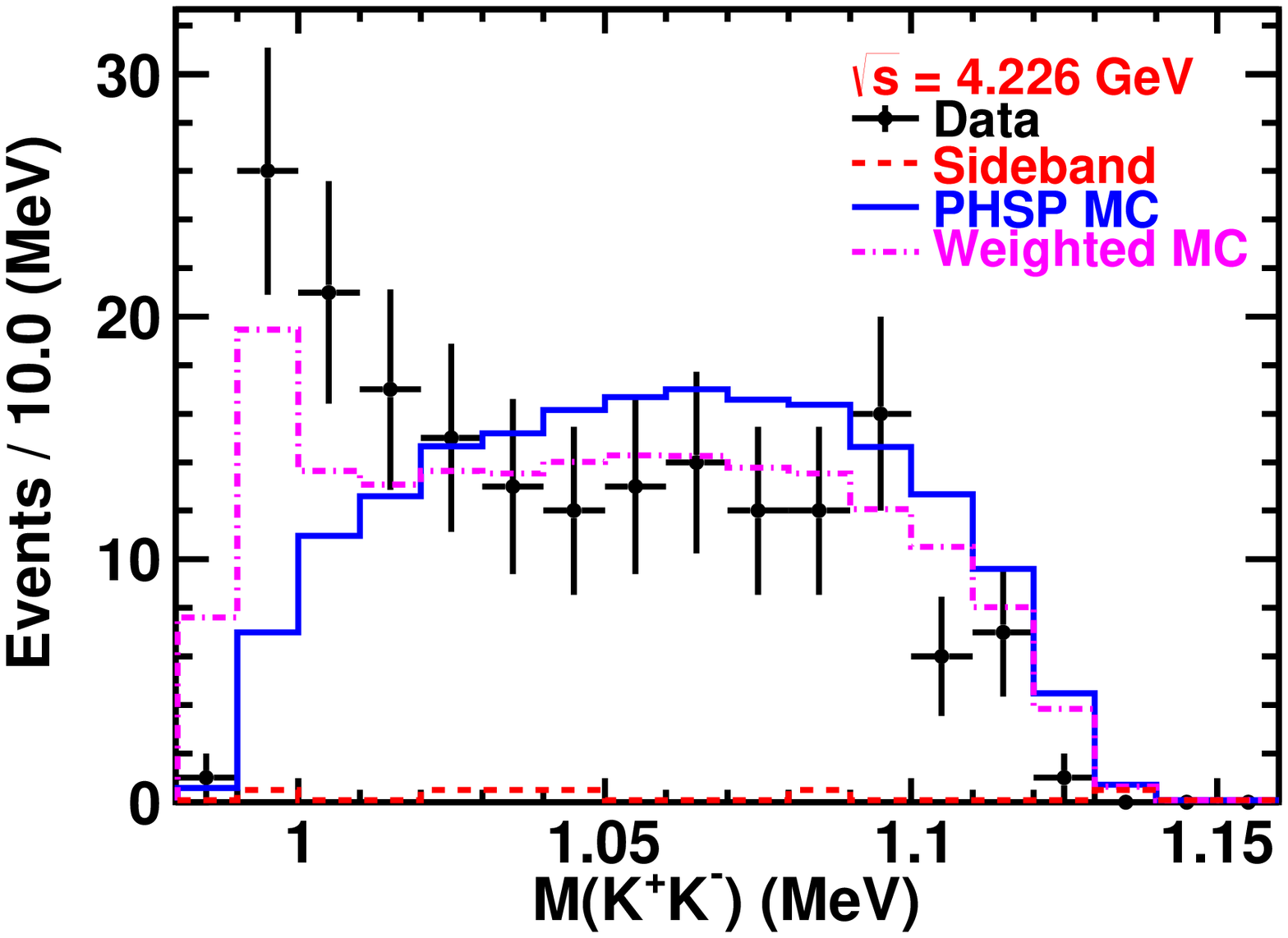}
	\includegraphics[angle=0,width=0.24\textwidth]{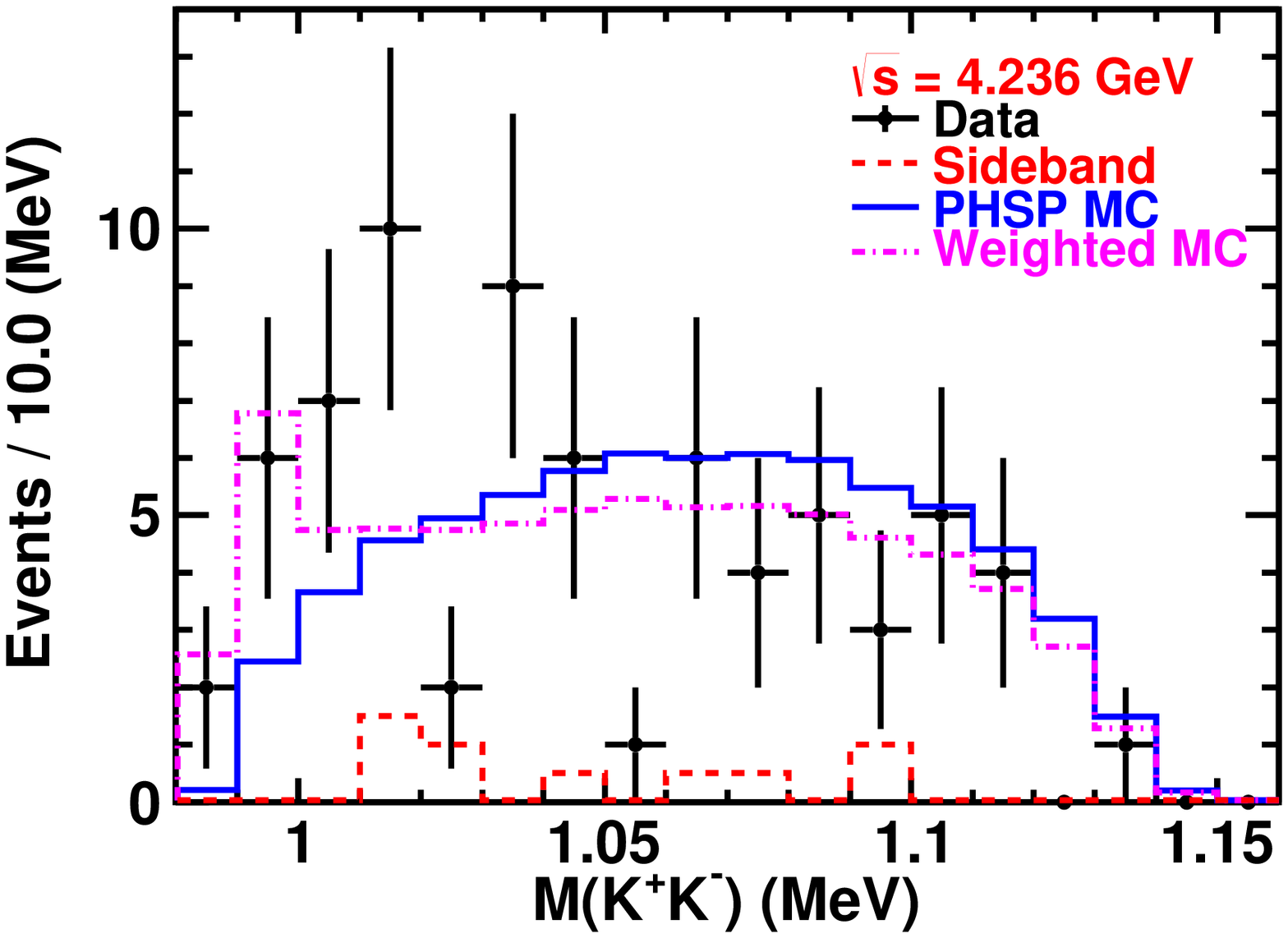}
	\includegraphics[angle=0,width=0.24\textwidth]{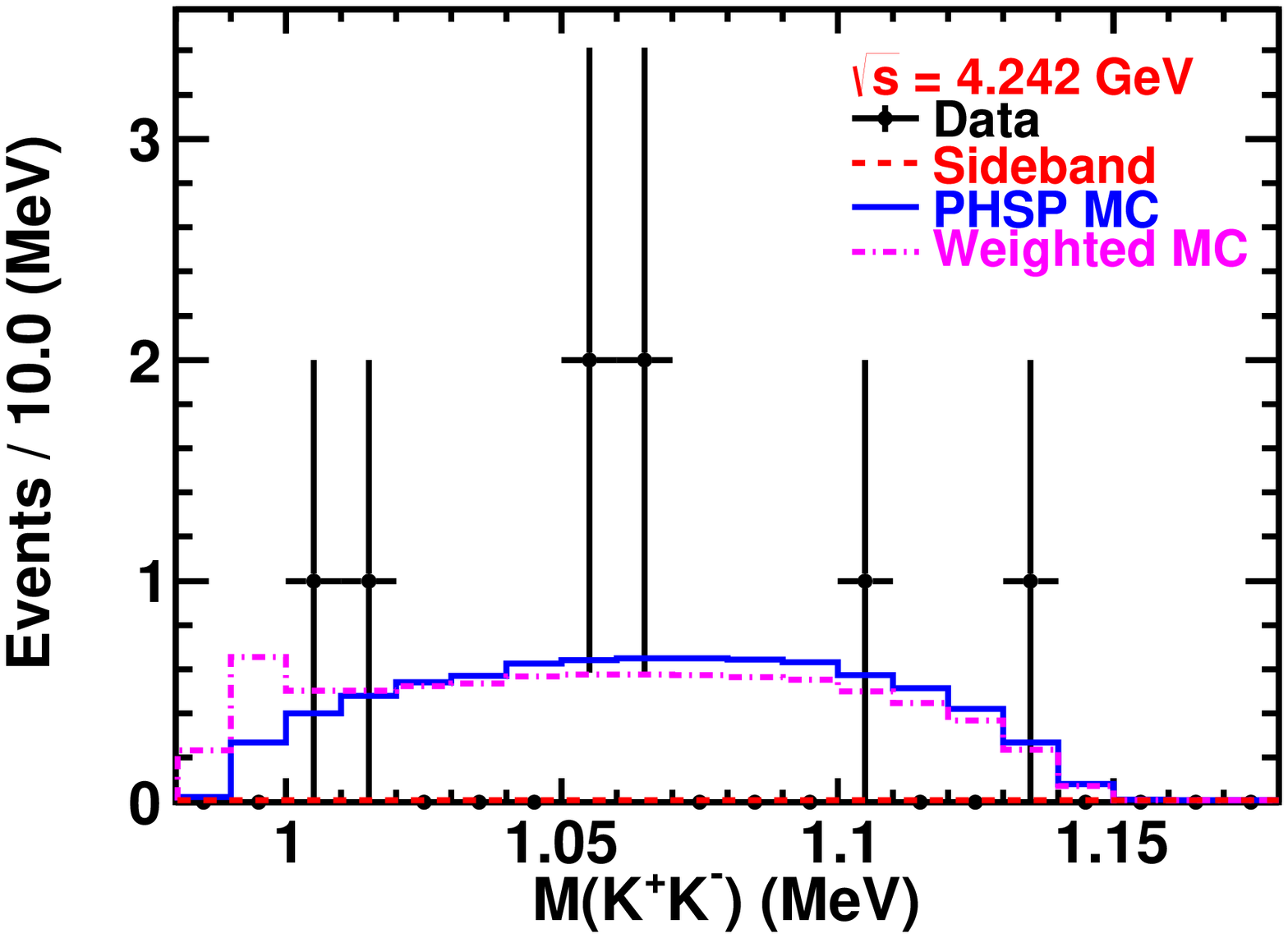}
	\includegraphics[angle=0,width=0.24\textwidth]{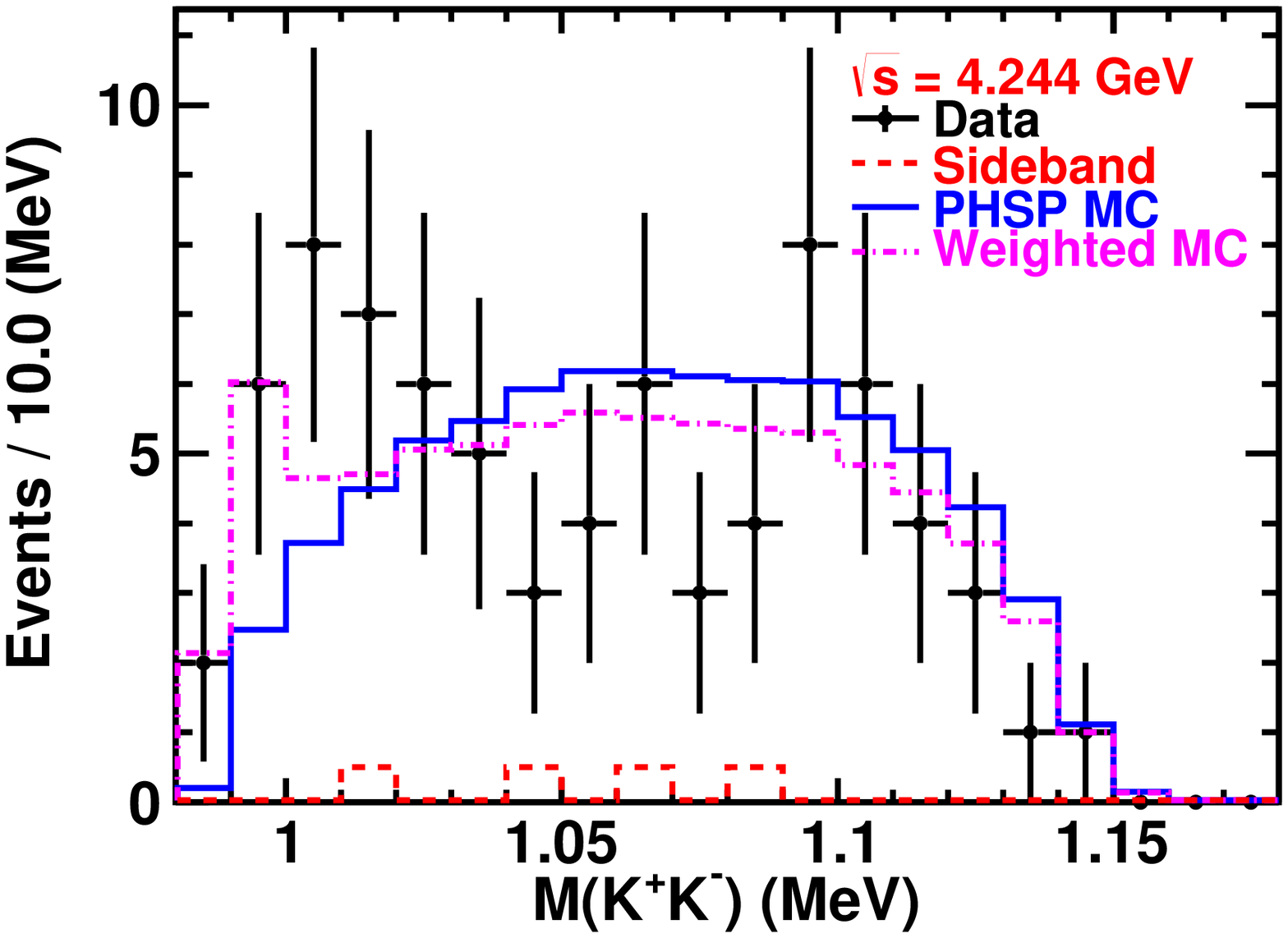}
	\includegraphics[angle=0,width=0.24\textwidth]{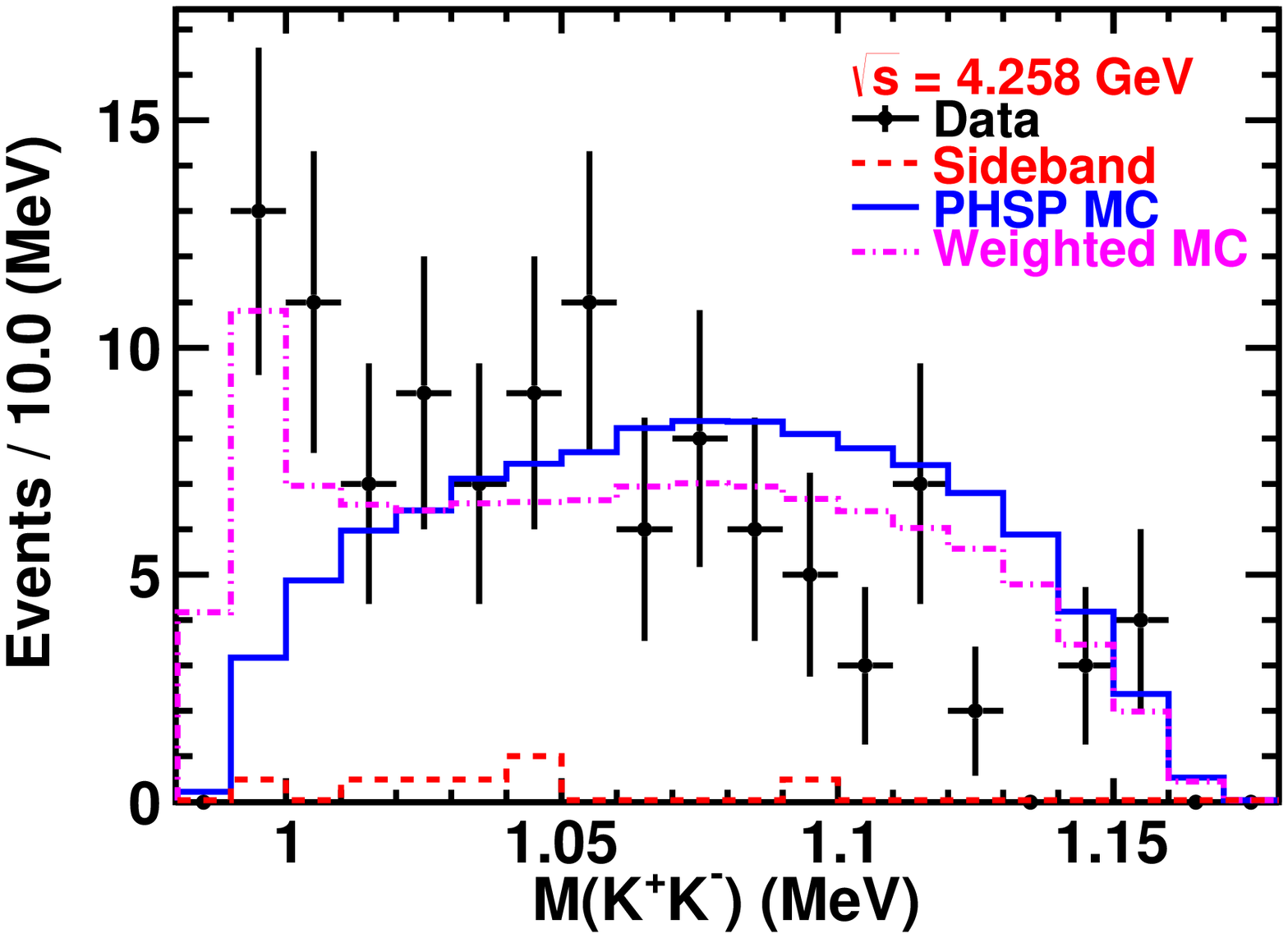}
 	\includegraphics[angle=0,width=0.24\textwidth]{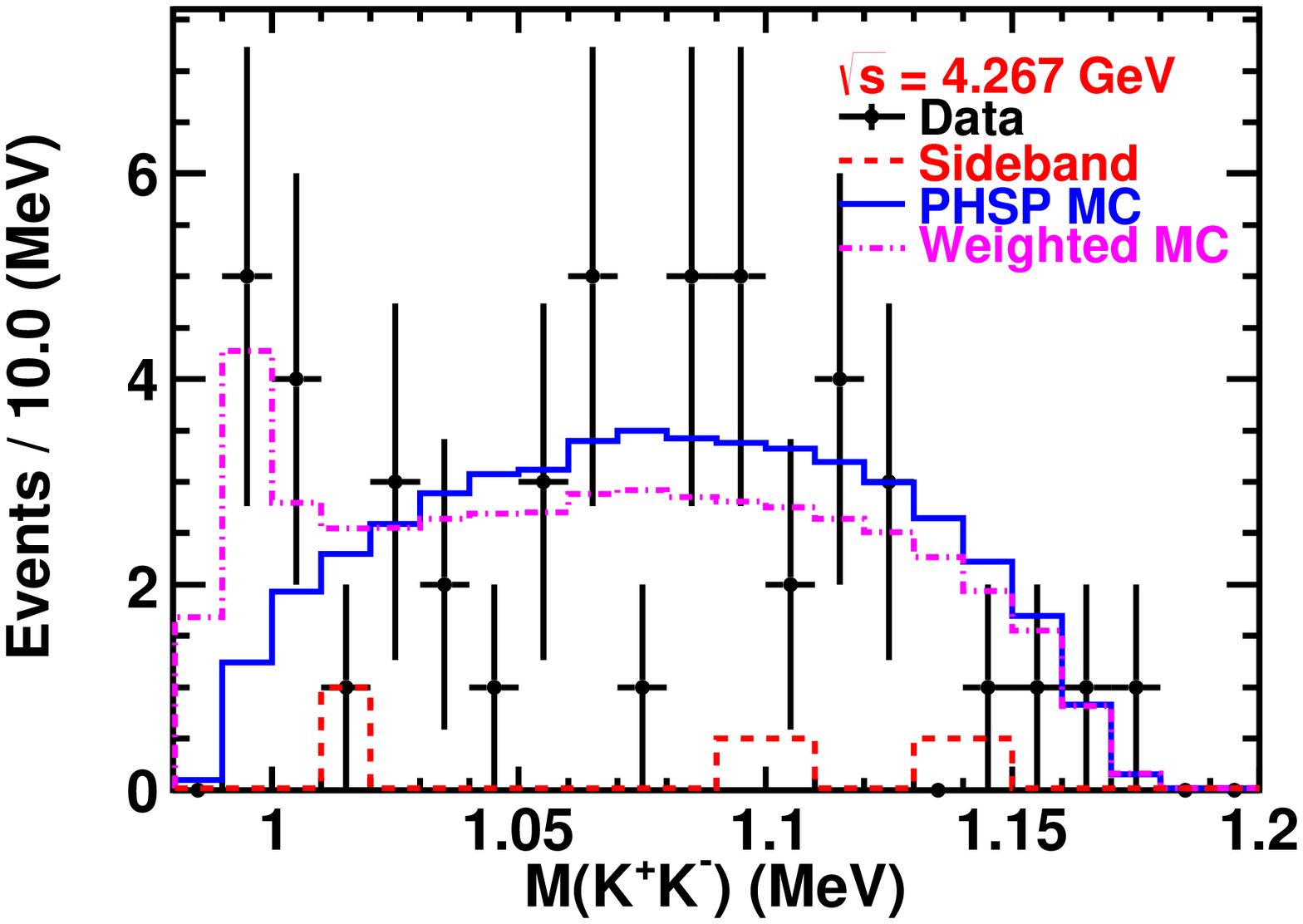}
	\includegraphics[angle=0,width=0.24\textwidth]{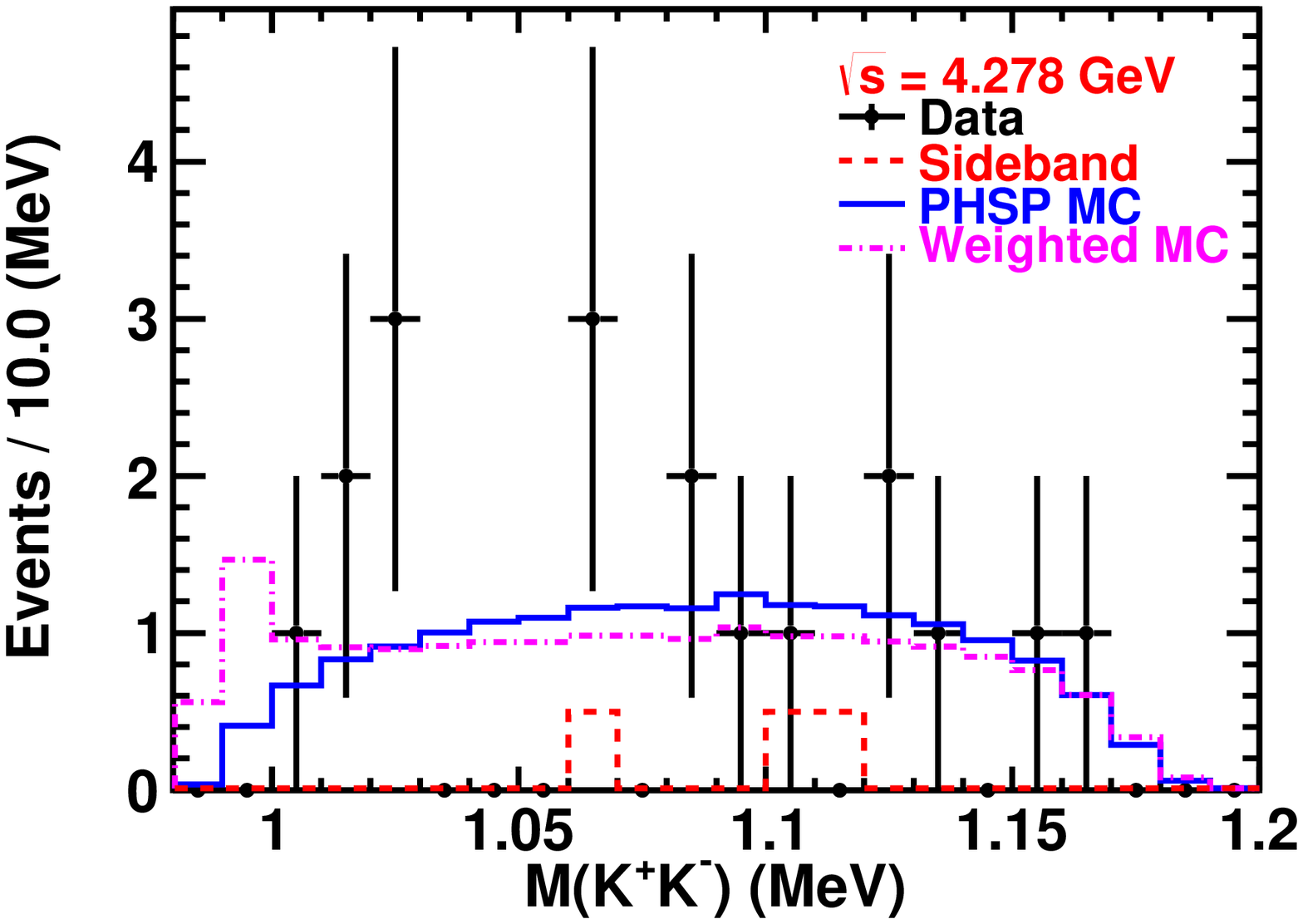}
	\includegraphics[angle=0,width=0.24\textwidth]{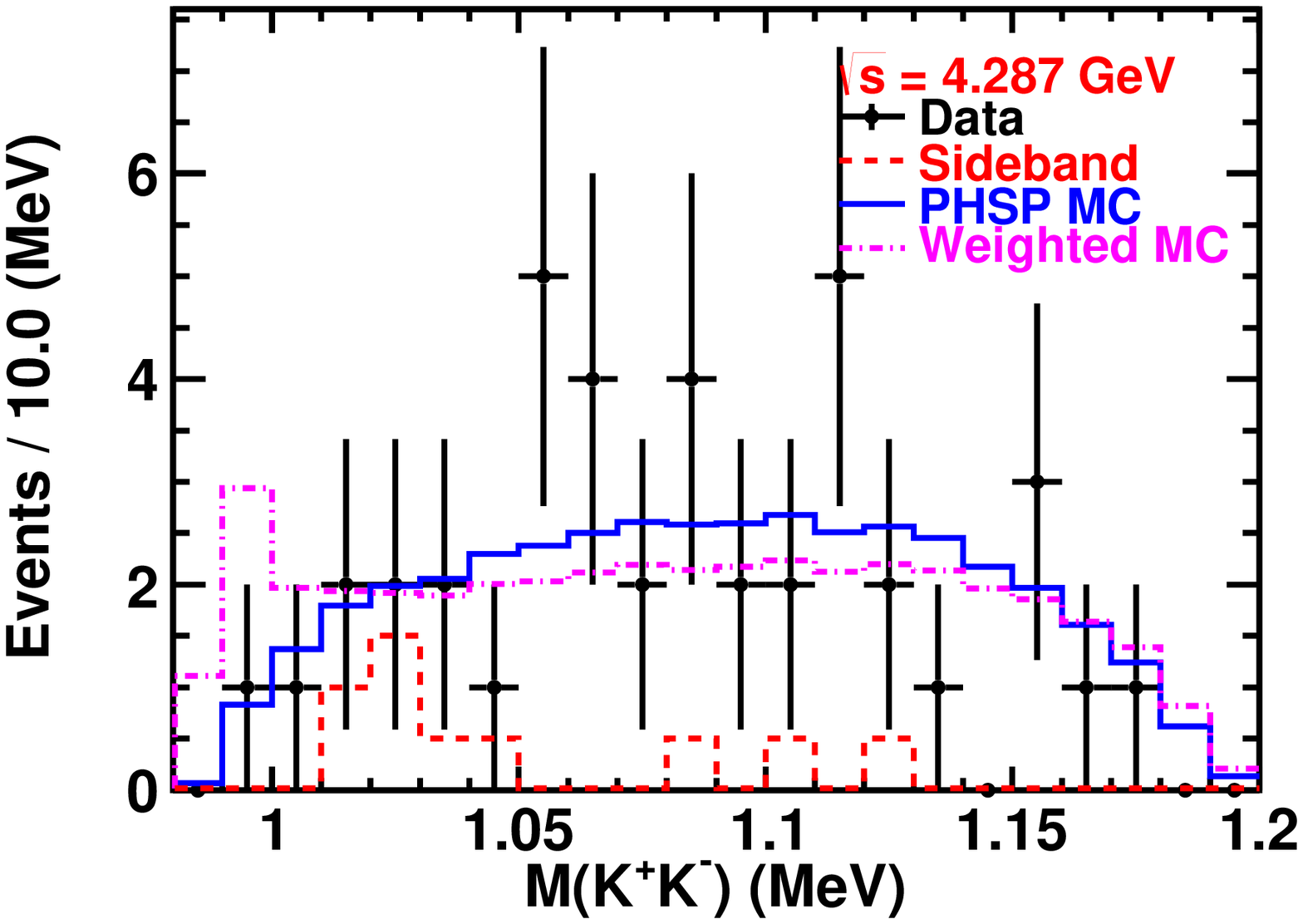}
	\includegraphics[angle=0,width=0.24\textwidth]{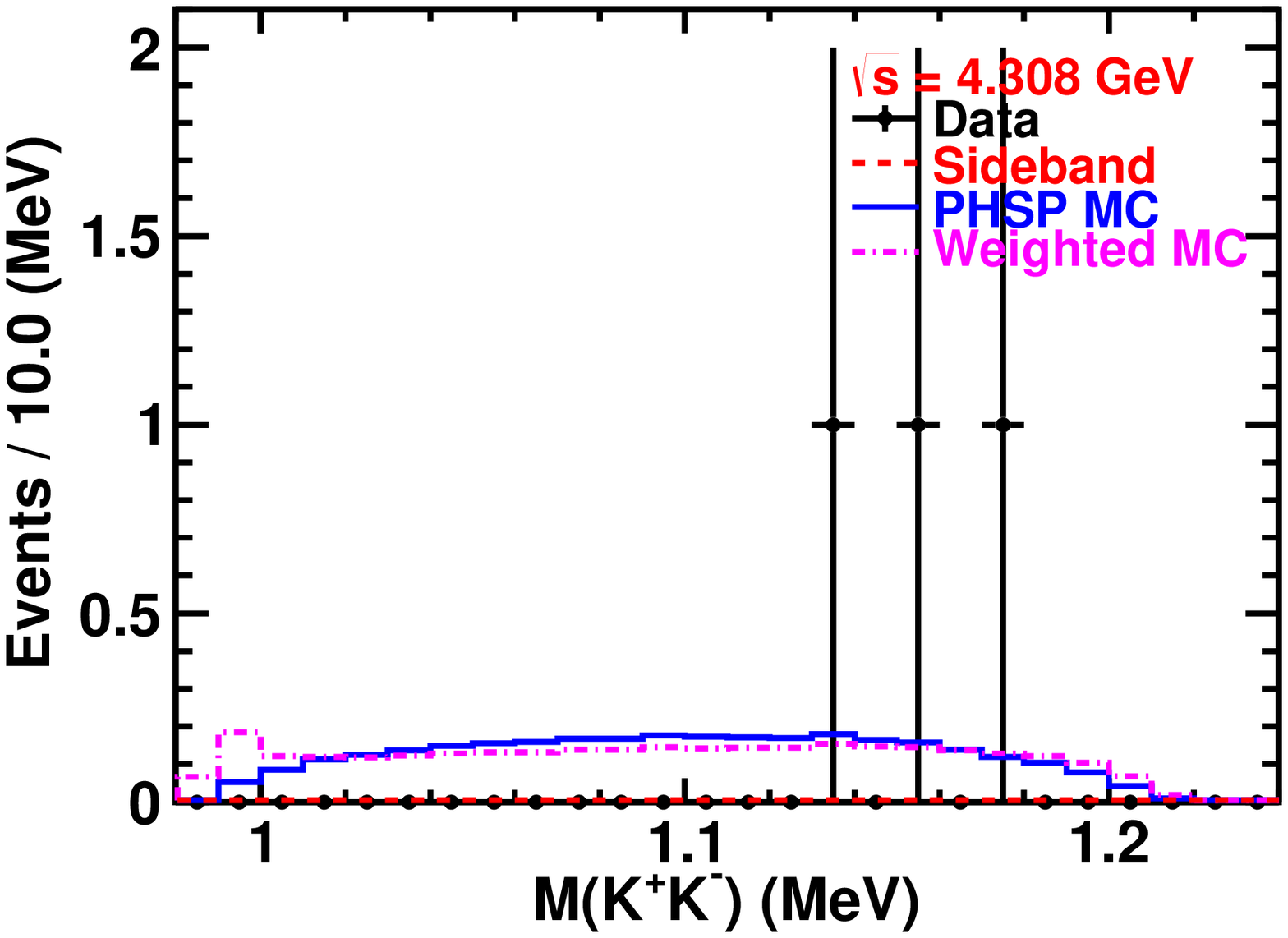}
	\includegraphics[angle=0,width=0.24\textwidth]{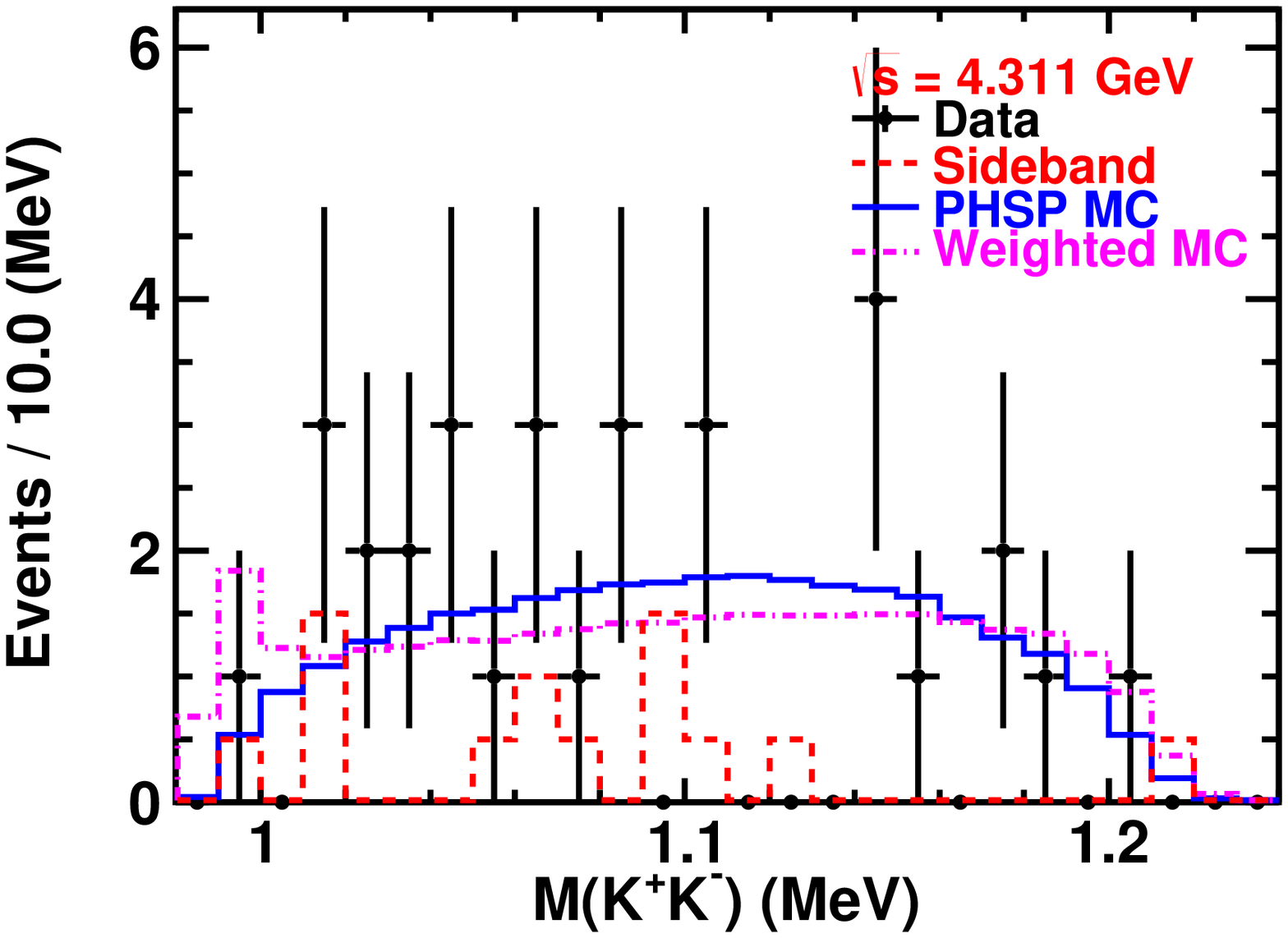}
	\includegraphics[angle=0,width=0.24\textwidth]{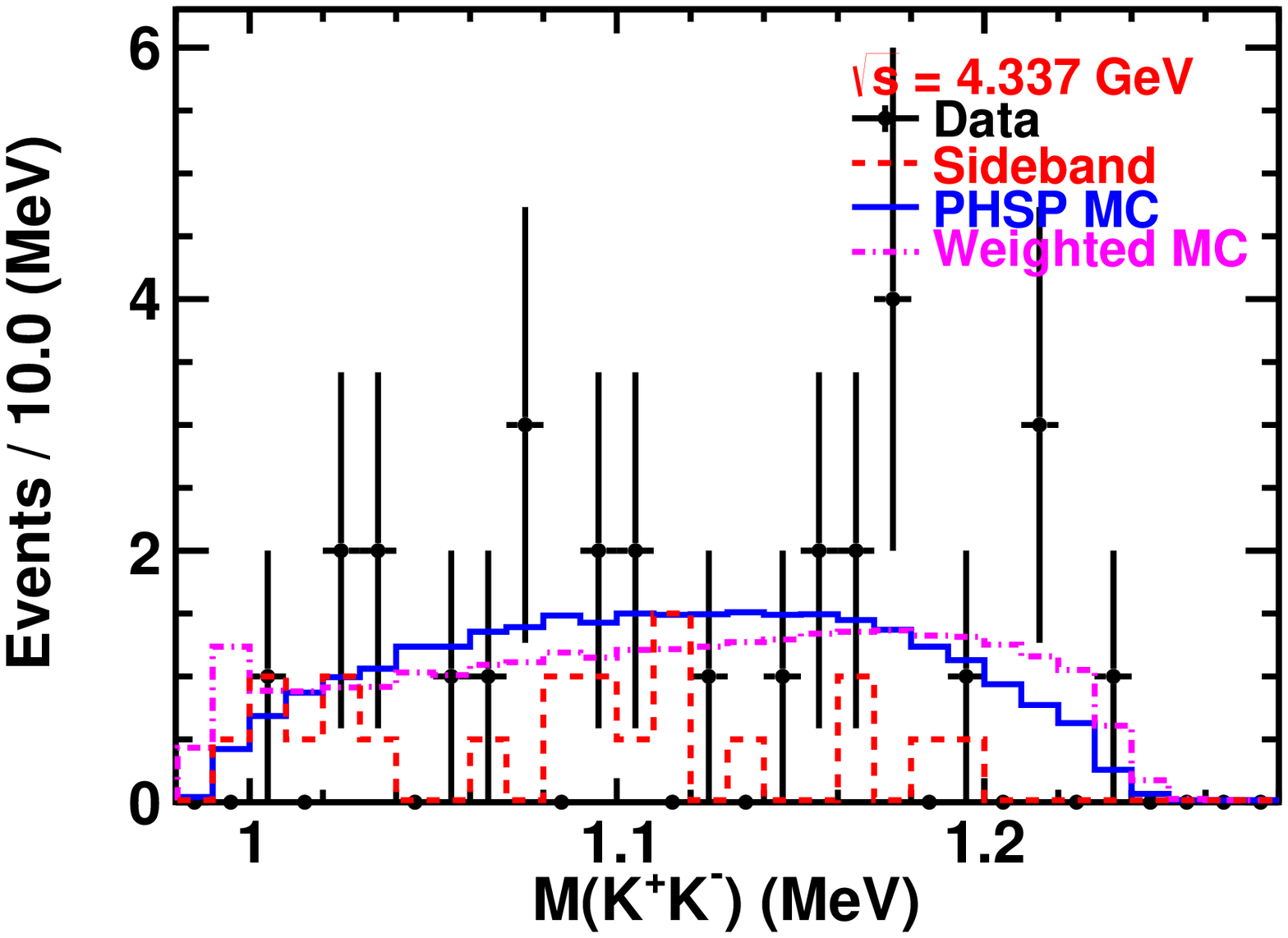}
 	\includegraphics[angle=0,width=0.24\textwidth]{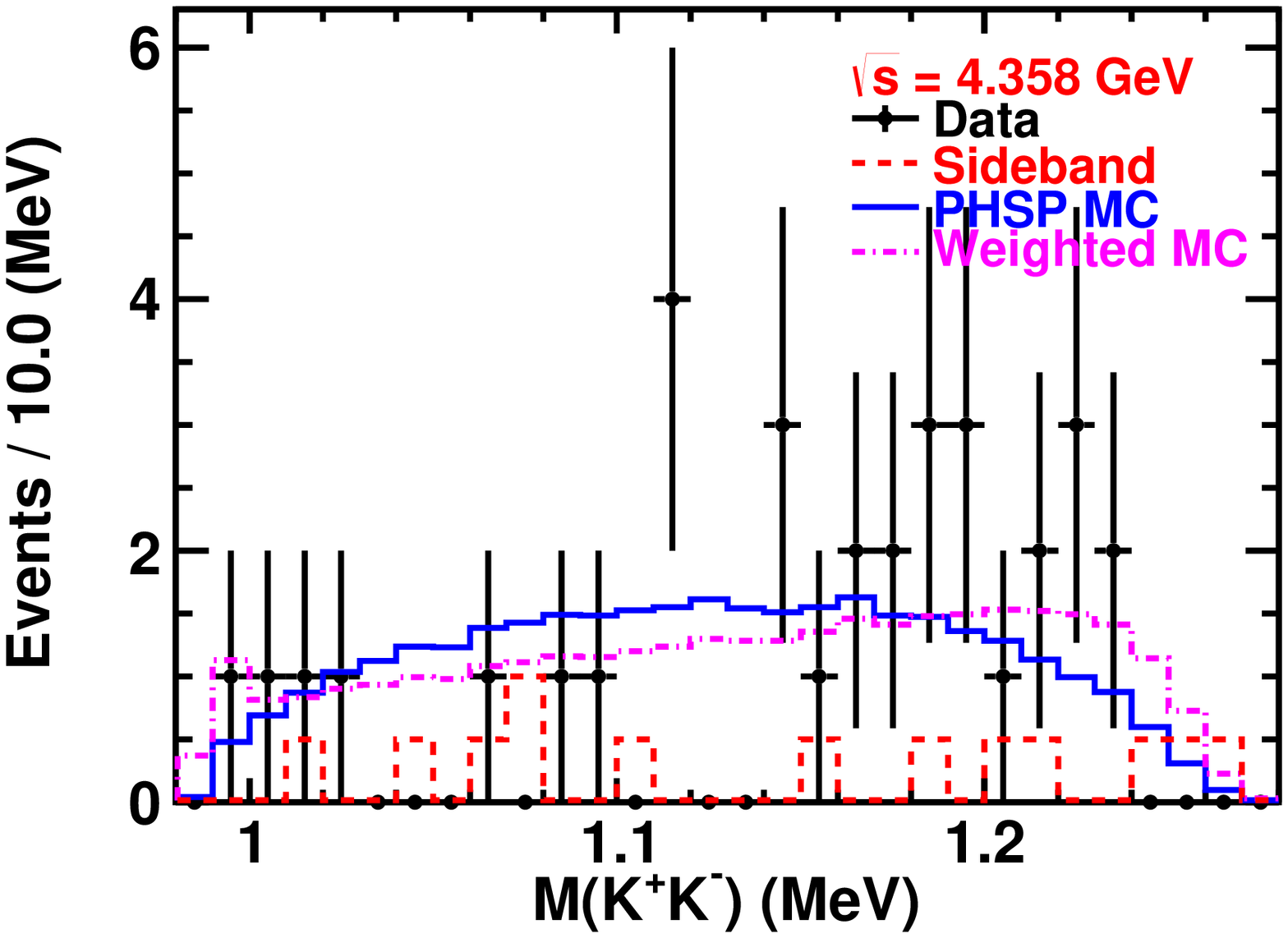}
	\includegraphics[angle=0,width=0.24\textwidth]{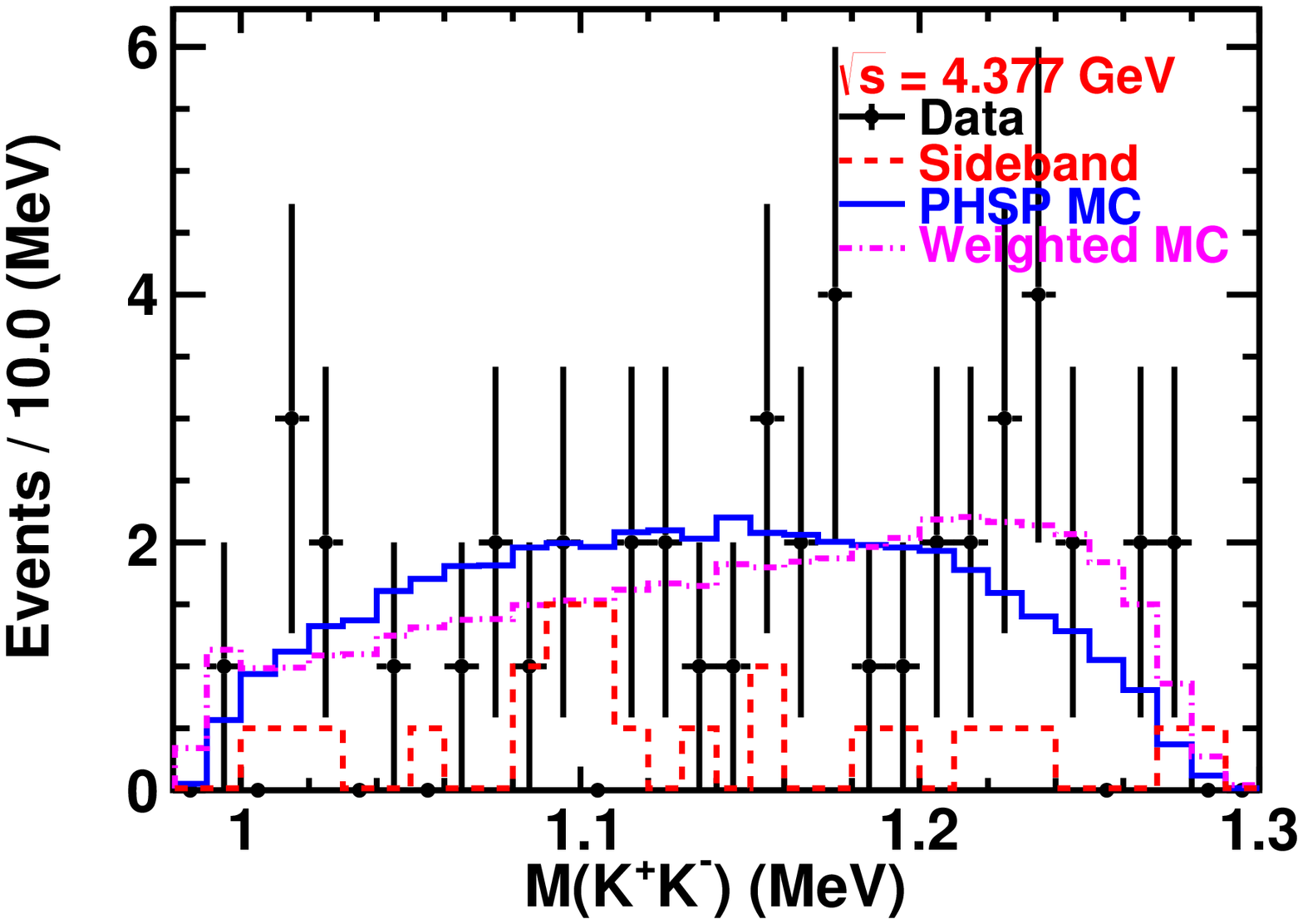}
	\includegraphics[angle=0,width=0.24\textwidth]{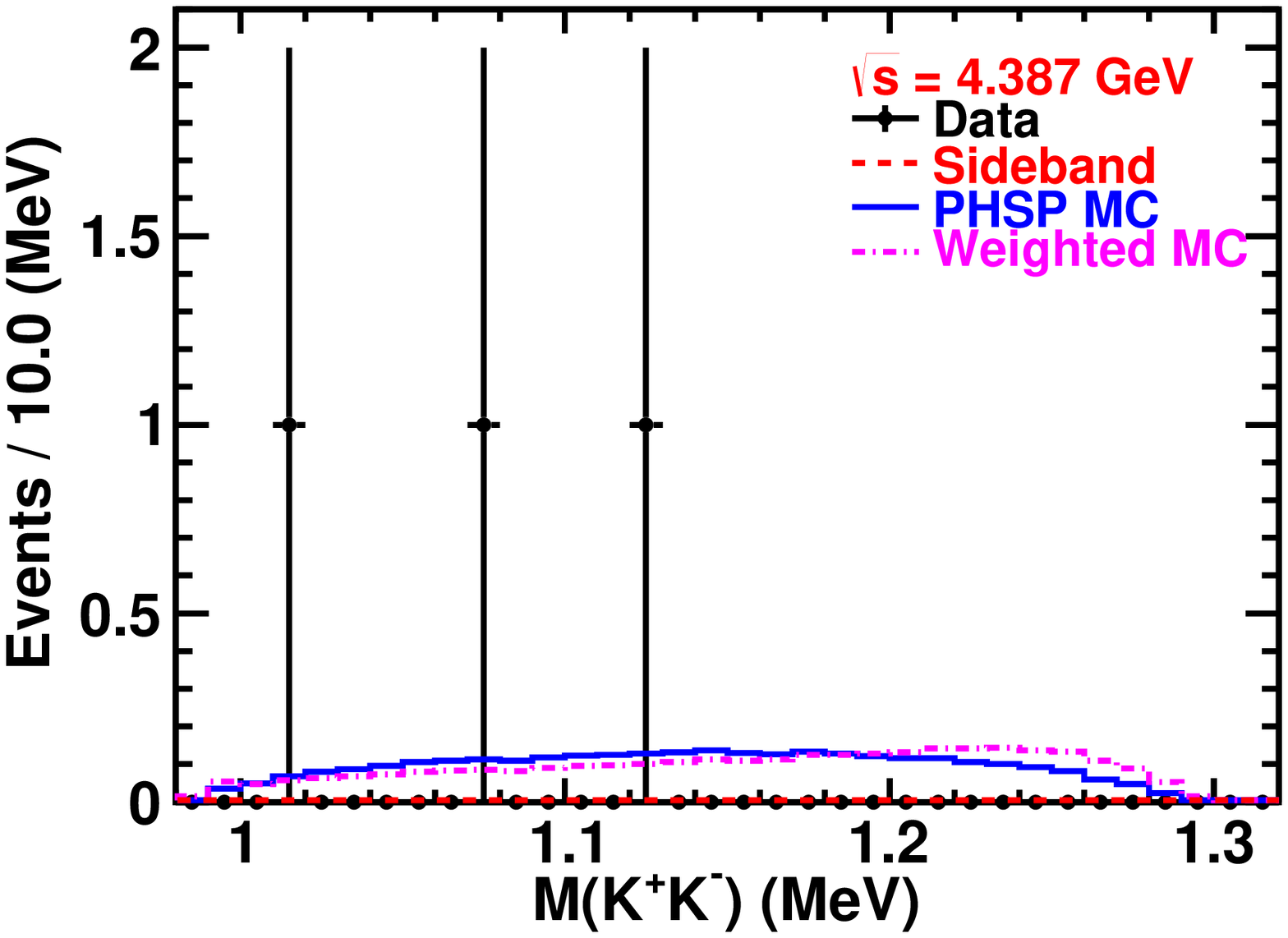}
	\includegraphics[angle=0,width=0.24\textwidth]{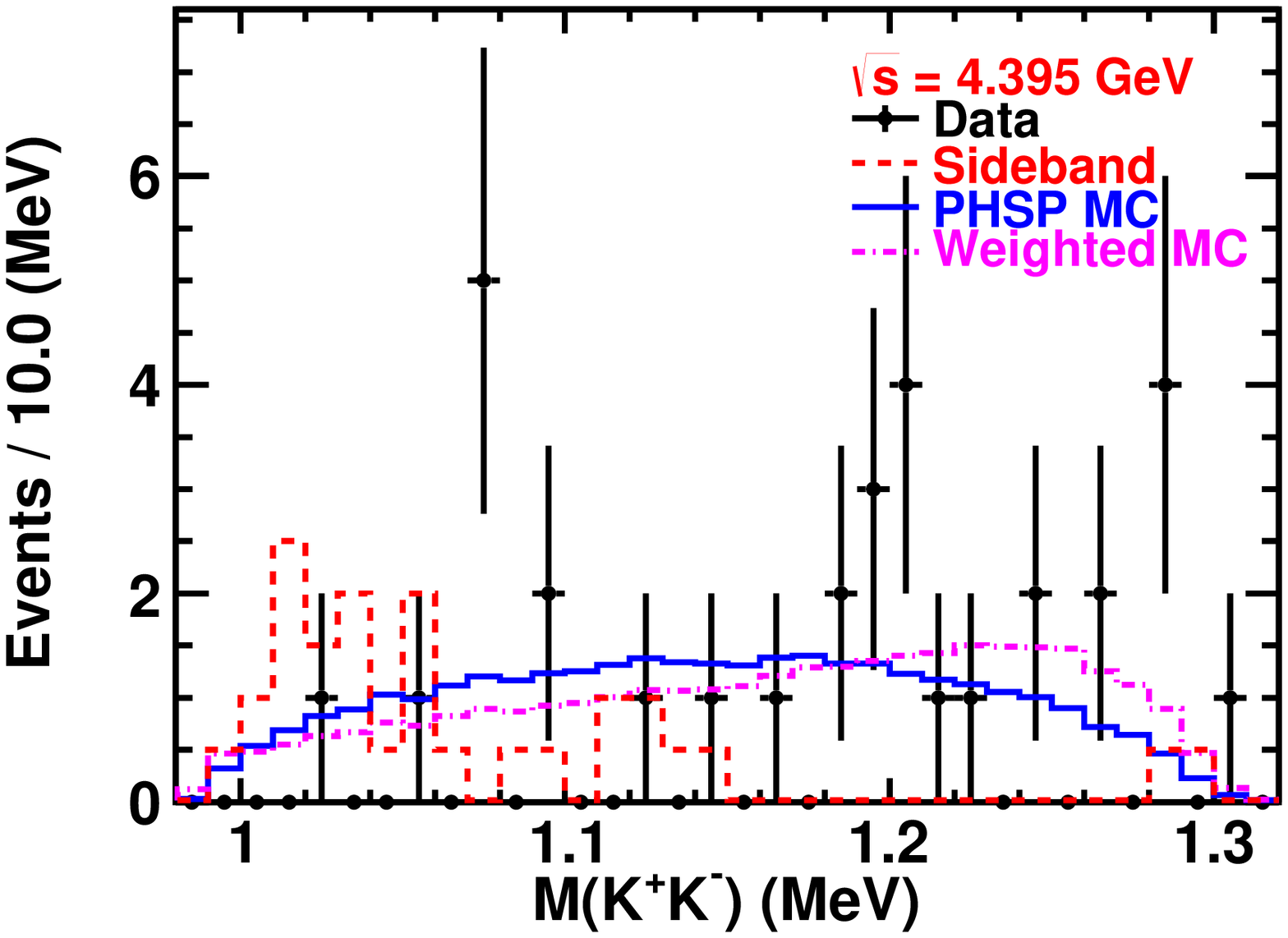}
	\includegraphics[angle=0,width=0.24\textwidth]{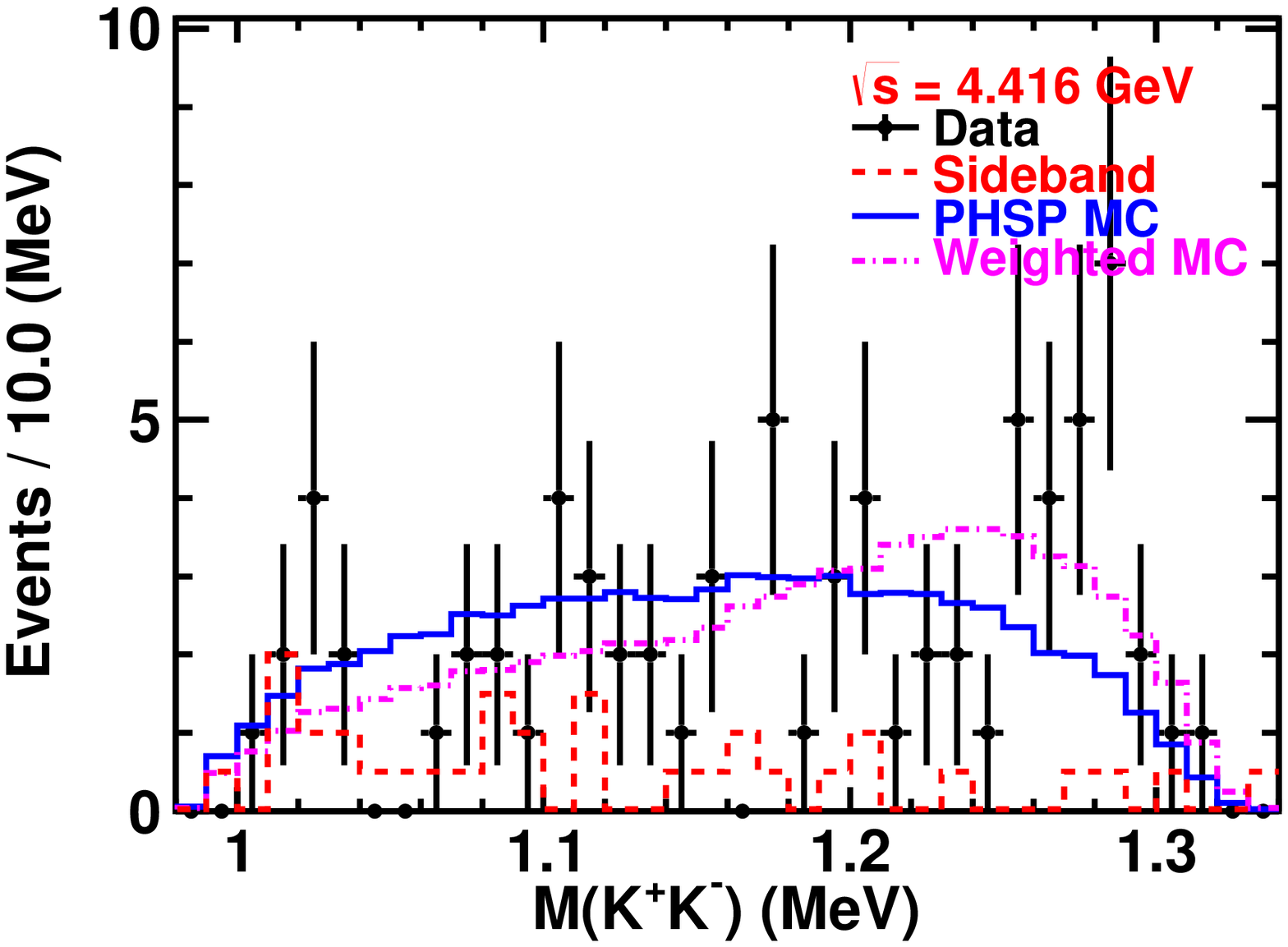}
	\includegraphics[angle=0,width=0.24\textwidth]{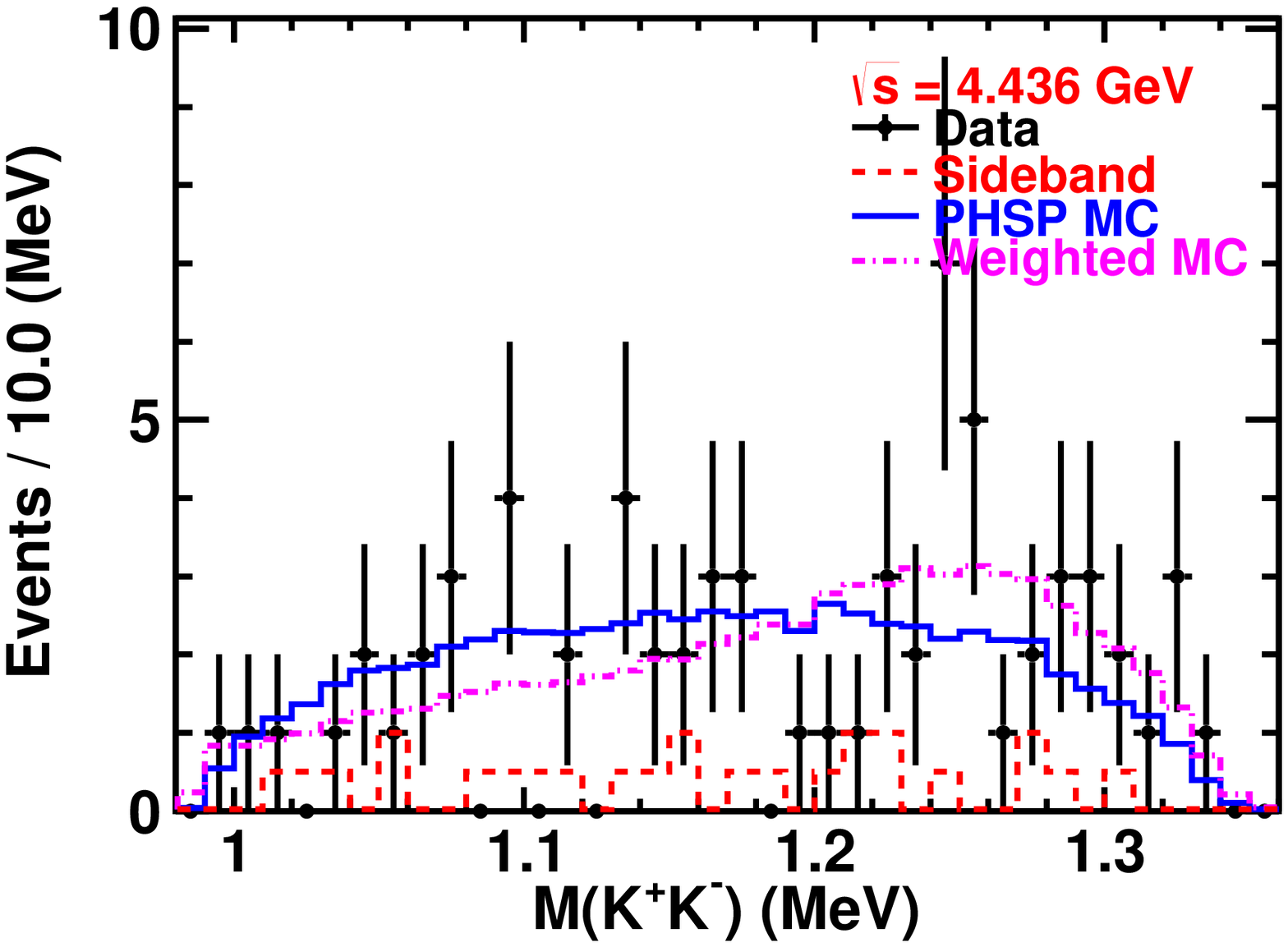}
 	\includegraphics[angle=0,width=0.24\textwidth]{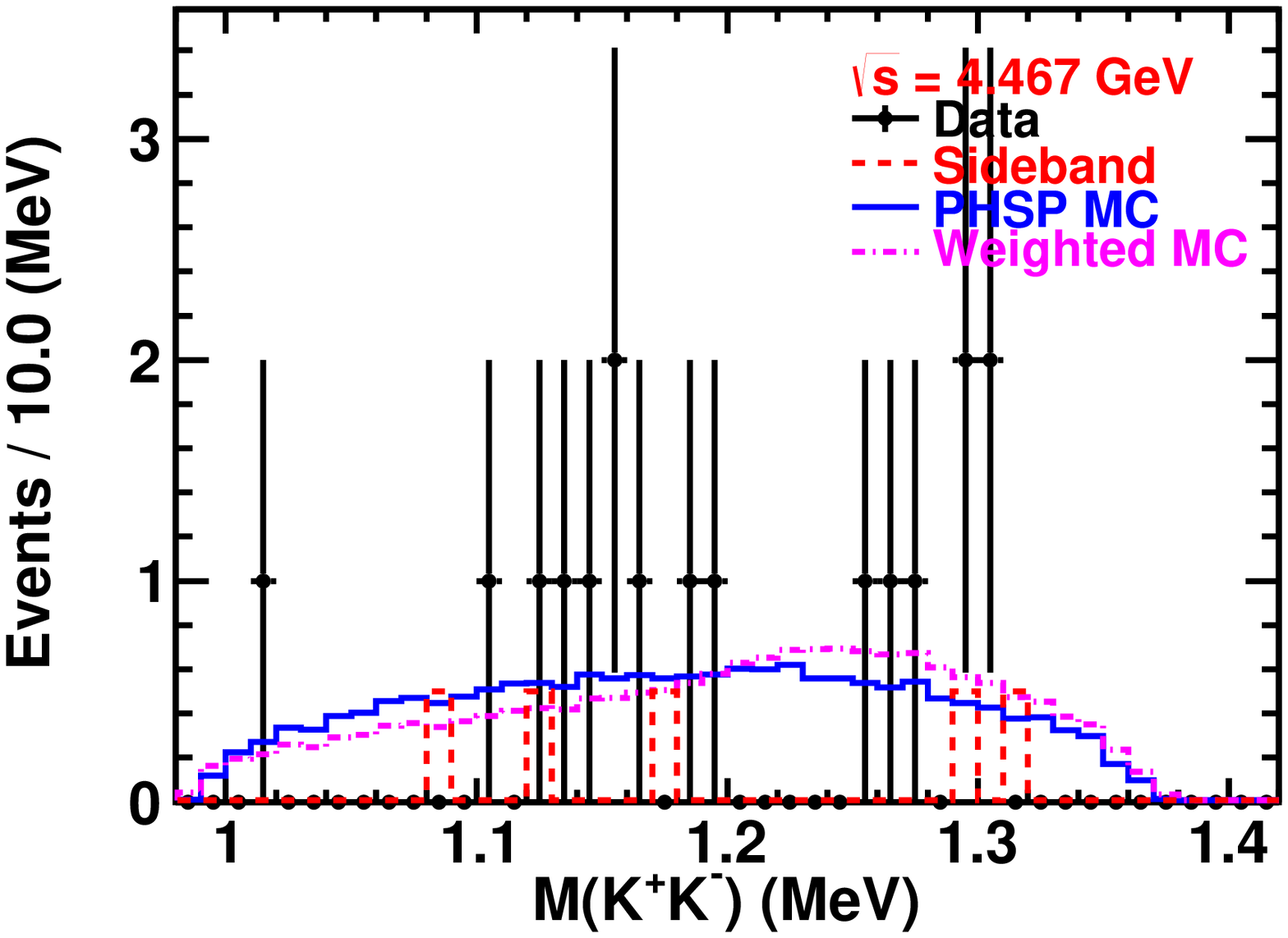}
	\includegraphics[angle=0,width=0.24\textwidth]{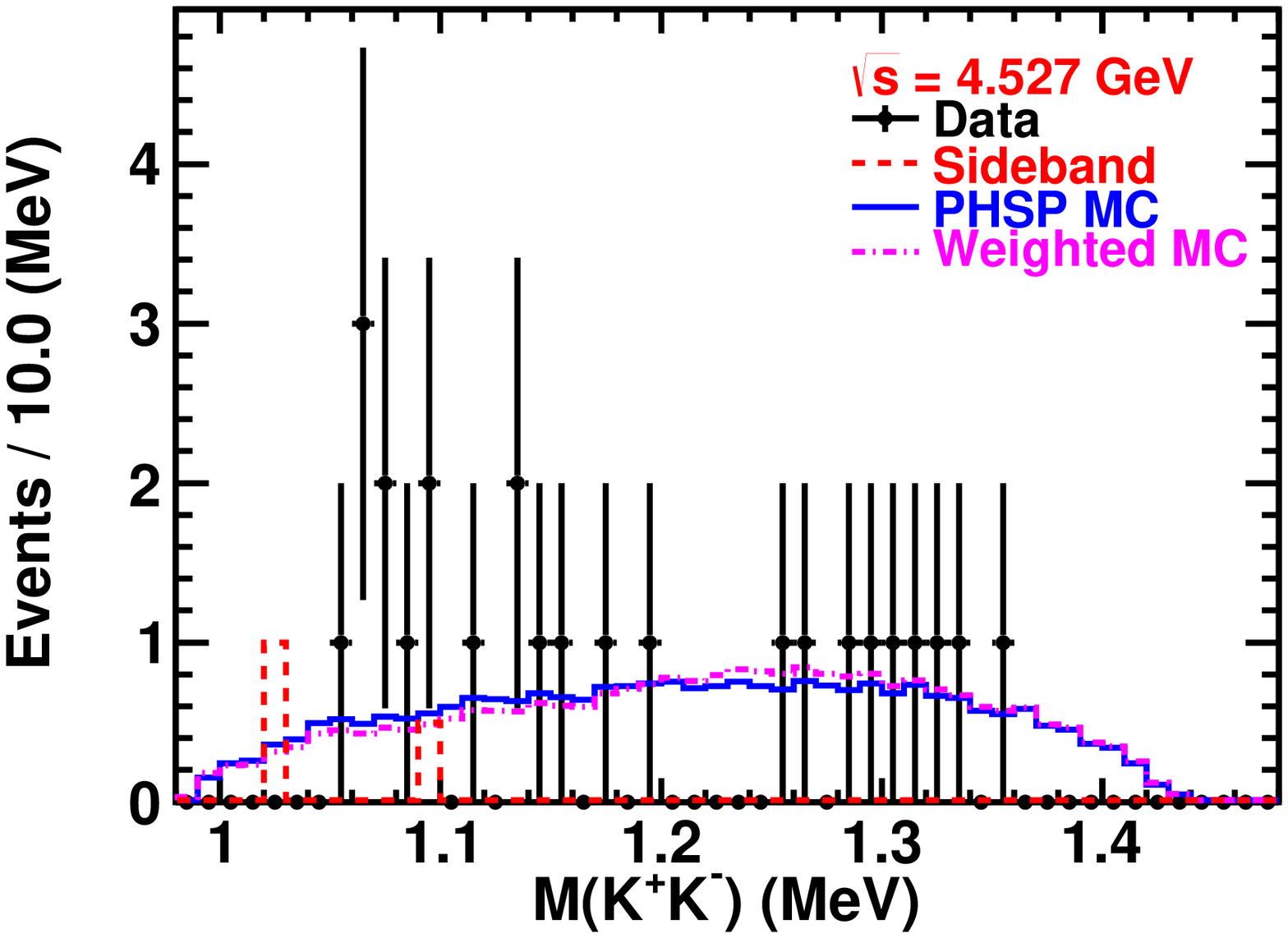}
	\includegraphics[angle=0,width=0.24\textwidth]{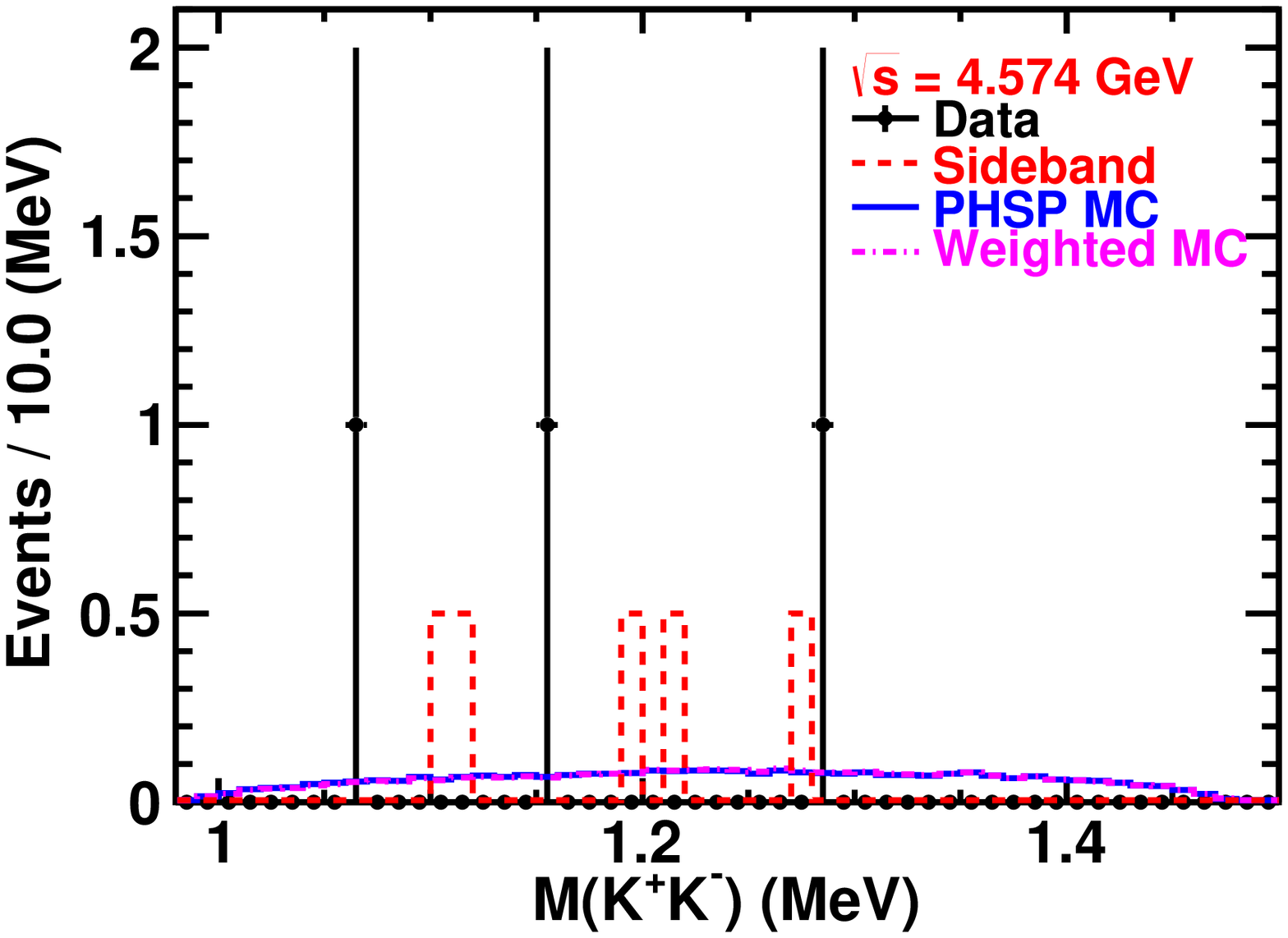}
	\includegraphics[angle=0,width=0.24\textwidth]{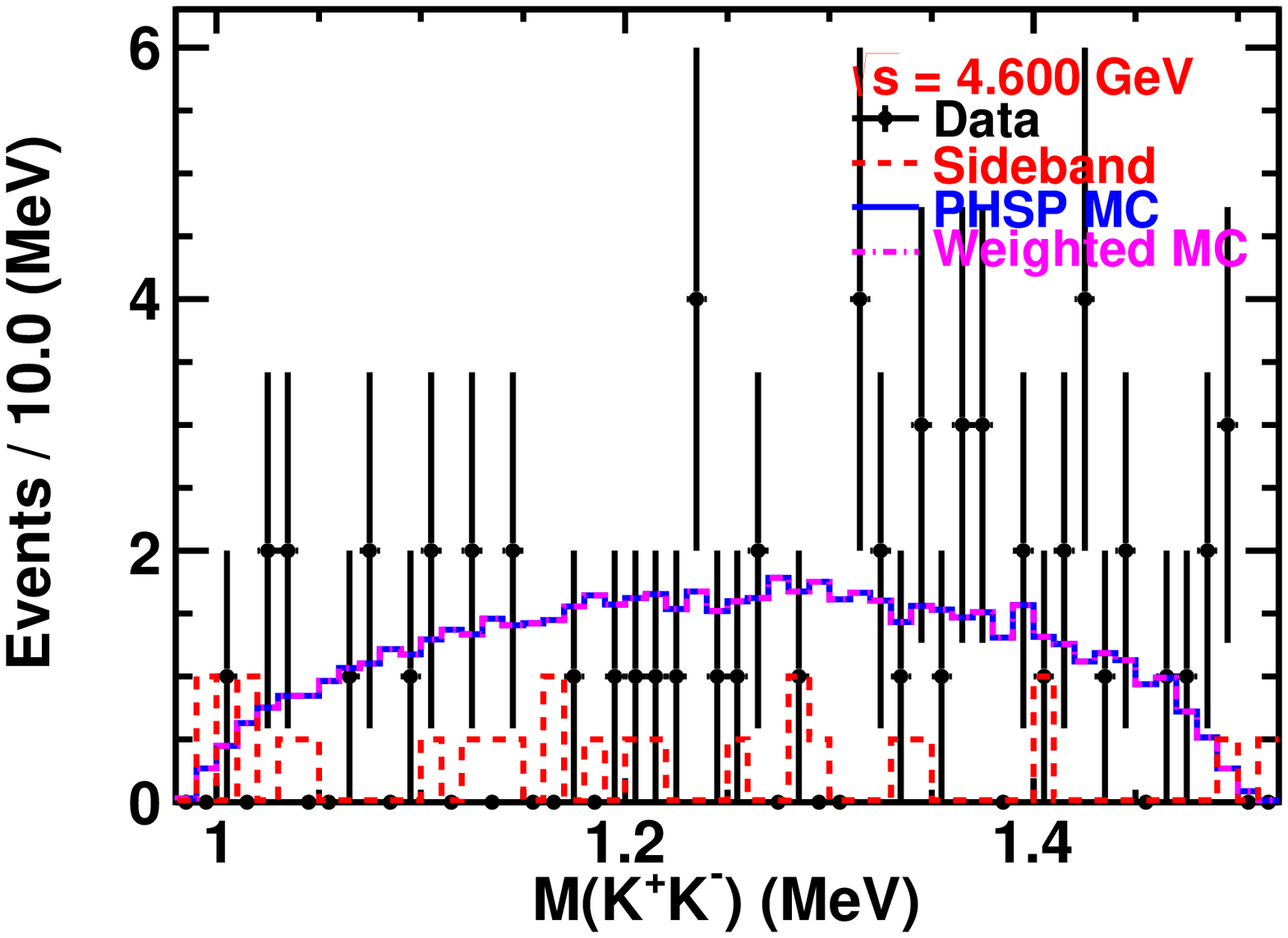}
 	\caption{The distributions of $K^+K^-$ invariant mass of the data, PHSP signal MC and weighted MC samples at each c.m. energy, where the black dots with error bars indicate data from the $J/\psi$ signal region, the red dashed curves indicate data from $J/\psi$ sideband regions, the blue histograms indicate PHSP signal MC sample (normalized to data) and the yank dashed-dot curves indicate the weighted signal MC sample (normalized to data). }
 	\label{Fig-MKK}
 \end{figure*}

Fig.~\ref{Fig-sig-yield} shows the distributions of the invariant mass of lepton pairs, $M(\ell^+\ell^-)$, for data and PHSP signal MC samples at various c.m. energies. Here, the $J/\psi$ signal and sideband regions are indicated with red and blue arrows, respectively. In addition, the distributions of invariant mass of kaons, $M(K^+K^-)$, for data, PHSP signal MC, and weighted MC samples at various c.m. energies are shown in Fig.~\ref{Fig-MKK}.

 \subsection{Born Cross Sections}
 \label{subs_a2}
 The Born cross sections of $e^{+}e^{-} \to K^{+}K^{-}J/\psi$ and related quantities such as the c.m. energy, the integrated luminosity, the numbers of observed signal events, the efficiencies, the radiative correction factors, the vacuum polarization factors are listed in Table.~\ref{Tab-Born-CS}.
 
 \begin{table*}[!htbp]
	\tableset
	\renewcommand{\arraystretch}{1.55}
	\caption{The Born cross sections of $e^{+}e^{-} \to K^{+}K^{-}J/\psi$ and related quantities. Here $\sqrt{s}$ is the c.m. energy, $\mathcal{L}_{int}$ is the integrated luminosity, $N^{obs}$ is the number of observed signal events, $\varepsilon$ is the efficiency, $(1+\delta)_{ISR}$ is the radiative correction factor, $\frac{1}{|1-\Pi|^2}$ is the vacuum polarization factor, and the $\sigma^{B}$ is the Born cross section, the first uncertainties are statistical, and the second systematic.}
	\label{Tab-Born-CS}
	\resizebox{\textwidth}{!}{
	\begin{tabular}{c cc cc cc c}
	\hline\hline
	Data sample	&	$\sqrt{s} $(GeV) & $\mathcal{L}_{int}$ (pb$^{-1}$)  & $N^{obs}$ & $\varepsilon$ & $(1+\delta)_{ISR}$ & $\frac{1}{|1-\Pi|^2}$ & $\sigma^{B}$(pb) \\ \hline
	4130   &  4.127    &  401.50   &  $6.5_{-2.4}^{+3.0}$ & 0.141    &  0.896    &  1.052    &  $1.02_{-0.38}^{+0.47} \pm 0.09$ ~\\ 
	4160   &  4.157    &  408.70   &  $6.5_{-2.4}^{+3.0}$ & 0.263    &  0.856    &  1.053    &  $0.56_{-0.21}^{+0.26} \pm 0.05$ ~\\ 
	4180   &  4.178    &  3194.50  &  $72.0_{-9.1}^{+9.7}$ &  0.315    &  0.820    &  1.054    &  $0.69_{-0.09}^{+0.09} \pm 0.06$ ~\\ 
	4190   &  4.189    &  570.10   &  $18.5_{-4.1}^{+4.7}$ &  0.335    &  0.792    &  1.056    &  $0.97_{-0.22}^{+0.25} \pm 0.08$ ~\\ 
	4200   &  4.199    &  526.00   &  $25.0_{-4.6}^{+5.2}$ &  0.357    &  0.787    &  1.056    &  $1.34_{-0.25}^{+0.28} \pm 0.12$ ~\\ 
	4210   &  4.209    &  572.10   &  $33.5_{-5.8}^{+6.4}$ &  0.358    &  0.762    &  1.057    &  $1.70_{-0.29}^{+0.33} \pm 0.15$ ~\\ 
	4220   &  4.219    &  569.20   &  $69.0_{-8.1}^{+8.7}$ &  0.379    &  0.771    &  1.056    &  $3.29_{-0.39}^{+0.41} \pm 0.37$ ~\\ 
	4230   &  4.226    &  1100.90  &  $183.0_{-13.3}^{+14.0}$ & 0.395    &  0.771    &  1.056    &  $4.33_{-0.31}^{+0.33} \pm 0.38$ ~\\ 
	4237   &  4.236    &  530.30   &  $65.5_{-8.3}^{+8.9}$ &  0.391    &  0.789    &  1.056    &  $3.18_{-0.40}^{+0.43} \pm 0.30$ ~\\ 
	4245   &  4.242    &  55.88    &  $8.0_{-2.5}^{+3.2}$ & 0.405    &  0.805    &  1.055    &  $3.49_{-1.09}^{+1.39} \pm 0.31$ ~\\ 
	4246   &  4.244    &  538.10   &  $74.5_{-8.5}^{+9.2}$ &  0.393    &  0.812    &  1.056    &  $3.44_{-0.39}^{+0.43} \pm 0.30$ ~\\ 
	4260   &  4.258    &  828.40   &  $107.5_{-10.3}^{+10.9}$ & 0.396    &  0.866    &  1.054    &  $3.01_{-0.29}^{+0.31} \pm 0.26$ ~\\ 
	4270   &  4.267    &  531.10   &  $45.0_{-6.7}^{+7.4}$ &  0.390    &  0.911    &  1.053    &  $1.90_{-0.28}^{+0.31} \pm 0.17$ ~\\ 
	4280   &  4.278    &  175.70   &  $16.5_{-4.0}^{+4.7}$ &  0.372    &  0.935    &  1.053    &  $2.15_{-0.52}^{+0.61} \pm 0.19$ ~\\ 
	4290   &  4.287    &  502.40   &  $35.0_{-6.3}^{+6.9}$ &  0.363    &  0.937    &  1.053    &  $1.63_{-0.29}^{+0.32} \pm 0.14$ ~\\ 
	4310   &  4.308    &  45.08    &  $3.0_{-1.4}^{+2.1}$ & 0.367    &  0.982    &  1.052    &  $1.47_{-0.69}^{+1.03} \pm 0.13$ ~\\ 
	4315   &  4.311    &  501.20   &  $24.0_{-5.6}^{+6.2}$ &  0.358    &  0.985    &  1.052    &  $1.08_{-0.25}^{+0.28} \pm 0.10$ ~\\ 
	4340   &  4.337    &  505.00   &  $18.5_{-5.6}^{+6.1}$ &  0.364    &  0.963    &  1.051    &  $0.83_{-0.25}^{+0.28} \pm 0.07$ ~\\ 
	4360   &  4.358    &  544.00   &  $26.5_{-5.7}^{+6.3}$ &  0.378    &  0.935    &  1.051    &  $1.10_{-0.24}^{+0.26} \pm 0.10$ ~\\ 
	4380   &  4.377    &  522.70   &  $33.0_{-7.1}^{+7.6}$ &  0.379    &  0.891    &  1.051    &  $1.49_{-0.32}^{+0.34} \pm 0.13$ ~\\ 
	4390   &  4.387    &  55.57    &  $3.0_{-1.4}^{+2.1}$ & 0.397    &  0.865    &  1.051    &  $1.25_{-0.58}^{+0.88} \pm 0.11$ ~\\ 
	4400   &  4.395    &  507.80   &  $16.0_{-6.1}^{+6.6}$ &  0.401    &  0.861    &  1.051    &  $0.73_{-0.28}^{+0.30} \pm 0.06$ ~\\ 
	4420   &  4.416    &  1090.70  &  $57.0_{-8.8}^{+9.4}$ &  0.417    &  0.815    &  1.052    &  $1.23_{-0.19}^{+0.20} \pm 0.11$ ~\\ 
	4440   &  4.436    &  569.90   &  $55.5_{-8.3}^{+8.9}$ &  0.436    &  0.780    &  1.054    &  $2.28_{-0.34}^{+0.37} \pm 0.20$ ~\\ 
	4470   &  4.467    &  111.09   &  $14.0_{-4.0}^{+4.6}$ &  0.450    &  0.738    &  1.055    &  $3.02_{-0.86}^{+0.99} \pm 0.26$ ~\\ 
	4530   &  4.527    &  112.12   &  $23.5_{-4.8}^{+5.4}$ &  0.460    &  0.751    &  1.054    &  $4.82_{-0.99}^{+1.11} \pm 0.41$ ~\\ 
	4575   &  4.574    &  48.93    &  $0.0_{-0.0}^{+2.3}$ & 0.430    &  0.856    &  1.054    &  $0.00_{-0.00}^{+1.01} \pm 0.00$ ~\\ 
	4600   &  4.600    &  586.90   &  $52.0_{-8.2}^{+8.7}$ &  0.400    &  0.899    &  1.055    &  $1.96_{-0.31}^{+0.33} \pm 0.17$ ~\\ 
	\hline \hline
	\end{tabular}
	}
\end{table*}

\subsection{Definition of likelihood function}
\label{subs_a3}
In the maximum likelihood fit to the dressed cross sections of $e^+e^- \rightarrow K^+ K^- J/\psi$, the likelihood is constructed as:

\begin{equation}
L(\mu; \vartheta) = \prod \limits_{i}^{6}L_i (\mu_i;~\vartheta_i)\prod \limits_j^{22}L_j (\mu_j; ~\vartheta_j),
\end{equation}
where $\mu_i$ and $\mu_j$ are the numbers of observed signal events, $\vartheta_i$ and $\vartheta_j$ are the parameters in the likelihood functions, and $L_i$ and $L_j$ are the likelihood functions for the data samples with $\mu \le 10$ and $\mu > 10$, respectively.

The likelihood functions are defined variously according to the different numbers of observed events. For the data samples with $\mu \le 10$, the likelihood function is described by a Poisson function: 
\begin{equation}
L_i(\mu_i; \vartheta_i) = P_i(\mu_i; \vartheta_i) = \frac{1}{\mu_i!}\vartheta_i^{\mu_i}e^{-\vartheta_i},
\end{equation}
While for the data samples with $\mu > 10$, the likelihood function is described by an asymmetric Gaussian function:
\begin{widetext}
\begin{equation}
	L_j (\mu_j, ~\sigma_{1j},~\sigma_{2j};~\vartheta_j)
	=  G(\mu_j,~\sigma_{1j},~\sigma_{2j};~\vartheta_j)
	= \left \{
	\begin{array}{rcl}
		\frac{1}{\sqrt{2\pi}~(\sigma_{1j} + \sigma_{2j})}~e^{-\frac{(\vartheta_j-\mu_j)^2}{2\sigma_{1j}^2}},	&	&{\vartheta_j > \mu_j;} ~\\~\\
		\frac{1}{\sqrt{2\pi}~(\sigma_{1j} + \sigma_{2j})}~e^{-\frac{(\vartheta_j-\mu_j)^2}{2\sigma_{2j}^2}},	&	&{\vartheta_j \leq \mu_j;} ~\\ ~\\
	\end{array}	
	\right.
\end{equation}
\end{widetext}
where $\sigma_{1}$ and $\sigma_{2}$ are the upper and lower statistical uncertainties of $\mu$, respectively. 
	
\subsection{The systematic uncertainties of the Born cross sections and the resonance parameters}\label{subs_a4}

\begin{figure*}[!htbp]
	\centering
	\begin{overpic}[scale=0.42]{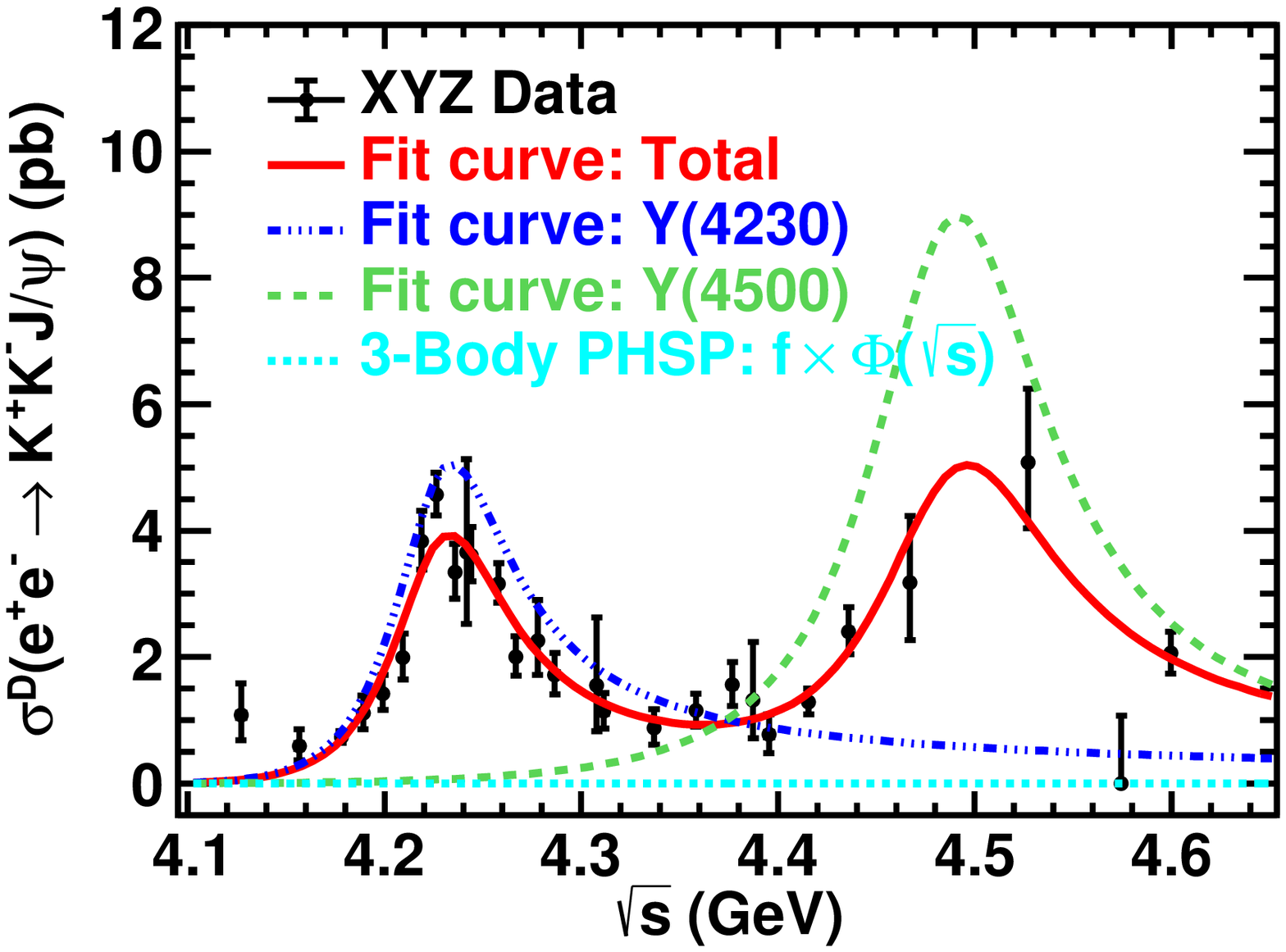}
	\put(75, 50){(a)}
	\end{overpic}
	\begin{overpic}[scale=0.42]{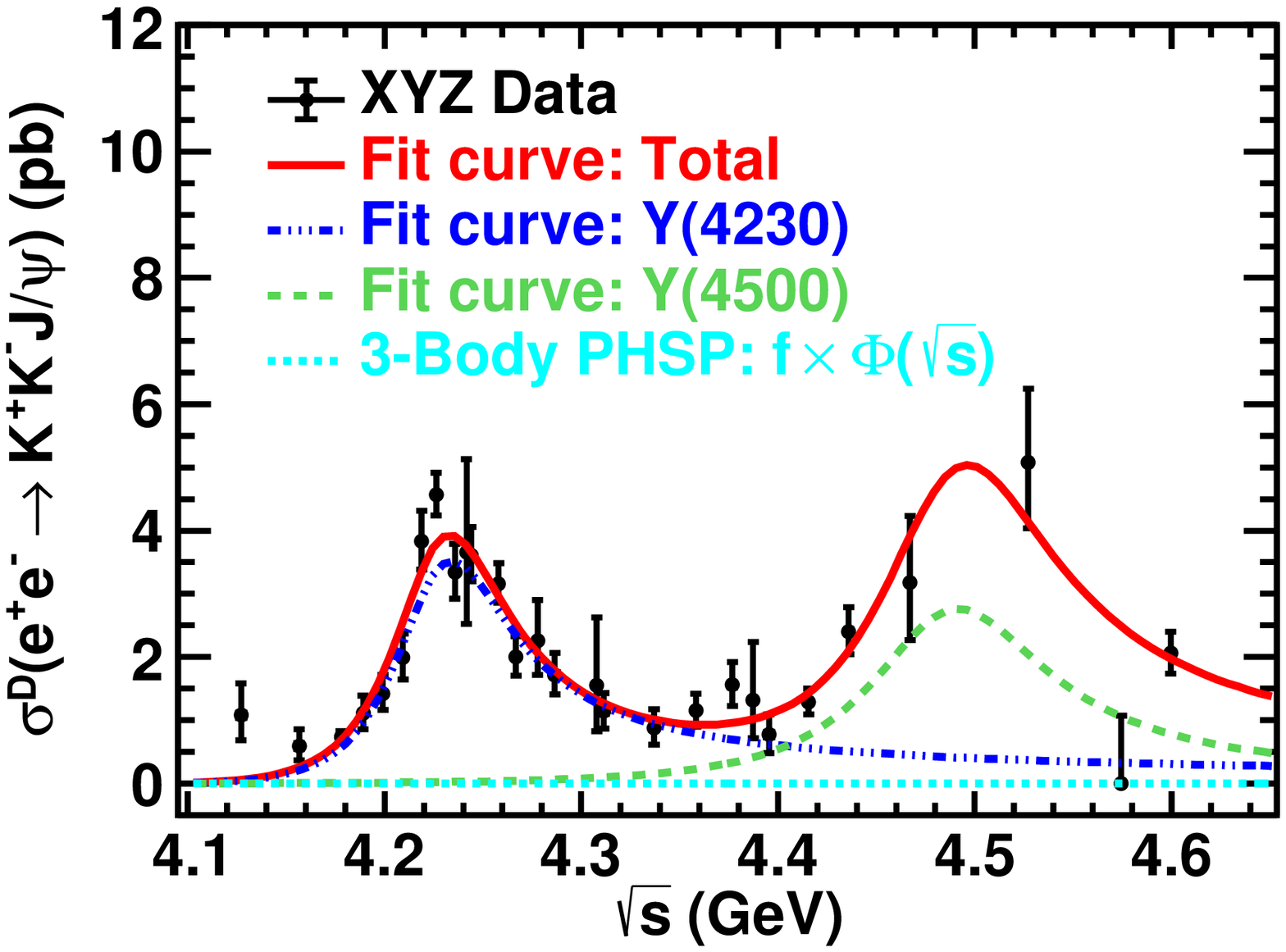}
	\put(75, 50){(b)}
	\end{overpic}
	\begin{overpic}[scale=0.42]{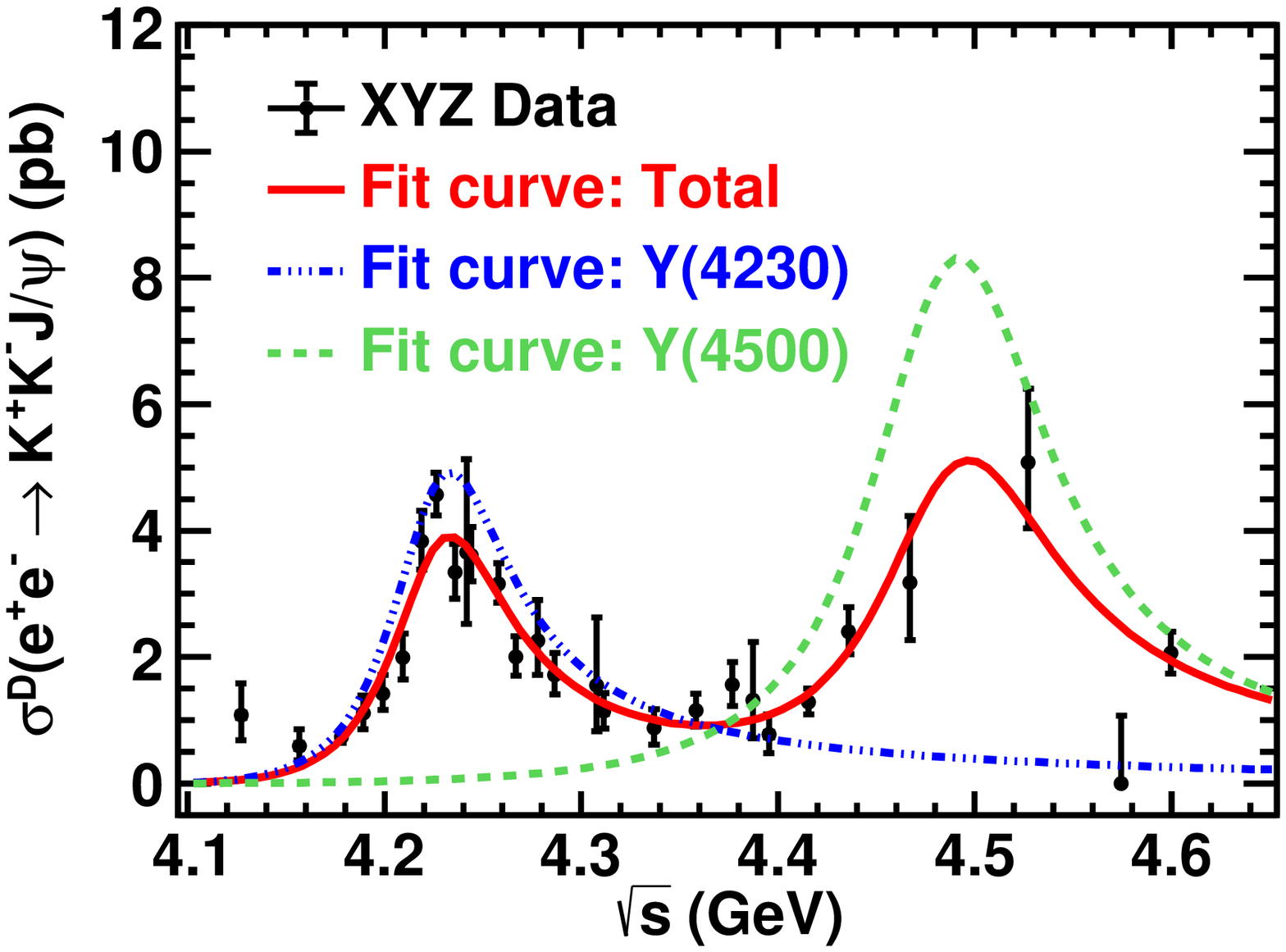}
	\put(75, 50){(c)}
	\end{overpic}
	\begin{overpic}[scale=0.42]{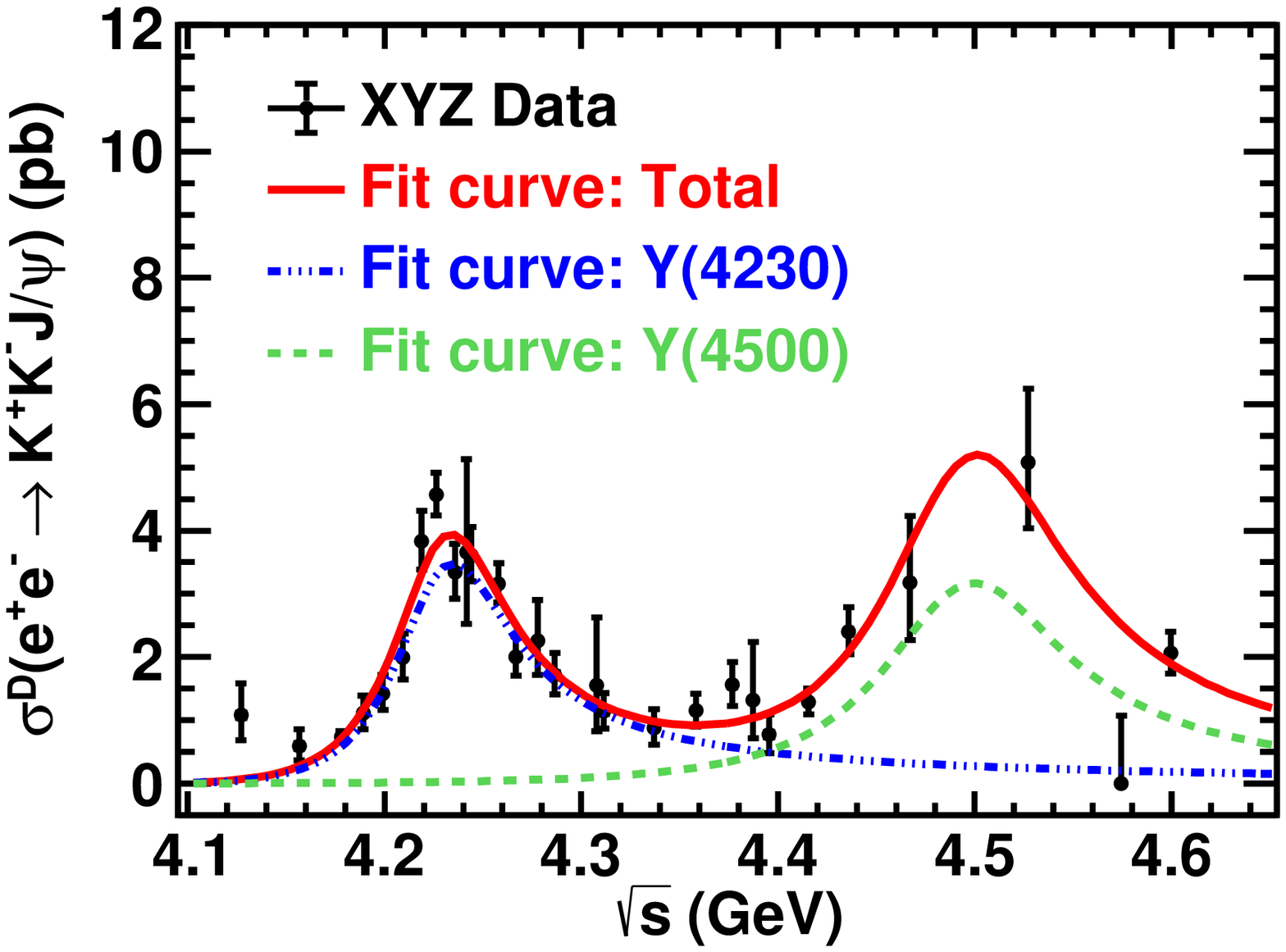}
	\put(75, 50){(d)}
	\end{overpic}
	\caption{The dressed cross sections fitting results with different function forms. (a), (b) are two sets of solutions to the sum of two coherent Breit-Wigner functions and 3-body phase space function fitting, respectively. (c), (d) are two sets of solutions to the two coherent Breit-Wigner functions with an energy-dependent full width fitting, respectively. }
	\label{Fig-sys-fit}
\end{figure*}

The systematic uncertainties of the Born cross sections are listed in Tab.~\ref{Tab-uncertainty} and the systematic uncertainties of the resonance parameters are listed in Tab.~\ref{Tab-Sys-resonance}. Fig.~\ref{Fig-sys-fit} shows dressed cross sections fitting results with different function forms. Generally, there should be four solutions when three coherent amplitudes are used to describe a lineshape of cross sections. However, only two have been found by us. We suppose that the other solutions are very close to the found ones, then cannot be separated by the scanning method. And one should notice that the fitting results, with the tentatively additional continuum term, only change slightly.

\begin{table*}[!htbp]
	\tableset
	\renewcommand{\arraystretch}{1.55}
	\caption{The systematic uncertainties (in units \%) in the measurement of Born cross sections. Here $\sqrt{s}$ is the c.m. energy of data samples, $\mathcal{L}_{int}$ is the integrated luminosity, $\mathcal{B}(J/\psi \rightarrow l^+l^-)$ is the branching fraction, $(1+\delta)_{ISR}$ is the radiation correction factor, $M(K^+K^-)$ is the intermediate structures in $K^+K^-$ system, $R(J/\psi)$ is the resolution of $J/\psi$, MUC is the criteria applied on the penetration depth in the muon counter.}
	\label{Tab-uncertainty}
	\resizebox{\textwidth}{!}{
	\begin{tabular}{c cc cc cc cc cc}
	\hline\hline
	Data sample	&	$\sqrt{s} $(GeV) & $\mathcal{L}_{int}$ & Tracking and PID & $\mathcal{B}(J/\psi \rightarrow l^+l^-)$ & kinematic fit	& $(1+\delta)_{ISR}$ & $M(K^+K^-)$ & $R(J/\psi)$ & MUC & Total  \\ \hline
	4130   & 4.127    & 1.0   & 4.5   & 0.4   & 0.4   & 1.2   & 6.9   & 0.3   & 2.1   & 8.7   ~\\ 
	4160   & 4.157    & 1.0   & 4.5   & 0.4   & 0.7   & 0.7   & 6.9   & 0.3   & 2.1   & 8.6   ~\\ 
	4180   & 4.178    & 1.0   & 4.5   & 0.4   & 0.9   & 2.6   & 6.9   & 0.2   & 2.1   & 9.0   ~\\ 
	4190   & 4.189    & 1.0   & 4.5   & 0.4   & 1.0   & 1.0   & 6.9   & 0.1   & 2.1   & 8.7   ~\\ 
	4200   & 4.199    & 1.0   & 4.5   & 0.4   & 1.1   & 0.6   & 6.9   & 0.3   & 2.1   & 8.7   ~\\ 
	4210   & 4.209    & 1.0   & 4.5   & 0.4   & 1.1   & 1.7   & 6.9   & 0.3   & 2.1   & 8.8   ~\\ 
	4220   & 4.219    & 1.0   & 4.5   & 0.4   & 1.1   & 6.9   & 6.9   & 0.2   & 2.1   & 11.1  ~\\ 
	4230   & 4.226    & 1.0   & 4.5   & 0.4   & 1.1   & 0.6   & 6.9   & 0.4   & 2.1   & 8.7   ~\\ 
	4237   & 4.236    & 1.0   & 4.5   & 0.4   & 1.2   & 3.5   & 6.9   & 0.4   & 2.1   & 9.3   ~\\ 
	4245   & 4.242    & 1.0   & 4.5   & 0.4   & 1.2   & 2.4   & 6.9   & 0.2   & 2.1   & 9.0   ~\\ 
	4246   & 4.244    & 1.0   & 4.5   & 0.4   & 1.2   & 1.5   & 6.9   & 0.2   & 2.1   & 8.8   ~\\ 
	4260   & 4.258    & 1.0   & 4.5   & 0.4   & 1.3   & 1.3   & 6.9   & 0.3   & 2.1   & 8.8   ~\\
	4270   & 4.267    & 1.0   & 4.5   & 0.4   & 1.3   & 2.7   & 6.9   & 0.3   & 2.1   & 9.1   ~\\ 
	4280   & 4.278    & 1.0   & 4.5   & 0.4   & 1.3   & 1.3   & 6.9   & 0.4   & 2.1   & 8.8   ~\\ 
	4290   & 4.287    & 1.0   & 4.5   & 0.4   & 1.2   & 1.2   & 6.9   & 0.3   & 2.1   & 8.7   ~\\ 
	4310   & 4.308    & 1.0   & 4.5   & 0.4   & 1.6   & 1.3   & 6.9   & 0.4   & 2.1   & 8.8   ~\\ 
	4315   & 4.312    & 1.0   & 4.5   & 0.4   & 1.6   & 0.7   & 6.9   & 0.4   & 2.1   & 8.8   ~\\ 
	4340   & 4.337    & 1.0   & 4.5   & 0.4   & 1.6   & 1.5   & 6.9   & 0.7   & 2.1   & 8.9   ~\\ 
	4360   & 4.358    & 1.0   & 4.5   & 0.4   & 1.4   & 0.5   & 6.9   & 0.5   & 2.1   & 8.7   ~\\ 
	4380   & 4.377    & 1.0   & 4.5   & 0.4   & 1.4   & 0.3   & 6.9   & 0.6   & 2.1   & 8.7   ~\\ 
	4390   & 4.387    & 1.0   & 4.5   & 0.4   & 1.2   & 1.0   & 6.9   & 0.5   & 2.1   & 8.7   ~\\ 
	4400   & 4.395    & 1.0   & 4.5   & 0.4   & 1.1   & 0.9   & 6.9   & 0.5   & 2.1   & 8.7   ~\\ 
	4420   & 4.416    & 1.0   & 4.5   & 0.4   & 0.9   & 1.0   & 6.9   & 0.7   & 2.1   & 8.7   ~\\ 
	4440   & 4.436    & 1.0   & 4.5   & 0.4   & 0.7   & 1.6   & 6.9   & 0.6   & 2.1   & 8.8   ~\\ 
	4470   & 4.467    & 1.0   & 4.5   & 0.4   & 0.4   & 0.2   & 6.9   & 0.4   & 2.1   & 8.6   ~\\ 
	4530   & 4.527    & 1.0   & 4.5   & 0.4   & 0.3   & 0.9   & 6.9   & 0.6   & 2.1   & 8.6   ~\\ 
	4575   & 4.574    & 1.0   & 4.5   & 0.4   & 0.2   & 0.0   & 6.9   & 0.4   & 2.1   & 8.6   ~\\ 
	4600   & 4.600    & 1.0   & 4.5   & 0.4   & 0.2   & 1.2   & 6.9   & 0.3   & 2.1   & 8.7   ~\\ 
	\hline\hline
	\end{tabular} }
\end{table*}

\begin{table*}[!htbp]
	\tableset
	\renewcommand{\arraystretch}{1.55}
	\caption{The systematic uncertainty in the measurement of the resonance parameters for solution I (solution II), where Non-Resonant indicated a three-body PHSP shape is added  in the default fit and considering the coherent between them, the cross section$_1$ is uncorrelated systematic uncertainties from the measurement of cross sections, while cross section$_2$ is correlated.}
	\label{Tab-Sys-resonance}
	\resizebox{\textwidth}{!}{
	\begin{tabular}{c|ccc| ccc|c}
	\hline\hline
	\multirow{2}{*}{\diagbox{Sources}{Parameters}} & \multicolumn{3}{c|}{$Y(4230)$} & \multicolumn{3}{c|}{$Y(4500)$}	& phase angle ~\\ \cline{2-8}
	~	&	$M$ (MeV)	&	$\Gamma_{tot}$ (MeV)	&	$\Gamma_{ee}\mathcal{B}$ (eV)	& $M$ (MeV)	&	$\Gamma_{tot}$ (MeV)	&	$\Gamma_{ee}\mathcal{B}$ (eV)	& $\varphi$ (rad)	~\\ \hline
	c.m. Energy		&	0.7		&	0.1		&	-- 				&	1.8		&	1.4 		&	-- 			& 0.00 (0.02)	~\\
	Non-Resonant		&	--		&	--		&	--				&	--		&	--		&	-- 			& 0.13 (0.51)	~\\
	$\Gamma_{tot}$	&	21.5		&	30.8		&	0.15 (0.10)		&	24.1		&	15.1		&	0.02 (0.13) 	& 0.50 (0.28)	~\\
	cross section$_1$	&	--		&	--		&	-- (--)			&	0.1 		&	0.1 		&	-- (--)		& -- 		~\\ 
	cross section$_2$	&	--		&	-- 		&	0.02 (0.01)		&	-- 		&	-- 		&	0.07 (0.02)	& --			~\\\hline \hline
	Total				&	21.5		&	30.8		&	0.15 (0.10)		&	24.1		&	15.2		&	0.07 (0.13)	& 0.52 (0.58)	~\\ \hline \hline
	\end{tabular}
	}
\end{table*}

\end{appendix}

\bibliography{apssamp}

\begin{thebibliography}{51}%
\makeatletter
\providecommand \@ifxundefined [1]{%
 \@ifx{#1\undefined}
}%
\providecommand \@ifnum [1]{%
 \ifnum #1\expandafter \@firstoftwo
 \else \expandafter \@secondoftwo
 \fi
}%
\providecommand \@ifx [1]{%
 \ifx #1\expandafter \@firstoftwo
 \else \expandafter \@secondoftwo
 \fi
}%
\providecommand \natexlab [1]{#1}%
\providecommand \enquote  [1]{``#1''}%
\providecommand \bibnamefont  [1]{#1}%
\providecommand \bibfnamefont [1]{#1}%
\providecommand \citenamefont [1]{#1}%
\providecommand \href@noop [0]{\@secondoftwo}%
\providecommand \href [0]{\begingroup \@sanitize@url \@href}%
\providecommand \@href[1]{\@@startlink{#1}\@@href}%
\providecommand \@@href[1]{\endgroup#1\@@endlink}%
\providecommand \@sanitize@url [0]{\catcode `\\12\catcode `\$12\catcode
  `\&12\catcode `\#12\catcode `\^12\catcode `\_12\catcode `\%12\relax}%
\providecommand \@@startlink[1]{}%
\providecommand \@@endlink[0]{}%
\providecommand \url  [0]{\begingroup\@sanitize@url \@url }%
\providecommand \@url [1]{\endgroup\@href {#1}{\urlprefix }}%
\providecommand \urlprefix  [0]{URL }%
\providecommand \Eprint [0]{\href }%
\providecommand \doibase [0]{https://doi.org/}%
\providecommand \selectlanguage [0]{\@gobble}%
\providecommand \bibinfo  [0]{\@secondoftwo}%
\providecommand \bibfield  [0]{\@secondoftwo}%
\providecommand \translation [1]{[#1]}%
\providecommand \BibitemOpen [0]{}%
\providecommand \bibitemStop [0]{}%
\providecommand \bibitemNoStop [0]{.\EOS\space}%
\providecommand \EOS [0]{\spacefactor3000\relax}%
\providecommand \BibitemShut  [1]{\csname bibitem#1\endcsname}%
\let\auto@bib@innerbib\@empty
\bibitem [{\citenamefont {Brambilla}\ \emph {et~al.}(2020)\citenamefont
  {Brambilla} \emph {et~al.}}]{BRAMBILLA20201}%
  \BibitemOpen
  \bibfield  {author} {\bibinfo {author} {\bibfnamefont {N.}~\bibnamefont
  {Brambilla}} \emph {et~al.},\ }\href
  {https://doi.org/https://doi.org/10.1016/j.physrep.2020.05.001} {\bibfield
  {journal} {\bibinfo  {journal} {Phys.\ Rept}\ }\textbf {\bibinfo {volume}
  {873}},\ \bibinfo {pages} {1} (\bibinfo {year} {2020})}\BibitemShut {NoStop}%
\bibitem [{\citenamefont {Aubert}\ \emph {et~al.}(2005)\citenamefont {Aubert}
  \emph {et~al.}}]{PhysRevLett.95.142001}%
  \BibitemOpen
  \bibfield  {author} {\bibinfo {author} {\bibfnamefont {B.}~\bibnamefont
  {Aubert}} \emph {et~al.} (\bibinfo {collaboration} {BABAR Collaboration}),\
  }\href {https://doi.org/10.1103/PhysRevLett.95.142001} {\bibfield  {journal}
  {\bibinfo  {journal} {Phys.\ Rev.\ Lett.}\ }\textbf {\bibinfo {volume}
  {95}},\ \bibinfo {pages} {142001} (\bibinfo {year} {2005})}\BibitemShut
  {NoStop}%
\bibitem [{\citenamefont {He}\ \emph {et~al.}(2006)\citenamefont {He} \emph
  {et~al.}}]{PhysRevD.74.091104}%
  \BibitemOpen
  \bibfield  {author} {\bibinfo {author} {\bibfnamefont {Q.}~\bibnamefont {He}}
  \emph {et~al.} (\bibinfo {collaboration} {CLEO Collaboration}),\ }\href
  {https://doi.org/10.1103/PhysRevD.74.091104} {\bibfield  {journal} {\bibinfo
  {journal} {Phys.\ Rev.\ D}\ }\textbf {\bibinfo {volume} {74}},\ \bibinfo
  {pages} {091104} (\bibinfo {year} {2006})}\BibitemShut {NoStop}%
\bibitem [{\citenamefont {Yuan}\ \emph {et~al.}(2007)\citenamefont {Yuan} \emph
  {et~al.}}]{PhysRevLett.99.182004}%
  \BibitemOpen
  \bibfield  {author} {\bibinfo {author} {\bibfnamefont {C.~Z.}\ \bibnamefont
  {Yuan}} \emph {et~al.} (\bibinfo {collaboration} {Belle Collaboration}),\
  }\href {https://doi.org/10.1103/PhysRevLett.99.182004} {\bibfield  {journal}
  {\bibinfo  {journal} {Phys. Rev. Lett.}\ }\textbf {\bibinfo {volume} {99}},\
  \bibinfo {pages} {182004} (\bibinfo {year} {2007})}\BibitemShut {NoStop}%
\bibitem [{\citenamefont {Ablikim}\ \emph
  {et~al.}(2017{\natexlab{a}})\citenamefont {Ablikim} \emph
  {et~al.}}]{PhysRevLett.118.092001}%
  \BibitemOpen
  \bibfield  {author} {\bibinfo {author} {\bibfnamefont {M.}~\bibnamefont
  {Ablikim}} \emph {et~al.} (\bibinfo {collaboration} {BESIII Collaboration}),\
  }\href {https://doi.org/10.1103/PhysRevLett.118.092001} {\bibfield  {journal}
  {\bibinfo  {journal} {Phys. Rev. Lett.}\ }\textbf {\bibinfo {volume} {118}},\
  \bibinfo {pages} {092001} (\bibinfo {year} {2017}{\natexlab{a}})}\BibitemShut
  {NoStop}%
\bibitem [{\citenamefont {Ablikim}\ \emph
  {et~al.}(2020{\natexlab{a}})\citenamefont {Ablikim} \emph
  {et~al.}}]{PhysRevD.102.012009}%
  \BibitemOpen
  \bibfield  {author} {\bibinfo {author} {\bibfnamefont {M.}~\bibnamefont
  {Ablikim}} \emph {et~al.} (\bibinfo {collaboration} {BESIII Collaboration}),\
  }\href {https://doi.org/10.1103/PhysRevD.102.012009} {\bibfield  {journal}
  {\bibinfo  {journal} {Phys. Rev. D}\ }\textbf {\bibinfo {volume} {102}},\
  \bibinfo {pages} {012009} (\bibinfo {year} {2020}{\natexlab{a}})}\BibitemShut
  {NoStop}%
\bibitem [{\citenamefont {Ablikim}\ \emph
  {et~al.}(2022{\natexlab{a}})\citenamefont {Ablikim} \emph
  {et~al.}}]{BESIII:2022jsj}%
  \BibitemOpen
  \bibfield  {author} {\bibinfo {author} {\bibfnamefont {M.}~\bibnamefont
  {Ablikim}} \emph {et~al.} (\bibinfo {collaboration} {BESIII Collaboration}),\
  }\href@noop {} {\  (\bibinfo {year} {2022}{\natexlab{a}})},\ \Eprint
  {https://arxiv.org/abs/2206.08554} {arXiv:2206.08554 [hep-ex]} \BibitemShut
  {NoStop}%
\bibitem [{\citenamefont {Ablikim}\ \emph
  {et~al.}(2017{\natexlab{b}})\citenamefont {Ablikim} \emph
  {et~al.}}]{PhysRevLett.118.092002}%
  \BibitemOpen
  \bibfield  {author} {\bibinfo {author} {\bibfnamefont {M.}~\bibnamefont
  {Ablikim}} \emph {et~al.} (\bibinfo {collaboration} {BESIII Collaboration}),\
  }\href {https://doi.org/10.1103/PhysRevLett.118.092002} {\bibfield  {journal}
  {\bibinfo  {journal} {Phys. Rev. Lett.}\ }\textbf {\bibinfo {volume} {118}},\
  \bibinfo {pages} {092002} (\bibinfo {year} {2017}{\natexlab{b}})}\BibitemShut
  {NoStop}%
\bibitem [{\citenamefont {Ablikim}\ \emph
  {et~al.}(2017{\natexlab{c}})\citenamefont {Ablikim} \emph
  {et~al.}}]{PhysRevD.96.032004}%
  \BibitemOpen
  \bibfield  {author} {\bibinfo {author} {\bibfnamefont {M.}~\bibnamefont
  {Ablikim}} \emph {et~al.} (\bibinfo {collaboration} {BESIII Collaboration}),\
  }\href {https://doi.org/10.1103/PhysRevD.96.032004} {\bibfield  {journal}
  {\bibinfo  {journal} {Phys. Rev. D}\ }\textbf {\bibinfo {volume} {96}},\
  \bibinfo {pages} {032004} (\bibinfo {year} {2017}{\natexlab{c}})}\BibitemShut
  {NoStop}%
\bibitem [{\citenamefont {Ablikim}\ \emph
  {et~al.}(2018{\natexlab{a}})\citenamefont {Ablikim} \emph
  {et~al.}}]{PhysRevD.97.052001}%
  \BibitemOpen
  \bibfield  {author} {\bibinfo {author} {\bibfnamefont {M.}~\bibnamefont
  {Ablikim}} \emph {et~al.} (\bibinfo {collaboration} {BESIII Collaboration}),\
  }\href {https://doi.org/10.1103/PhysRevD.97.052001} {\bibfield  {journal}
  {\bibinfo  {journal} {Phys. Rev. D}\ }\textbf {\bibinfo {volume} {97}},\
  \bibinfo {pages} {052001} (\bibinfo {year} {2018}{\natexlab{a}})}\BibitemShut
  {NoStop}%
\bibitem [{\citenamefont {Ablikim}\ \emph
  {et~al.}(2021{\natexlab{a}})\citenamefont {Ablikim} \emph
  {et~al.}}]{PhysRevD.104.052012}%
  \BibitemOpen
  \bibfield  {author} {\bibinfo {author} {\bibfnamefont {M.}~\bibnamefont
  {Ablikim}} \emph {et~al.} (\bibinfo {collaboration} {BESIII Collaboration}),\
  }\href {https://doi.org/10.1103/PhysRevD.104.052012} {\bibfield  {journal}
  {\bibinfo  {journal} {Phys. Rev. D}\ }\textbf {\bibinfo {volume} {104}},\
  \bibinfo {pages} {052012} (\bibinfo {year} {2021}{\natexlab{a}})}\BibitemShut
  {NoStop}%
\bibitem [{\citenamefont {Ablikim}\ \emph
  {et~al.}(2015{\natexlab{a}})\citenamefont {Ablikim} \emph
  {et~al.}}]{PhysRevLett.114.092003}%
  \BibitemOpen
  \bibfield  {author} {\bibinfo {author} {\bibfnamefont {M.}~\bibnamefont
  {Ablikim}} \emph {et~al.} (\bibinfo {collaboration} {BESIII Collaboration}),\
  }\href {https://doi.org/10.1103/PhysRevLett.114.092003} {\bibfield  {journal}
  {\bibinfo  {journal} {Phys. Rev. Lett.}\ }\textbf {\bibinfo {volume} {114}},\
  \bibinfo {pages} {092003} (\bibinfo {year} {2015}{\natexlab{a}})}\BibitemShut
  {NoStop}%
\bibitem [{\citenamefont {Ablikim}\ \emph
  {et~al.}(2019{\natexlab{a}})\citenamefont {Ablikim} \emph
  {et~al.}}]{PhysRevD.99.091103}%
  \BibitemOpen
  \bibfield  {author} {\bibinfo {author} {\bibfnamefont {M.}~\bibnamefont
  {Ablikim}} \emph {et~al.} (\bibinfo {collaboration} {BESIII Collaboration}),\
  }\href {https://doi.org/10.1103/PhysRevD.99.091103} {\bibfield  {journal}
  {\bibinfo  {journal} {Phys. Rev. D}\ }\textbf {\bibinfo {volume} {99}},\
  \bibinfo {pages} {091103} (\bibinfo {year} {2019}{\natexlab{a}})}\BibitemShut
  {NoStop}%
\bibitem [{\citenamefont {Ablikim}\ \emph
  {et~al.}(2019{\natexlab{b}})\citenamefont {Ablikim} \emph
  {et~al.}}]{PhysRevLett.122.102002}%
  \BibitemOpen
  \bibfield  {author} {\bibinfo {author} {\bibfnamefont {M.}~\bibnamefont
  {Ablikim}} \emph {et~al.} (\bibinfo {collaboration} {BESIII Collaboration}),\
  }\href {https://doi.org/10.1103/PhysRevLett.122.102002} {\bibfield  {journal}
  {\bibinfo  {journal} {Phys. Rev. Lett.}\ }\textbf {\bibinfo {volume} {122}},\
  \bibinfo {pages} {102002} (\bibinfo {year} {2019}{\natexlab{b}})}\BibitemShut
  {NoStop}%
\bibitem [{\citenamefont {Ablikim}\ \emph
  {et~al.}(2019{\natexlab{c}})\citenamefont {Ablikim} \emph
  {et~al.}}]{PhysRevLett.122.232002}%
  \BibitemOpen
  \bibfield  {author} {\bibinfo {author} {\bibfnamefont {M.}~\bibnamefont
  {Ablikim}} \emph {et~al.} (\bibinfo {collaboration} {BESIII Collaboration}),\
  }\href {https://doi.org/10.1103/PhysRevLett.122.232002} {\bibfield  {journal}
  {\bibinfo  {journal} {Phys. Rev. Lett.}\ }\textbf {\bibinfo {volume} {122}},\
  \bibinfo {pages} {232002} (\bibinfo {year} {2019}{\natexlab{c}})}\BibitemShut
  {NoStop}%
\bibitem [{\citenamefont {Zhu}(2021)}]{doi:10.1142/S0217751X21501268}%
  \BibitemOpen
  \bibfield  {author} {\bibinfo {author} {\bibfnamefont {K.}~\bibnamefont
  {Zhu}},\ }\href {https://doi.org/10.1142/S0217751X21501268} {\bibfield
  {journal} {\bibinfo  {journal} {Inter. Jour. Mod. Phys. A}\ }\textbf
  {\bibinfo {volume} {36}},\ \bibinfo {pages} {2150126} (\bibinfo {year}
  {2021})}\BibitemShut {NoStop}%
\bibitem [{\citenamefont {Ablikim}\ \emph
  {et~al.}(2020{\natexlab{b}})\citenamefont {Ablikim} \emph
  {et~al.}}]{PhysRevD.102.031101}%
  \BibitemOpen
  \bibfield  {author} {\bibinfo {author} {\bibfnamefont {M.}~\bibnamefont
  {Ablikim}} \emph {et~al.} (\bibinfo {collaboration} {BESIII Collaboration}),\
  }\href {https://doi.org/10.1103/PhysRevD.102.031101} {\bibfield  {journal}
  {\bibinfo  {journal} {Phys. Rev. D}\ }\textbf {\bibinfo {volume} {102}},\
  \bibinfo {pages} {031101} (\bibinfo {year} {2020}{\natexlab{b}})}\BibitemShut
  {NoStop}%
\bibitem [{\citenamefont {Ablikim}\ \emph
  {et~al.}(2020{\natexlab{c}})\citenamefont {Ablikim} \emph
  {et~al.}}]{PhysRevD.101.012008}%
  \BibitemOpen
  \bibfield  {author} {\bibinfo {author} {\bibfnamefont {M.}~\bibnamefont
  {Ablikim}} \emph {et~al.} (\bibinfo {collaboration} {BESIII Collaboration}),\
  }\href {https://doi.org/10.1103/PhysRevD.101.012008} {\bibfield  {journal}
  {\bibinfo  {journal} {Phys. Rev. D}\ }\textbf {\bibinfo {volume} {101}},\
  \bibinfo {pages} {012008} (\bibinfo {year} {2020}{\natexlab{c}})}\BibitemShut
  {NoStop}%
\bibitem [{\citenamefont {Zhu}(2022)}]{PhysRevD.105.L031506}%
  \BibitemOpen
  \bibfield  {author} {\bibinfo {author} {\bibfnamefont {K.}~\bibnamefont
  {Zhu}},\ }\href {https://doi.org/10.1103/PhysRevD.105.L031506} {\bibfield
  {journal} {\bibinfo  {journal} {Phys. Rev. D}\ }\textbf {\bibinfo {volume}
  {105}},\ \bibinfo {pages} {L031506} (\bibinfo {year} {2022})}\BibitemShut
  {NoStop}%
\bibitem [{\citenamefont {Coan}\ \emph {et~al.}(2006)\citenamefont {Coan} \emph
  {et~al.}}]{PhysRevLett.96.162003}%
  \BibitemOpen
  \bibfield  {author} {\bibinfo {author} {\bibfnamefont {T.~E.}\ \bibnamefont
  {Coan}} \emph {et~al.} (\bibinfo {collaboration} {CLEO Collaboration}),\
  }\href {https://doi.org/10.1103/PhysRevLett.96.162003} {\bibfield  {journal}
  {\bibinfo  {journal} {Phys. Rev. Lett.}\ }\textbf {\bibinfo {volume} {96}},\
  \bibinfo {pages} {162003} (\bibinfo {year} {2006})}\BibitemShut {NoStop}%
\bibitem [{\citenamefont {Yuan}\ \emph {et~al.}(2008)\citenamefont {Yuan} \emph
  {et~al.}}]{PhysRevD.77.011105}%
  \BibitemOpen
  \bibfield  {author} {\bibinfo {author} {\bibfnamefont {C.~Z.}\ \bibnamefont
  {Yuan}} \emph {et~al.} (\bibinfo {collaboration} {Belle Collaboration}),\
  }\href {https://doi.org/10.1103/PhysRevD.77.011105} {\bibfield  {journal}
  {\bibinfo  {journal} {Phys. Rev. D}\ }\textbf {\bibinfo {volume} {77}},\
  \bibinfo {pages} {011105} (\bibinfo {year} {2008})}\BibitemShut {NoStop}%
\bibitem [{\citenamefont {Shen}\ \emph {et~al.}(2014)\citenamefont {Shen} \emph
  {et~al.}}]{PhysRevD.89.072015}%
  \BibitemOpen
  \bibfield  {author} {\bibinfo {author} {\bibfnamefont {C.~P.}\ \bibnamefont
  {Shen}} \emph {et~al.} (\bibinfo {collaboration} {Belle Collaboration}),\
  }\href {https://doi.org/10.1103/PhysRevD.89.072015} {\bibfield  {journal}
  {\bibinfo  {journal} {Phys. Rev. D}\ }\textbf {\bibinfo {volume} {89}},\
  \bibinfo {pages} {072015} (\bibinfo {year} {2014})}\BibitemShut {NoStop}%
\bibitem [{\citenamefont {Ablikim}\ \emph
  {et~al.}(2018{\natexlab{b}})\citenamefont {Ablikim} \emph
  {et~al.}}]{PhysRevD.97.071101}%
  \BibitemOpen
  \bibfield  {author} {\bibinfo {author} {\bibfnamefont {M.}~\bibnamefont
  {Ablikim}} \emph {et~al.} (\bibinfo {collaboration} {BESIII Collaboration}),\
  }\href {https://doi.org/10.1103/PhysRevD.97.071101} {\bibfield  {journal}
  {\bibinfo  {journal} {Phys. Rev. D}\ }\textbf {\bibinfo {volume} {97}},\
  \bibinfo {pages} {071101} (\bibinfo {year} {2018}{\natexlab{b}})}\BibitemShut
  {NoStop}%
\bibitem [{\citenamefont {Wang}\ \emph {et~al.}(2019)\citenamefont {Wang},
  \citenamefont {Chen}, \citenamefont {Liu},\ and\ \citenamefont
  {Matsuki}}]{PhysRevD.99.114003}%
  \BibitemOpen
  \bibfield  {author} {\bibinfo {author} {\bibfnamefont {J.~Z.}\ \bibnamefont
  {Wang}}, \bibinfo {author} {\bibfnamefont {D.~Y.}\ \bibnamefont {Chen}},
  \bibinfo {author} {\bibfnamefont {X.}~\bibnamefont {Liu}},\ and\ \bibinfo
  {author} {\bibfnamefont {T.}~\bibnamefont {Matsuki}},\ }\href
  {https://doi.org/10.1103/PhysRevD.99.114003} {\bibfield  {journal} {\bibinfo
  {journal} {Phys. Rev. D}\ }\textbf {\bibinfo {volume} {99}},\ \bibinfo
  {pages} {114003} (\bibinfo {year} {2019})}\BibitemShut {NoStop}%
\bibitem [{\citenamefont {Dong}\ \emph {et~al.}(2021)\citenamefont {Dong},
  \citenamefont {Guo},\ and\ \citenamefont {Zou}}]{Dong:2021juy}%
  \BibitemOpen
  \bibfield  {author} {\bibinfo {author} {\bibfnamefont {X.~K.}\ \bibnamefont
  {Dong}}, \bibinfo {author} {\bibfnamefont {F.~K.}\ \bibnamefont {Guo}},\ and\
  \bibinfo {author} {\bibfnamefont {B.~S.}\ \bibnamefont {Zou}},\ }\href
  {https://doi.org/10.13725/j.cnki.pip.2021.02.001} {\bibfield  {journal}
  {\bibinfo  {journal} {Progr. Phys.}\ }\textbf {\bibinfo {volume} {41}},\
  \bibinfo {pages} {65} (\bibinfo {year} {2021})},\ \Eprint
  {https://arxiv.org/abs/2101.01021} {arXiv:2101.01021 [hep-ph]} \BibitemShut
  {NoStop}%
\bibitem [{\citenamefont {Chiu}\ and\ \citenamefont
  {Hsieh}(2006)}]{PhysRevD.73.094510}%
  \BibitemOpen
  \bibfield  {author} {\bibinfo {author} {\bibfnamefont {T.~W.}\ \bibnamefont
  {Chiu}}\ and\ \bibinfo {author} {\bibfnamefont {T.~H.}\ \bibnamefont {Hsieh}}
  (\bibinfo {collaboration} {TWQCD Collaboration}),\ }\href
  {https://doi.org/10.1103/PhysRevD.73.094510} {\bibfield  {journal} {\bibinfo
  {journal} {Phys. Rev. D}\ }\textbf {\bibinfo {volume} {73}},\ \bibinfo
  {pages} {094510} (\bibinfo {year} {2006})}\BibitemShut {NoStop}%
\bibitem [{\citenamefont {Ablikim}\ \emph {et~al.}(2016)\citenamefont {Ablikim}
  \emph {et~al.}}]{Ablikim_2016}%
  \BibitemOpen
  \bibfield  {author} {\bibinfo {author} {\bibfnamefont {M.}~\bibnamefont
  {Ablikim}} \emph {et~al.} (\bibinfo {collaboration} {BESIII Collaboration}),\
  }\href {https://doi.org/10.1088/1674-1137/40/6/063001} {\bibfield  {journal}
  {\bibinfo  {journal} {Chin. Phys. C}\ }\textbf {\bibinfo {volume} {40}},\
  \bibinfo {pages} {063001} (\bibinfo {year} {2016})}\BibitemShut {NoStop}%
\bibitem [{\citenamefont {Ablikim}\ \emph
  {et~al.}(2021{\natexlab{b}})\citenamefont {Ablikim} \emph
  {et~al.}}]{BESIII:2020eyu}%
  \BibitemOpen
  \bibfield  {author} {\bibinfo {author} {\bibfnamefont {M.}~\bibnamefont
  {Ablikim}} \emph {et~al.} (\bibinfo {collaboration} {BESIII Collaboration}),\
  }\href {https://doi.org/10.1088/1674-1137/ac1575} {\bibfield  {journal}
  {\bibinfo  {journal} {Chin. Phys. C}\ }\textbf {\bibinfo {volume} {45}},\
  \bibinfo {pages} {103001} (\bibinfo {year} {2021}{\natexlab{b}})}\BibitemShut
  {NoStop}%
\bibitem [{\citenamefont {Ablikim}\ \emph
  {et~al.}(2015{\natexlab{b}})\citenamefont {Ablikim} \emph
  {et~al.}}]{Ablikim_2015}%
  \BibitemOpen
  \bibfield  {author} {\bibinfo {author} {\bibfnamefont {M.}~\bibnamefont
  {Ablikim}} \emph {et~al.} (\bibinfo {collaboration} {BESIII Collaboration}),\
  }\href {https://doi.org/10.1088/1674-1137/39/9/093001} {\bibfield  {journal}
  {\bibinfo  {journal} {Chin.\ Phys.\ C}\ }\textbf {\bibinfo {volume} {39}},\
  \bibinfo {pages} {093001} (\bibinfo {year} {2015}{\natexlab{b}})}\BibitemShut
  {NoStop}%
\bibitem [{\citenamefont {Ablikim}\ \emph
  {et~al.}(2022{\natexlab{b}})\citenamefont {Ablikim} \emph
  {et~al.}}]{BESIII:2022xii}%
  \BibitemOpen
  \bibfield  {author} {\bibinfo {author} {\bibfnamefont {M.}~\bibnamefont
  {Ablikim}} \emph {et~al.} (\bibinfo {collaboration} {{BESIII
  Collaboration}}),\ }\href@noop {} {} (\bibinfo {year} {2022}{\natexlab{b}}),\
  \Eprint {https://arxiv.org/abs/2203.03133} {arXiv:2203.03133 [hep-ex]}
  \BibitemShut {NoStop}%
\bibitem [{\citenamefont {Ablikim}\ \emph
  {et~al.}(2020{\natexlab{d}})\citenamefont {Ablikim} \emph
  {et~al.}}]{BESIII:2020nme}%
  \BibitemOpen
  \bibfield  {author} {\bibinfo {author} {\bibfnamefont {M.}~\bibnamefont
  {Ablikim}} \emph {et~al.} (\bibinfo {collaboration} {{BESIII
  Collaboration}}),\ }\href {https://doi.org/10.1088/1674-1137/44/4/040001}
  {\bibfield  {journal} {\bibinfo  {journal} {Chin. Phys. C}\ }\textbf
  {\bibinfo {volume} {44}},\ \bibinfo {pages} {040001} (\bibinfo {year}
  {2020}{\natexlab{d}})}\BibitemShut {NoStop}%
\bibitem [{\citenamefont {Ablikim}\ \emph {et~al.}(2010)\citenamefont {Ablikim}
  \emph {et~al.}}]{ABLIKIM2010345}%
  \BibitemOpen
  \bibfield  {author} {\bibinfo {author} {\bibfnamefont {M.}~\bibnamefont
  {Ablikim}} \emph {et~al.} (\bibinfo {collaboration} {{BESIII
  Collaboration}}),\ }\href
  {https://doi.org/https://doi.org/10.1016/j.nima.2009.12.050} {\bibfield
  {journal} {\bibinfo  {journal} {Nucl.\ Instrum.\ Meth.\ A}\ }\textbf
  {\bibinfo {volume} {614}},\ \bibinfo {pages} {345} (\bibinfo {year}
  {2010})}\BibitemShut {NoStop}%
\bibitem [{\citenamefont {Agostinelli}\ \emph {et~al.}(2003)\citenamefont
  {Agostinelli} \emph {et~al.}}]{AGOSTINELLI2003250}%
  \BibitemOpen
  \bibfield  {author} {\bibinfo {author} {\bibfnamefont {S.}~\bibnamefont
  {Agostinelli}} \emph {et~al.},\ }\href
  {https://doi.org/https://doi.org/10.1016/S0168-9002(03)01368-8} {\bibfield
  {journal} {\bibinfo  {journal} {Nucl.\ Instrum.\ Meth.\ A}\ }\textbf
  {\bibinfo {volume} {506}},\ \bibinfo {pages} {250} (\bibinfo {year}
  {2003})}\BibitemShut {NoStop}%
\bibitem [{\citenamefont {Deng}\ \emph {et~al.}(2006)\citenamefont {Deng} \emph
  {et~al.}}]{2005-0159}%
  \BibitemOpen
  \bibfield  {author} {\bibinfo {author} {\bibfnamefont {Z.~Y.}\ \bibnamefont
  {Deng}} \emph {et~al.},\ }\href
  {http://hepnp.ihep.ac.cn//article/id/283d17c0-e8fa-4ad7-bfe3-92095466def1}
  {\bibfield  {journal} {\bibinfo  {journal} {Chin.\ Phys.\ C}\ }\textbf
  {\bibinfo {volume} {30}},\ \bibinfo {pages} {371} (\bibinfo {year}
  {2006})}\BibitemShut {NoStop}%
\bibitem [{\citenamefont {Jadach}\ \emph {et~al.}(2001)\citenamefont {Jadach},
  \citenamefont {Ward},\ and\ \citenamefont {Was}}]{PhysRevD.63.113009}%
  \BibitemOpen
  \bibfield  {author} {\bibinfo {author} {\bibfnamefont {S.}~\bibnamefont
  {Jadach}}, \bibinfo {author} {\bibfnamefont {B.~F.~L.}\ \bibnamefont
  {Ward}},\ and\ \bibinfo {author} {\bibfnamefont {Z.}~\bibnamefont {Was}},\
  }\href {https://doi.org/10.1103/PhysRevD.63.113009} {\bibfield  {journal}
  {\bibinfo  {journal} {Phys.\ Rev.\ D}\ }\textbf {\bibinfo {volume} {63}},\
  \bibinfo {pages} {113009} (\bibinfo {year} {2001})}\BibitemShut {NoStop}%
\bibitem [{\citenamefont {Ping}(2014)}]{2014-08-01-083001}%
  \BibitemOpen
  \bibfield  {author} {\bibinfo {author} {\bibfnamefont {R.~G.}\ \bibnamefont
  {Ping}},\ }\href {https://doi.org/10.1088/1674-1137/38/8/083001} {\bibfield
  {journal} {\bibinfo  {journal} {Chin.\ Phys.\ C}\ }\textbf {\bibinfo {volume}
  {38}},\ \bibinfo {pages} {083001} (\bibinfo {year} {2014})}\BibitemShut
  {NoStop}%
\bibitem [{\citenamefont {Ping}(2008)}]{2007-0205}%
  \BibitemOpen
  \bibfield  {author} {\bibinfo {author} {\bibfnamefont {R.~G.}\ \bibnamefont
  {Ping}},\ }\href {https://doi.org/10.1088/1674-1137/32/8/001} {\bibfield
  {journal} {\bibinfo  {journal} {Chin.\ Phys.\ C}\ }\textbf {\bibinfo {volume}
  {32}},\ \bibinfo {pages} {599} (\bibinfo {year} {2008})}\BibitemShut
  {NoStop}%
\bibitem [{\citenamefont {Lange}(2001)}]{LANGE2001152}%
  \BibitemOpen
  \bibfield  {author} {\bibinfo {author} {\bibfnamefont {D.~J.}\ \bibnamefont
  {Lange}},\ }\href
  {https://doi.org/https://doi.org/10.1016/S0168-9002(01)00089-4} {\bibfield
  {journal} {\bibinfo  {journal} {Nucl.\ Instrum.\ Meth.\ A}\ }\textbf
  {\bibinfo {volume} {462}},\ \bibinfo {pages} {152} (\bibinfo {year}
  {2001})}\BibitemShut {NoStop}%
\bibitem [{\citenamefont {Golonka}\ and\ \citenamefont
  {Was}(2006)}]{Golonka2006}%
  \BibitemOpen
  \bibfield  {author} {\bibinfo {author} {\bibfnamefont {P.}~\bibnamefont
  {Golonka}}\ and\ \bibinfo {author} {\bibfnamefont {Z.}~\bibnamefont {Was}},\
  }\href {https://doi.org/10.1140/epjc/s2005-02396-4} {\bibfield  {journal}
  {\bibinfo  {journal} {Eur.\ Phys.\ J.\ C}\ ,\ \bibinfo {pages} {97}}
  (\bibinfo {year} {2006})}\BibitemShut {NoStop}%
\bibitem [{\citenamefont {Zyla}\ \emph {et~al.}(2020)\citenamefont {Zyla} \emph
  {et~al.}}]{Zyla:2020zbs}%
  \BibitemOpen
  \bibfield  {author} {\bibinfo {author} {\bibfnamefont {P.}~\bibnamefont
  {Zyla}} \emph {et~al.} (\bibinfo {collaboration} {Particle Data Group}),\
  }\href {https://doi.org/10.1093/ptep/ptaa104} {\bibfield  {journal} {\bibinfo
   {journal} {PTEP}\ }\textbf {\bibinfo {volume} {2020}},\ \bibinfo {pages}
  {083C01} (\bibinfo {year} {2020})}\BibitemShut {NoStop}%
\bibitem [{\citenamefont {Chen}\ \emph {et~al.}(2000)\citenamefont {Chen},
  \citenamefont {Huang}, \citenamefont {Qi}, \citenamefont {Zhang},\ and\
  \citenamefont {Zhu}}]{PhysRevD.62.034003}%
  \BibitemOpen
  \bibfield  {author} {\bibinfo {author} {\bibfnamefont {J.~C.}\ \bibnamefont
  {Chen}}, \bibinfo {author} {\bibfnamefont {G.~S.}\ \bibnamefont {Huang}},
  \bibinfo {author} {\bibfnamefont {X.~R.}\ \bibnamefont {Qi}}, \bibinfo
  {author} {\bibfnamefont {D.~H.}\ \bibnamefont {Zhang}},\ and\ \bibinfo
  {author} {\bibfnamefont {Y.~S.}\ \bibnamefont {Zhu}},\ }\href
  {https://doi.org/10.1103/PhysRevD.62.034003} {\bibfield  {journal} {\bibinfo
  {journal} {Phys. Rev. D}\ }\textbf {\bibinfo {volume} {62}},\ \bibinfo
  {pages} {034003} (\bibinfo {year} {2000})}\BibitemShut {NoStop}%
\bibitem [{\citenamefont {Yang}\ \emph {et~al.}(2014)\citenamefont {Yang},
  \citenamefont {Ping},\ and\ \citenamefont {Chen}}]{YANG-Rui-Ling:61301}%
  \BibitemOpen
  \bibfield  {author} {\bibinfo {author} {\bibfnamefont {R.~L.}\ \bibnamefont
  {Yang}}, \bibinfo {author} {\bibfnamefont {R.~G.}\ \bibnamefont {Ping}},\
  and\ \bibinfo {author} {\bibfnamefont {H.}~\bibnamefont {Chen}},\ }\href
  {https://doi.org/10.1088/0256-307X/31/6/061301} {\bibfield  {journal}
  {\bibinfo  {journal} {Chin. Phys. Lett.}\ }\textbf {\bibinfo {volume} {31}},\
  \bibinfo {eid} {061301} (\bibinfo {year} {2014})}\BibitemShut {NoStop}%
\bibitem [{\citenamefont {Rolke}\ \emph {et~al.}(2005)\citenamefont {Rolke},
  \citenamefont {López},\ and\ \citenamefont {Conrad}}]{ROLKE2005493}%
  \BibitemOpen
  \bibfield  {author} {\bibinfo {author} {\bibfnamefont {W.~A.}\ \bibnamefont
  {Rolke}}, \bibinfo {author} {\bibfnamefont {A.~M.}\ \bibnamefont {López}},\
  and\ \bibinfo {author} {\bibfnamefont {J.}~\bibnamefont {Conrad}},\ }\href
  {https://doi.org/https://doi.org/10.1016/j.nima.2005.05.068} {\bibfield
  {journal} {\bibinfo  {journal} {Nucl.\ Instrum.\ Meth.\ A}\ }\textbf
  {\bibinfo {volume} {551}},\ \bibinfo {pages} {493} (\bibinfo {year}
  {2005})}\BibitemShut {NoStop}%
\bibitem [{\citenamefont {Danilkin}\ \emph {et~al.}(2020)\citenamefont
  {Danilkin}, \citenamefont {Molnar},\ and\ \citenamefont
  {Vanderhaeghen}}]{PhysRevD.102.016019}%
  \BibitemOpen
  \bibfield  {author} {\bibinfo {author} {\bibfnamefont {I.}~\bibnamefont
  {Danilkin}}, \bibinfo {author} {\bibfnamefont {D.~A.~S.}\ \bibnamefont
  {Molnar}},\ and\ \bibinfo {author} {\bibfnamefont {M.}~\bibnamefont
  {Vanderhaeghen}},\ }\href {https://doi.org/10.1103/PhysRevD.102.016019}
  {\bibfield  {journal} {\bibinfo  {journal} {Phys. Rev. D}\ }\textbf {\bibinfo
  {volume} {102}},\ \bibinfo {pages} {016019} (\bibinfo {year}
  {2020})}\BibitemShut {NoStop}%
\bibitem [{\citenamefont {Actis}\ \emph {et~al.}(2010)\citenamefont {Actis}
  \emph {et~al.}}]{Actis2010}%
  \BibitemOpen
  \bibfield  {author} {\bibinfo {author} {\bibfnamefont {S.}~\bibnamefont
  {Actis}} \emph {et~al.} (\bibinfo {collaboration} {Working Group on Radiative
  Corrections and Monte Carlo Generators for Low Energies}),\ }\href
  {https://doi.org/10.1140/epjc/s10052-010-1251-4} {\bibfield  {journal}
  {\bibinfo  {journal} {Eur.\ Phys.\ Jour.\ C}\ }\textbf {\bibinfo {volume}
  {66}},\ \bibinfo {pages} {585} (\bibinfo {year} {2010})}\BibitemShut
  {NoStop}%
\bibitem [{\citenamefont {Sun}\ \emph {et~al.}(2021)\citenamefont {Sun} \emph
  {et~al.}}]{sun2021iterative}%
  \BibitemOpen
  \bibfield  {author} {\bibinfo {author} {\bibfnamefont {W.}~\bibnamefont
  {Sun}} \emph {et~al.},\ }\href@noop {} {} (\bibinfo {year} {2021}),\ \Eprint
  {https://arxiv.org/abs/2011.07889} {arXiv:2011.07889 [hep-ex]} \BibitemShut
  {NoStop}%
\bibitem [{\citenamefont {Wilks}(1938)}]{Wilks:1938dza}%
  \BibitemOpen
  \bibfield  {author} {\bibinfo {author} {\bibfnamefont {S.~S.}\ \bibnamefont
  {Wilks}},\ }\href {https://doi.org/10.1214/aoms/1177732360} {\bibfield
  {journal} {\bibinfo  {journal} {Annals Math. Statist.}\ }\textbf {\bibinfo
  {volume} {9}},\ \bibinfo {pages} {60} (\bibinfo {year} {1938})}\BibitemShut
  {NoStop}%
\bibitem [{\citenamefont {Zhu}\ \emph {et~al.}(2011)\citenamefont {Zhu},
  \citenamefont {Mo}, \citenamefont {Yuan},\ and\ \citenamefont
  {Wang}}]{S0217751X11054589}%
  \BibitemOpen
  \bibfield  {author} {\bibinfo {author} {\bibfnamefont {K.}~\bibnamefont
  {Zhu}}, \bibinfo {author} {\bibfnamefont {X.~H.}\ \bibnamefont {Mo}},
  \bibinfo {author} {\bibfnamefont {C.~Z.}\ \bibnamefont {Yuan}},\ and\
  \bibinfo {author} {\bibfnamefont {P.}~\bibnamefont {Wang}},\ }\href
  {https://doi.org/10.1142/S0217751X11054589} {\bibfield  {journal} {\bibinfo
  {journal} {Int.\ J.\ Mod.\ Phys.\ A}\ }\textbf {\bibinfo {volume} {26}},\
  \bibinfo {pages} {4511} (\bibinfo {year} {2011})}\BibitemShut {NoStop}%
\bibitem [{\citenamefont {Ablikim}\ \emph
  {et~al.}(2015{\natexlab{c}})\citenamefont {Ablikim} \emph
  {et~al.}}]{PhysRevD.91.112005}%
  \BibitemOpen
  \bibfield  {author} {\bibinfo {author} {\bibfnamefont {M.}~\bibnamefont
  {Ablikim}} \emph {et~al.} (\bibinfo {collaboration} {BESIII Collaboration}),\
  }\href {https://doi.org/10.1103/PhysRevD.91.112005} {\bibfield  {journal}
  {\bibinfo  {journal} {Phys. Rev. D}\ }\textbf {\bibinfo {volume} {91}},\
  \bibinfo {pages} {112005} (\bibinfo {year} {2015}{\natexlab{c}})}\BibitemShut
  {NoStop}%
\bibitem [{\citenamefont {Ablikim}\ \emph {et~al.}(2013)\citenamefont {Ablikim}
  \emph {et~al.}}]{PhysRevD.87.012002}%
  \BibitemOpen
  \bibfield  {author} {\bibinfo {author} {\bibfnamefont {M.}~\bibnamefont
  {Ablikim}} \emph {et~al.} (\bibinfo {collaboration} {BESIII Collaboration}),\
  }\href {https://doi.org/10.1103/PhysRevD.87.012002} {\bibfield  {journal}
  {\bibinfo  {journal} {Phys. Rev. D}\ }\textbf {\bibinfo {volume} {87}},\
  \bibinfo {pages} {012002} (\bibinfo {year} {2013})}\BibitemShut {NoStop}%
\bibitem [{\citenamefont {Qiao}(2006)}]{QIAO2006263}%
  \BibitemOpen
  \bibfield  {author} {\bibinfo {author} {\bibfnamefont {C.~F.}\ \bibnamefont
  {Qiao}},\ }\href
  {https://doi.org/https://doi.org/10.1016/j.physletb.2006.06.038} {\bibfield
  {journal} {\bibinfo  {journal} {Phys. Lett. B}\ }\textbf {\bibinfo {volume}
  {639}},\ \bibinfo {pages} {263} (\bibinfo {year} {2006})}\BibitemShut
  {NoStop}%
\end{thebibliography}%

\end{document}